\documentclass[10pt]{article}

\usepackage{amsmath,amsfonts,epsfig,paperyannis}

\newcommand{\ggain}{\hbox{${\cal G}$}}
\def\dt{{\delta t}}
\def\dx{{\delta x}}

\begin{document}

\title{An equation-free computational approach for extracting 
population-level behavior from individual-based models 
of biological dispersal}

\author{
  Radek Erban\thanks{School of Mathematics, 
  University of Minnesota, Minneapolis, MN 55455, USA; {\bf present address:}
Mathematical Institute, University of Oxford, 24-29 St. Giles',
Oxford, OX1 3LB, United Kingdom, {\it e-mail:  erban@maths.ox.ac.uk}.
Research supported in part by NSF grant
DMS 0317372 and the Minnesota Supercomputing Institute.}
\and
Ioannis G. Kevrekidis\thanks{Princeton University,
Department Of Chemical Engineering, PACM \& Mathematics,
Engineering Quadrangle,
Olden Street, Princeton, NJ 08544, USA;
{\it e-mail: yannis@princeton.edu}.  Research  supported in part
by  an NSF/ITR grant (CTS 0205484).}
\and
Hans G. Othmer\thanks{School of Mathematics and Digital Technology Center,
	 University of Minnesota, Minneapolis, MN 55455, USA;
{\it e-mail: othmer@math.umn.edu}. Research supported in part by NIH grant GM
        29123, NSF grant DMS 0317372, and the  Minnesota Supercomputing Institute.} 
}

\date{\today}

\maketitle 

{\small \par \noindent {\it Abstract:} The movement of many organisms
can be described as a random walk at either or both the individual and
population level.  The rules for this random walk are based on complex
biological processes and it may be difficult to develop a tractable,
quantitatively-accurate, individual-level model.  However, important
problems in areas ranging from ecology to medicine involve large
collections of individuals, and a further intellectual challenge is to
model population-level behavior based on a detailed individual-level
model.  Because of the large number of interacting individuals and
because the individual-level model is complex, classical direct Monte
Carlo simulations can be very slow, and often of little practical use.
In this case, an equation-free approach \cite{Kevrekidis:2003:EFM} may
provide effective methods for the analysis and simulation of
individual-based models.  In this paper we analyze equation-free
coarse projective integration.  For analytical purposes, we start with
known partial differential equations describing biological random
walks and we study the projective integration of these equations.  In
particular, we illustrate how to accelerate explicit numerical methods for
solving these equations.  Then we present illustrative kinetic Monte
Carlo simulations of these random walks and show a decrease in
computational time by as much as a factor of a thousand can be
obtained by exploiting the ideas developed by analysis of the closed
form PDEs.  The illustrative biological example here is chemotaxis,
but it could be any random walker which biases its movement in
response to environmental cues.
\par}

\bigskip

\noindent

\section{Introduction}

In current complex systems modeling practice, we are often presented
with a model at a {\it fine} level of description (atomistic,
stochastic, individual-based), while we want to study the behavior at
a macroscopic {\it coarse-grained} (continuum, population) level.
This situation frequently arises in the modeling of biological dispersal,
where significant progress is being made in modeling at the individual
organism/cell level, while the derivation of the corresponding closed, 
macroscopic population level equations remains very difficult, and lags far behind 
in development. 
The example here is bacterial chemotaxis, for which 
much is known about signal transduction and motor behavior of individual
cells, but only in a limited number of cases can one rigorously 
derive equations describing behavior of bacterial populations 
\cite{Erban:2004:ICB,Erban:2004:STS}. 
Usually one can develop a suitable cell-based stochastic model, and
 would like to obtain population-level information without having a
 coarse-grained evolution equation. Computational methods for obtaining an
approximation to the macroscopic evolution without explicitly obtaining 
equations have been developed 
\cite{Kevrekidis:2003:EFM,Gear:2001:PIM,Gear:2002:CIB,Siettos:2003:CBD}. 
The main idea is to use short bursts of appropriately-initialized
computations using the detailed, fine-scale model, followed by
processing of the results to obtain estimates of the desired
macroscopic quantities such as the spatial distribution of the number
density, time derivatives, and various measures of the sensitivity of
the solution with respect to parameters.

The first, and probably most important step of this equation-free
approach is to determine what are the appropriate variables in terms
of which one could hope to close macroscopic evolution equations.
Typically these variables are a few slowly-evolving lower moments of
the many-particle distribution function ({\em e.g.,} cell density for
chemotactic movement \cite{Erban:2004:ICB}, species concentrations for
reaction-diffusion problems \cite{Gardiner:1985:HSM}, or density and
momentum fields, the zeroth and first moments of the distribution of
molecules in velocity space, for the Navier Stokes equations
\cite{Cercignani:1988:BEA,Chapman:1991:MTN}).
In most cases, knowledge of the level of closure (the number and
identity of variables with which one  can write a deterministic
model for the process) comes from extensive experimental experience
and observation, long before it is rigorously justified by theory.
In the equation-free approach, the simplest conceptual path for selecting
the appropriate {\it observables} as state variables is to perform a
homotopy between conditions at which an accurate closed equation is known and
validated, and the conditions of interest, when this is possible. 
Model reduction in general is based on the assumption that, after
rapid initial transients, higher order moments of the evolving
distributions can be approximately-represented as functionals of the
slow, ``master" ones - the ones in terms of which we write the closed
equations.
The closure is thus embodied in a ``slow manifold": a graph of a function
(in moment space) which, given the values of the few governing 
lower moments, provides the ``slaved" higher order moment values.
Separation of time scales between the rapid equilibration of the
higher, slaved moments, and the slow evolution of the ``master" ones
underpins the derivation of closed, reduced, macroscopic equations.
The idea then is to design computational experiments with the
fine scale simulator that test for this separation of time scales,
and suggest variables capable of parametrizing 
the slow manifold \cite{Siettos:2003:CBD,Nadler:2005:DMS}.

As is discussed in more detail in \cite{Siettos:2003:CBD}, it is
possible, using matrix-free iterative linear algebra methods, to
estimate, using direct simulation, characteristic relaxation time scales for the problem (at
least in the neighborhood of a particular equilibrium).
These time scales, and the eigendirections corresponding to them,
can guide the modeler in deciding the {\it number}, and even
 the selection of variables capable of parametrizing
(at least locally) this slow manifold.
In this process, homotopy and knowledge of the appropriate 
parametrizing variables in some region of operating parameter
space, gives us a starting point for variable selection.
In the more general, and much more difficult case in which  we
begin  a completely new problem and  have no initial
knowledge of what might be good ``order parameters" in terms
of which to attempt to close macroscopic equations, the 
alternative is to use data processing techniques on extensive
experimental (or computational experimental) runs, to try and
develop a reasonable reduction hypothesis.
Algorithms for data compression, from the more traditional 
principal component analysis to the more modern sparse kernel 
and diffusion map feature analysis
may be useful here \cite{Smola:1999:SKF,Nadler:2005:DMS}.
This is, however, a separate and active research subject in
itself, and we will not pursue here.

In this paper we will assume that we have enough knowledge
of the problem to identify a  set of variables in terms of which to write
a closed equation. In that spirit we study the coarse integration
of simple models for chemotaxis of cells, and we assume
that the slow dynamics of the system are parametrized by 
cellular density. 
The main goal is to illustrate the computational gain of equation-free
methods, by which we mean a large speed up of the stochastic
simulation for a class of biologically-motivated problems involving
slow dispersal of organisms/cells.

The paper is organized as follows.  In Section \ref{seccoarseint}, we
present a brief overview of equation-free methods with emphasis on
coarse projective integration.  We present the main strategy which we
will use for the analysis of coarse integration -- namely the
deterministic projective integration of partial differential equations
(PDEs).  Moreover, we show how the results of this paper can be
interpreted in terms of equation-free {\it coarse} projective
integration for kinetic Monte Carlo (kMC) simulations of random walks;
and we define the gain of these methods.  In Section
\ref{secchemotaxis} we present partial differential equations modeling
the dispersal of cells, and we provide two biological motivations of
the chemotaxis system studied later.  We also discuss the main
mathematical properties of these equations.  Finally, we introduce a
test family of spatial signal profiles which are used in the
computational examples in Sections 3, 4 and 5.  In Section
\ref{secprojint}, we study the efficiency of projective integration
for different discretizations of the macroscopic PDE equations.  We
obtain a measure of efficiency (gain) of the method for different
choices of the ``inner integrator". We demonstrate a stable,
signal-independent method, {\em i.e.,} a method which has the same
gain for all mathematically admissible environmental changes. We also
study more accurate inner integrators, for which the efficiency
depends on the size of the environmental signal (concentration
gradients).  Section \ref{secnumexamples} contains illustrative
numerical results; here we provide computations illustrating the
analysis in Section
\ref{secprojint} and give examples for which the method leads to a
significant reduction in the computational time required.  In Section
\ref{secmontecarlo} we return to the original random walk problem. We
discuss the application of our approach to accelerating the Monte
Carlo simulations and present a case in which the computational time
is reduced by a factor of $10^3$.  Finally, in Section
\ref{secdiscussion} we summarize the results, and mention significant
generalizations.  We conclude by reiterating the main elements of the
equation-free approach as a ``wrapper" around a usually slow, cell- or
organism-based stochastic simulator, aimed at assisting in the
efficient study of emergent, population-level behavior.

\section{Equation free methods - coarse integration}

\label{seccoarseint}

Consider a large collection of randomly-walking 
individuals for which we have a microscopic model, and suppose that we want to know the time
evolution of the macroscopic density $N$ of the individuals.
One approach is to derive partial differential equation(s) for
 macroscopic observables, such as the density $N$, and then compute
 the solution of the PDE(s) using standard numerical methods. This
 entails a choice of algorithm, a time step $\Delta t$, and a routine
 which computes the density $N(t+\Delta t)$ from the density $N(t).$

If explicit macroscopic equations are not available, we can still compute
the density of individuals at time $t+\Delta t$ from the density
of individuals at time $t$ using Monte Carlo simulation of the
microscopic model.
This can be done as follows.

\leftskip 1cm

\noindent
{\bf (a)} Given the macroscopic initial density $N(t),$ 
construct consistent microscopic initial conditions 
(initialize each individual so that the density is $N(t)$).
 
\noindent
{\bf (b)} Evolve the system using the microscopic Monte Carlo simulator for 
time $\Delta t.$

\noindent
{\bf (c)} Compute the density of individuals $N(t + \Delta t)$ from
the microscopic data at time $t + \Delta t.$

\leftskip 0cm

\begin{figure}[here]
\centerline{
\epsfxsize=2.8in\epsfbox{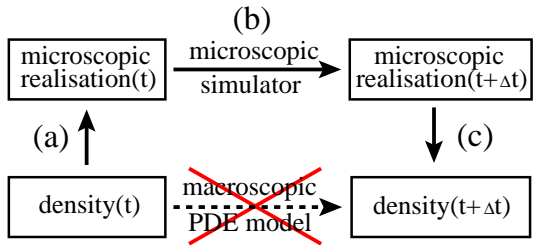}
}
\caption{{\it Schematic of microscopic timestepper (a) -- (c).}}
\label{fig1}
\end{figure}

\noindent
Steps (a) -- (c) provide an alternative path to computing  $N(t + \Delta t)$
from $N(t)$ as illustrated in Figure \ref{fig1}.
The main goal is to compute the long time evolution of cellular density $N,$
and to that end we could simply use step $(b)$ many times,
{\em i.e.,} we could in principle run the microscopic simulator only. 
However, since the biological models are often complex,
step (b) can be very computationally intensive. 
Thus the key constraint
is that we are in fact able to run the microscopic simulator
only for short times. 
Since, we seek the long time evolution,
we have to combine (a) -- (c) with another step which can
be formulated in many ways, {\em e.g.,}

\leftskip 1cm

\noindent
{\bf (d)} Using the macroscopic data for $N$ computed in (a) -- (c), 
estimate the time derivative $\frac{\partial N}{\partial t}(t + \Delta t)$.
Because of fluctuations due to the stochastic nature of the simulation,
we may require several
independent microscopic realizations of $N(t)$ in part (a) to
be able to accurately estimate the expected density $N(t+ \Delta t)$
and its time derivative.
We then take advantage of the assumed smoothness (in time) of the trajectory of
the unavailable macroscopic evolution, and take a large projective 
step by estimating the density $N(t + \Delta t + T)$ for some 
$T>0$ 
as 
$$
N(t + \Delta t + T) \approx N(t + \Delta t) + T 
\frac{\partial N}{\partial t} (t + \Delta t).
$$
The density  $N(t + \Delta t + T)$ is then used 
as a new initial condition in (a).

\leftskip 0cm

\noindent
The algorithm (a) -- (d) is called {\it coarse projective integration},
specifically, coarse projective forward Euler, and it can be formulated in many ways 
\cite{Gear:2001:PIM,Gear:2002:CIB,Kevrekidis:2003:EFM}.
For example we can use different methods to estimate
the time derivative of $N$ in (d), or we can extrapolate using other
macroscopic variables in part (d), {\em e.g.,} with flux profiles as opposed to
density
profiles. In any case, the actual projective step is performed on some spatial
discretization of these macroscopic variable profiles {\em e.g.,} finite 
difference, finite element, or spectral decompositions of the profiles.

\noindent
The algorithm (a) -- (d) can speed up the computations
provided that we can safely (in terms of stability and
accuracy) choose $T \gg \Delta t$ and 
provided that the so called ``lift--run-restrict procedure" (a) -- (c)
does not require excessive computation to estimate
the time derivative of $N$. 
In particular, the more time is spent in part (a) -- (c) of the algorithm,  
the larger  $T/\Delta t$ in part (d) must be chosen to have the potential
for computational gain. 
Since we also study modifications of (a) -- (d), we will define the 
gain $\ggain$ of the coarse projective integration method as
\begin{equation}
\ggain = \frac{
\mbox{time to compute the evolution of the system by running a Monte Carlo
simulator only}}
{
\mbox{time to compute the evolution of the system by coarse projective
integration (a) -- (d)}}.
\label{generalgain}
\end{equation}
For example, if we need $k$ realizations of the Monte Carlo evolution
in steps (a) -- (c) to compute the evolution of the system
in the interval $[t, t + \Delta t],$ and if we assume that the 
computational time of step (d) is negligible, then the gain $\ggain$ 
can be simply estimated as $\ggain = \frac{T + \Delta t}{k \Delta t}.$ 
On the other hand, one might argue that scaling by $k$ may
be too severe, since the equation we are evolving is not one
for the single fluctuating realization, but for the expected
density profile estimated, for example, as the average of
$k$ copies.

\begin{figure}[here]
\centerline{
\epsfxsize=4in\epsfbox{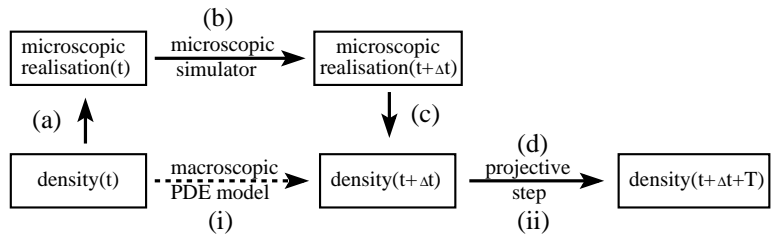}
}
\caption{{\it We first analyse (i) -- (ii)
in parameter regimes where macroscopic equations are available.}}
\label{fig2}
\end{figure}

As a first, {\it illustrative}  step in the analysis of the gain of coarse
integration,
we will replace the stochastic part (a) -- (c) by a deterministic 
operator as shown in Figure \ref{fig2}. 
This means that we assume that we {\it do} know, at least 
for some parameter regime, a closed macroscopic
equation for the expected density profile of the particular 
kinetic Monte Carlo simulation. 
We then replace steps (a) -- (c) by a short deterministic
integration (i). We run this
deterministic integrator  only for a short time $\Delta t$,
and process its results to obtain an extrapolation in time (ii);
we then repeat the process.
In this deterministic setup, we can more easily study
the dependence of the gain on the parameters of the model and, in particular,
the gap between slow and fast eigenvalues in the spectrum of the
equation.
Assuming that most of the computational time is spent in part 
(i), we can rewrite the definition (\ref{generalgain}) in the deterministic
setting (i) -- (ii) as follows
\begin{equation}
\ggain = \frac{T+\Delta t}
{\Delta t}.
\label{determgain}
\end{equation}
In the following section we introduce biologically-motivated problems
for which the corresponding macroscopic equations are known for some,
or for all, parameter regimes.

\begin{figure}[here]
\centerline{
\epsfxsize=4in\epsfbox{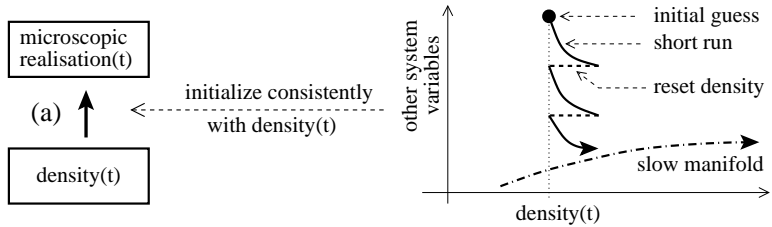}
}
\caption{{\it Initialization of other state variables in step (a).}}
\label{figinitial}
\end{figure}

Finally, let us mention that step (a) requires that initialization
of all system variables be done consistently with the density profile $N(t)$. 
This means that we initialize all individuals in such a way that 
the macroscopic density is equal to $N(t).$ There are many ways to 
accomplish this. Ideally, we would like to initialize the 
remaining macroscopic system observables ({\em e.g.} higher
moments of the cell distribution function than density, the $0$-th 
moment) on a slow manifold parametrized by the density profile $N(t)$ 
- that is, we would like to initialize them ``slaved\footnote{ 
The underlying idea is that
the set of moments of the cell distribution constitutes a singularly
perturbed system, characterized by time scale separation: higher order
moments are assumed to quickly become functionals of the low, slow,
governing ones, like density (i.e. they quickly approach a slow manifold
parametrized by density). The same  key assumption also underlies the
analytical derivation of coarse-grained, macroscopic equations.} to" $N(t)$.
One possible procedure (which we use in Section 
\ref{secmontecarlo}) is schematically illustrated in Figure
\ref{figinitial}. 
Here we make an initial guess for other state variables
and we run the Monte Carlo simulator for a short time only, then
we reset the position of each individual to its initial value,
keeping all other state variables unchanged. Repeating
this procedure several times, we can find the initial condition
close to the slow manifold of the system \cite{Gear:2004:PSM,Gear:2004:CDM}. 

\section{Chemotaxis}

\label{secchemotaxis}

Many organisms that move in a random walk respond to environmental signals by 
biasing their rules of movement. If we consider chemical signals in 
the environment, the corresponding motility behavior is called 
{\it chemotaxis} or {\it chemokinesis}, depending on whether the organism 
senses the direction of signal gradients directly, or responds by changing 
its speed or the frequency of turning.
We will not distinguish
between different terminologies, and we will call {\it chemotaxis}
any alteration of behavior caused by the environmental 
cues; chemotaxis will be the illustrative biological example
in this paper.
At the population level, chemotaxis can lead
to aggregation, travelling waves and pattern formation 
(see {\em e.g.,} \cite{Budrene:1991:CPF}
for {\it E. coli}, 
\cite{Alcantara:1974:SPD,Dallon:1997:DCM} for {\it Dictyostelium
discoideum}) and an important task is to explain
population-level behavior in terms of individual-based models.
To do that, equation free methods may be suitable \cite{Setayeshgar:2003:ACI}.
However our purpose here is to use the strategy for analysis outlined 
in Figure \ref{fig2}, and to this end we choose a chemotactic example 
for which the macroscopic equations are known.
First, in Section \ref{secbacchem},
we describe the simplified model of bacterial chemotaxis
for which the macroscopic equations were derived in some parameter
regimes \cite{Erban:2004:ICB,Erban:2004:STS}. 
Next, in Section 
\ref{secdirchem}, we present an even simpler random walk,
which involves directional sensing, and is more suitable
for modeling of certain eukaryotic organisms. 
Here the macroscopic equations 
can be derived for any choice of parameters. 
These equations have the same structure as in the bacterial case.
In Section \ref{secprojint},
we report the results of projective integration of the chemotaxis equations.

\subsection{Bacterial chemotaxis}

\label{secbacchem}

Flagellated bacteria, the best studied
of which is {\it E.coli}, have two different modes of motile behavior
that are determined by the rotation of their flagella.  When rotated
counterclockwise, the flagella coalesce into a propulsive bundle that
produces a relatively straight ``run".  When rotated clockwise they fly
apart and the bacterium ``tumbles'' without significant
translocation.  Hence, a bacterium runs at a constant velocity for a
random length of time, then tumbles for a random length of time,
chooses a new direction at random, and repeats the process.  In order to 
find food or avoid noxious substances, a bacterium increases its runs
in favorable directions and decreases them when going in an unfavorable
direction.  The run length is controlled by a complex biochemical
network \cite{Barkai:1997:RSB, Spiro:1997:MEA} that involves signal 
transduction and 
alteration of an intracellular protein called CheY that controls the 
direction of rotation of the flagellar motors, and consequently 
changes the movement of the bacterium.
 
In the absence of an extracellular signal the duration of both runs
and tumbles are exponentially distributed, with means of 1 s and
$10^{-1}$ s, respectively \cite{Berg:1990:BM}, and in a gradient of
attractant the cell increases or decreases the run time according as
it moves in a favorable or unfavorable direction.  Since the tumbling
time is small compared to the typical running time, we can decribe the
motion of {\it E. coli}\, as a {\it velocity jump process}
\cite{Othmer:1988:MDB},
which means that a bacterium runs in some direction and, at random instants
of time changes its velocity according to a Poisson process with
mean turning rate $\gamma.$ The turning rate is altered 
by CheY \cite{Cluzel:2000:UBM}, so we can write $\gamma = \gamma(y_1)$
where $y_1$ denotes the concentration of the phosphorylated form
of CheY.

Let $y = (y_1, y_2, \dots, y_m) \in \er^m$ denote the intracellular 
variables, which can include the concentration of proteins, receptors, etc., 
and let $S(x,t) = (S_1, S_2, \dots, S_M) \in \er^M$  denote the signals in 
the environment. Then existing deterministic models of bacterial signal
transduction pathways can be cast in the form of a system of ordinary 
differential equations that describe the evolution of the intracellular 
state, forced by the extracellular signal. Thus
\begin{equation}
\dfrac{\mbox{d}y}{\mbox{d}t} = f(y,S)
\label{rom14}
\end{equation}
where $f: \er^m \times \er^M \to \er^m$ describes the particular
model. The equation (\ref{rom14}) is integrated along the trajectory
of each cell, and the $y_1$  component of the solution 
together with $\gamma = \gamma(y_1)$
defines the random walk of each bacterium.

As was noted in \cite{Erban:2004:ICB, Erban:2004:STS}, 
existing models for signal transduction and models of flagellar 
motor behavior involve tens of chemical species, which makes the
problem very complicated for analysis.
However, the essential aspects of the dynamics can
be captured by a much simpler ``cartoon" model which involves just
two variables. 
For the ``cartoon" model, one can derive closed macroscopic
equations for some parameter regimes (see \cite{Erban:2004:ICB} in 1D,
see \cite{Erban:2004:STS} in 2D/3D). In 
\cite{Erban:2004:ICB,Erban:2004:STS}, equation (\ref{rom14}) and equation
for $\gamma(y_1)$ read as follows
\begin{equation}
\dfrac{\mbox{d}y_1}{\mbox{d}t} = \frac{g(S(x, t)) - (y_1 + y_2)}{t_e},
\qquad
\dfrac{\mbox{d}y_2}{\mbox{d}t} = \frac{g(S(x, t)) - y_2}{t_a},
\qquad
\gamma = \gamma_0 - \beta y_1,
\label{internmodel}
\end{equation}
where $t_e \ll t_a$ are constants, $x$ is the current position of a
cell, $S: \er^n \times [0,\infty) \to [0,\infty)$ is the concentration 
of the chemoattractant, and $g: [0,\infty) \to [0,\infty)$ models the first 
step of signal transduction. The constant $\gamma_0 > 0$ is
the turning rate if no chemoattractant is present, which is changed
by the linear function $-\beta y_1,$ with $\beta > 0,$ if attractant 
gradients are present. 

In this paper, we restrict the random walks to {\it movement
along the real line}, which means that individuals move to the left or
to the right with constant speed $s$ and,
at random instants of time, change their direction with 
turning frequency $\gamma.$ 
In this case, using 
(\ref{internmodel}) in suitable parameter 
regimes, one can derive a macroscopic partial differential
equation for the density
of individuals $N \equiv N(x,t)$ of the following form \cite{Erban:2004:ICB}.
\begin{equation}
\label{hyperbchemotequation}
\frac{\partial^2 N}{\partial t^2} 
+ 2 \gamma_0 \frac{\partial N}{\partial t}
=
\frac{\partial}{\partial x}
\left(
s^2
\frac{\partial N}{\partial x}
-  
g^\prime(S(x)) \frac{2 \beta s^2 t_a}
{(1 + 2 \gamma_0 t_a)(1 + 2 \gamma_0 t_e)}
S^\prime(x) N
\right)
\end{equation}
The macroscopic equation (\ref{hyperbchemotequation}) is valid for 
shallow gradients of the signal (small $S^\prime(x)$) and for a suitable
order of magnitude of  the parameters involved 
(see \cite{Erban:2004:ICB} for details).
%

Since, bacteria are too small to sense spatial gradients of 
the chemoattractant over their body lengths, 
they alter their turning rates as described above, to achieve
the desired response to changes in chemoattractant concentration. 
On the other hand, eukaryotic unicellular organisms like 
{\it Dictyostelium discoideum} are large enough to sense
directly the chemical gradients and respond to them appropriately.
Motivated by this observation, in the following section we present a
simple example of a 1D random walk of individuals such that a
cell can sense directly the gradient of chemoattractant $S^\prime(x)$ and 
respond with changes of its direction according to the gradient seen
by the cell.

\subsection{Chemotaxis with directional sensing}

\label{secdirchem}

We  consider the random movement
of individuals which reduce their probability of changing direction
when moving in a favorable direction, {\em e.g.,} in the 
direction of increasing attractant. 
We suppose as earlier that a particle  moves along the 
$x$ axis at a constant speed $s$, but 
that at random instants of time it reverses its 
direction according to a Poisson process with turning frequency 
\begin{equation}
\gamma = \gamma_0 \pm b S^\prime(x)
\label{turnfreq}
\end{equation}
where $b$ is a positive constant and the sign depends on the direction of the 
particle movement:  plus  for particles moving to the left,
and minus  for  particles moving to the right. 
Let $R(x,t)$ ({\it resp}.$L(x,t)$) be the density of particles at $(x,t)$ which
 are moving to the right ({\it resp}. left): then $R(x,t)$ and $L(x,t)$ satisfy
 the equations
\begin{equation}
\frac{\partial R}{\partial t}
+
s
\frac{\partial R}{\partial x}
=
- (\gamma_0 - b S^\prime(x)) R + (\gamma_0 + b S^\prime(x)) L,
\label{pplussig}
\end{equation}
\begin{equation}
\frac{\partial L}{\partial t}
-
s
\frac{\partial L}{\partial x}
=
(\gamma_0 - b S^\prime(x)) R - (\gamma_0 + b S^\prime(x)) L.
\label{pminussig}
\end{equation}
Equations of this type
have been studied by many authors, and for a discussion of previous work see
\cite{Othmer:1998:OCS,Hillen:2000:HMC}.

The density of particles at $(x,t)$ is given by
the sum $N(x,t) = R(x,t) + L(x,t),$ and the flux is $s R(x,t) - s L(x,t)$.
We are primarily interested in the evolution 
of the macroscopic density $N,$ and therefore we  rewrite the equations
(\ref{pplussig}) and (\ref{pminussig}) as the equations for
the variables $N$ and $J$ given by 
\begin{equation}
N = R + L, \quad
J = R - L
\qquad \qquad \Leftrightarrow \qquad \qquad 
R = \frac{N + J}{2}, 
\quad
L = \frac{N - J}{2},
\label{variablesnjsig}
\end{equation}
where $J$ is a rescaled flux. 
Then adding and subtracting
(\ref{pplussig}) and (\ref{pminussig}), gives
\begin{equation}
\frac{\partial N}{\partial t} + s \frac{\partial J}{\partial x}
= 0,
\label{contineqsig}
\end{equation}
\begin{equation}
\frac{\partial J}{\partial t}
+
s \frac{\partial N}{\partial x}
=
- 2 \gamma_0 J  + 2 b S^\prime(x) N.
\label{fluxeqsig}
\end{equation}
Thus the random walk can be described by the closed
system of two equations (\ref{contineqsig}) and (\ref{fluxeqsig}) with given 
initial conditions $N(\cdot,0)$ and $J(\cdot,0).$

Finally, assuming sufficient smoothness, we can convert (\ref{contineqsig}) -- (\ref{fluxeqsig}) into a second order damped hyperbolic equation for $N,$ namely
\begin{equation}
\frac{\partial^2 N}{\partial t^2}
+
2 \gamma_0 \frac{\partial N}{\partial t}
= 
s^2 \frac{\partial^2 N}{\partial x^2}
-
2 b s \frac{\partial}{\partial x} \left(S^\prime (x) N \right).
\label{telnsig}
\end{equation}
This is a hyperbolic version of the classical Keller-Segel 
equation \cite{Keller:1971:MC,Keller:1971:TBC}. 
Note that (\ref{telnsig}) has the same structure as  
(\ref{hyperbchemotequation}), which 
can also be written as a system of two equations of the form 
(\ref{contineqsig}) -- (\ref{fluxeqsig}). Therefore, the
system (\ref{contineqsig}) -- (\ref{fluxeqsig}) can also be viewed
as a macroscopic description of bacterial chemotaxis.

\subsection{Scaling and mathematical formulation of main problems}

\label{secscaling}

If we consider the system (\ref{contineqsig}) -- (\ref{fluxeqsig})
as a description of the collective movement of bacteria {\it E. coli}, then we can give 
biologically realistic values for the parameters $s$ and $\gamma_0.$
The speed of a bacterium is $s \simeq 10 \mu$m/sec
and the turning frequency is $\gamma_0 \simeq 1$ $\mbox{sec}^{-1}$.
To nondimensionalize equations
(\ref{contineqsig}) -- (\ref{fluxeqsig}), we
choose the characteristic time scale $T_0 = \gamma_0^{-1}$, we
denote the characteristic space scale as $L_0$, and 
the characteristic concentration as $N_0.$ Define
\begin{equation}
\hat{s} = \frac{s T_0}{L_0},
\qquad
\hat{S}^\prime(x) = \frac{b S^\prime (x) T_0}{L_0},
\qquad
\hat{N} = \frac{N}{N_0},
\qquad
\hat{J} = \frac{J}{N_0},
\qquad
\hat{t} = \frac{t}{T_0},
\qquad
\hat{x} = \frac{x}{L_0}.
\label{scalinghat}
\end{equation}
Then the nondimensionalized equations 
(\ref{contineqsig}) -- (\ref{fluxeqsig}) have the form
\begin{equation}
\frac{\partial \hat{N}}{\partial \hat{t}} + 
\hat{s} \frac{\partial \hat{J}}{\partial \hat{x}}
= 0,
\qquad \qquad
\frac{\partial \hat{J}}{\partial \hat{t}}
+
\hat{s} \frac{\partial \hat{N}}{\partial \hat{x}}
=
- 2  \hat{J}  + 2 \hat{S}^\prime(x) \hat{N},
\label{continfluxeqsig}
\end{equation}
and to simplify notation, we drop the hats in (\ref{continfluxeqsig}) 
and obtain the nondimensionalized system
\begin{modelsystem}{NJ}
\qquad
\frac{\partial N}{\partial t} + s \frac{\partial J}{\partial x}
 = 0 & \hskip 3.8cm
\label{contineqsig2}
\\[5pt]
\qquad
\frac{\partial J}{\partial t}
+
s \frac{\partial N}{\partial x}
=
- 2 J  + 2 S^\prime(x) N
\label{fluxeqsig2}
\end{modelsystem}
Here we have one dimensionless parameter $s$ and one dimensionless
function $S^\prime(x)$, and we 
estimate the orders of them as follows. In a typical macroscopic
bacterial experiment the characteristic length scale $L_0$ is 1 cm,
and since the 
characteristic time scale is $T_0 = \gamma_0^{-1}=1$ sec, we have
$s \simeq 10^{-3}.$ If the characteristic length scale is $10$ cm
then $s = 10^{-4}$, and in either case $s$
is a small parameter. A realistic choice of $S^\prime(x)$ must ensure 
that the turning rate (\ref{turnfreq}) is positive, {\em i.e.,}
$|S^\prime(x)| \le 1.$ Hence, we will assume throughout that
\begin{equation}
s \ll 1,
\qquad \quad \qquad
|S^\prime(x)| \le 1.
\label{sSxassume}
\end{equation}
The system (NJ) is a linear hyperbolic system
of two equations with nonconstant coefficients, which can be rewritten
in diagonal form as a system of two equations for the right
and left fluxes ({\em cf.}  (\ref{pplussig}) and (\ref{pminussig})). 
Thus, (NJ) can be rewritten as
\begin{modelsystem}{RL}
\qquad
\frac{\partial R}{\partial t}
+
s
\frac{\partial R}{\partial x}
=
- [1 - S^\prime(x)] R + [1 + S^\prime(x)] L
& \hskip 1.7cm
\label{Req}
\\[5pt]
\qquad
\frac{\partial L}{\partial t}
-
s
\frac{\partial L}{\partial x}
=
[1 - S^\prime(x)] R - [1 + S^\prime(x)] L
\label{Leq}
\end{modelsystem}
We also know that the system (NJ) can be written as 
a single second order equation for $N$ (compare with (\ref{telnsig})),  
or as the following system for  the  variables $N$ and $U$. 
\begin{modelsystem}{NU}
\qquad
\frac{\partial N}{\partial t} = U
& \hskip 2.5cm
\label{Neq}
\\[5pt]
\qquad
\frac{\partial U}{\partial t}
=
s^2 \frac{\partial^2 N}{\partial x^2}
-
2 s \frac{\partial}{\partial x} \left(S^\prime(x) N \right)
- 2 U
\label{Deq}
\end{modelsystem}
In the following sections we study the system (NJ) or its equivalent 
formulations
(RL) and (NU). We will restrict our computations to the finite interval
$[0,2]$ with no flux boundary conditions which, in the formulation
(NJ), can be written in the form
\begin{equation}
J(0,t) = \frac{\partial N}{\partial x} (0,t) = S^\prime(0) = 0,
\quad
\mbox{and}
\quad
J(2,t) = \frac{\partial N}{\partial x} (2,t) 
= S^\prime(2) = 0, \qquad \qquad
\mbox{for} \; t \ge 0.
\label{boundarycondNJ}
\end{equation}
As indicated here we also impose no-flux boundary conditions on the signal.

Finally, let us identify the dimensionless times of interest. The
characteristic time scale was set as to the mean turning time, {\em
i.e.,} $T_0=1$ sec, since that characterizes the microscopic dynamics,
but the macroscopic times of interest in pattern formation experiments
are several hours or days. From the mathematical point of view we are
interested in the long term dynamics and steady states, and therefore
we want to develop methods to compute the density profile $N(x,t)$ for
dimensionless times $t \gg 1$.

\subsection{Slow and fast variables and the slow manifold}

In this section we consider spatial regions where the signal derivative
is either zero or maximal possible (to assure a nonnegative turning rate).
We show that in
such regions the fluxes relax to  functionals of the density for large times,
{\em i.e.} the memory of the initial  flux decays quickly.
Thus the long-term dynamics can be described
by a single first-order in time equation for the density $N$. 
Similar conclusions
can be also made about systems (RL) and (NU). 
For example,
in the case  of (RL), we could characterize the long-term dynamics using only
the right flux $R$. 
or only the left flux $L$,
or any linear combination of $R$ and $L$ ({\em e.g.,} the density $N$).
Knowing the density $N$, we can compute either (or both of) the
right and left fluxes - alternatively, these fluxes quickly evolve
to functionals of the density field; this constitutes our ``slow
manifold".
The choice of the ``right" observables can be made by the modeler;
for historical (as well as practical) reasons we will use the density $N$
in the following as a description of the slow variables.

\subsubsection{Special choices of $S^\prime(x)$}

If $S^\prime(x)=0,$ then system (NJ) can be rewritten
as a second order damped wave equation 
\begin{equation}
\frac{\partial^2 N}{\partial t^2}
+
2 \frac{\partial N}{\partial t}
= 
s^2 \frac{\partial^2 N}{\partial x^2}.
\label{dampedwave}
\end{equation}
It is well-known \cite{Zauderer:1983:PDE} that the asymptotic
behavior of the solution of (\ref{dampedwave}) under the boundary conditions 
(\ref{boundarycondNJ}) is given by the corresponding diffusion equation
\begin{equation}
\frac{\partial N}{\partial t}
= 
\frac{s^2}{2} \frac{\partial^2 N}{\partial x^2}.
\label{diffeq}
\end{equation}
Consequently, the long-term,  slow dynamics can be described by 
this first order in time equation for density only.

Next consider a spatial region where the
signal gradient is the maximum possible, {\em i.e.,} $S^\prime(x)=1$.
If the region with maximal signal gradient is large enough,
then (RL) in this region reduces to
\begin{equation}
\frac{\partial R}{\partial t}
+
s
\frac{\partial R}{\partial x}
=
2 L,
\qquad
\frac{\partial L}{\partial t}
-
s
\frac{\partial L}{\partial x}
=
- 2 L.
\end{equation}
Seen the leftward flux $L$  decays exponentially
according to the second equation, 
the long-term behavior (in large spatial regions with 
$S^\prime(x) = 1$) is given
by the rightward flux only. Since, $N$= $R$ + $L$ and $L \to 0$, the long
time dynamics is simply described by the first order transport equation
\begin{equation}
\frac{\partial N}{\partial t}
+
s
\frac{\partial N}{\partial x}
=
0.
\label{ntransport}
\end{equation}
A similar transport equation holds for the minimal possible signal
gradient $S^\prime(x) = -1.$ Of course the boundary conditions
(\ref{boundarycondNJ}) require that we cannot choose
$S^\prime(x)=1$ in the whole domain of interest, and 
consequently, (\ref{ntransport})
only gives a good approximation of the behavior of cellular
density in large spatial regions with maximal signal gradient. 
On the other hand, if we consider the random walk in a finite domain 
and we look for long term dynamics/stationary state then the no-flux boundary 
conditions (\ref{boundarycondNJ}) have to be taken into account.

\subsubsection{(NJ) for general signals}

For general signals, the behavior is just a combination
of transport and diffusion as given by the second order
equation (\ref{telnsig}). The steady state of (NJ) under no-flux
boundary conditions is given by
$$
s^2 \frac{\partial^2 N}{\partial x^2}
-
2 s \frac{\partial}{\partial x} \left(S^\prime (x) N \right)
=  0,
$$
and it follows that
\begin{equation}
N_s(x) = C \exp \left( \frac{2}{s} S(x) \right)
\label{steadyequation}
\end{equation}
where the constant $C$ is given by the initial condition for $N$.
The interesting question is whether the behavior of (NJ) can indeed
be described by a single first order equation for large times.
The simplest choice is to use a parabolic counterpart
of (\ref{telnsig}), given in  dimensionless form as
\begin{equation}
\frac{\partial N}{\partial t}
= 
\frac{s^2}{2} \frac{\partial^2 N}{\partial x^2}
-
s \frac{\partial}{\partial x} \left(S^\prime (x) N \right).
\label{parabchem}
\end{equation}
Equation (\ref{parabchem}) has the same steady state
as (NJ), and moreover it reduces to (\ref{diffeq})
for constant signals. 
On the other hand, if $S^\prime (x) =  1,$
then equation (\ref{parabchem}) differs from (\ref{ntransport})
by the term $\frac{s^2}{2} \frac{\partial^2 N}{\partial x^2}$ 
which adds artificical diffusion to the system \cite{Strikwerda:1989:FDS}.
Consequently, if we have extended spatial regions where 
$S^\prime (x) =  1,$ then the equation (\ref{parabchem})
gives different transient behavior than (NJ) but finally leads 
to the same steady state as (NJ).
It is important to note that a major issue in equation-free
computation is how many independent variables are needed in order to
close with a {\it first order in time} system, because it may be
difficult to initialize microscopic variables consistently with given
macroscopic observables  and {\it their history} ({\em e.g.} their
first order time derivatives).
In regimes where at least a second-order-in-time equation is needed
for closure,  initializing the density is not enough; the
time derivative of density must also be prescribed.
In such a case we would use an alternative initialization
for equation-free computations: we would prescribe
right- and left- going fluxes $R$ and $L$, which would be sufficient
to start a particle-based simulation, because it is much easier
to initialize particles based on more than one independent
variables rather than based on the ``history" of a single 
variable.\footnote{In doing projective integration based on 
simulations over the  entire spatial domain, the spatial order of
the equation does not play a crucial role. If, however, one tries to
use equation-free techniques such as the gaptooth scheme and patch
dynamics
\cite{Kevrekidis:2003:EFM,Gear:2003:GTM,Samaey:2003:GTS}, implementing
effective matching conditions between patches becomes important, and
that is crucially affected by the  spatial order of the effective
evolution equation. The design of computational experiments to determine the
spatial order of an unknown (in closed form) equation is
an interesting subject, discussed in part in \cite{Li:2003:DNC}.}

\subsection{Test family of signal functions}

\begin{figure}
\picturesABC{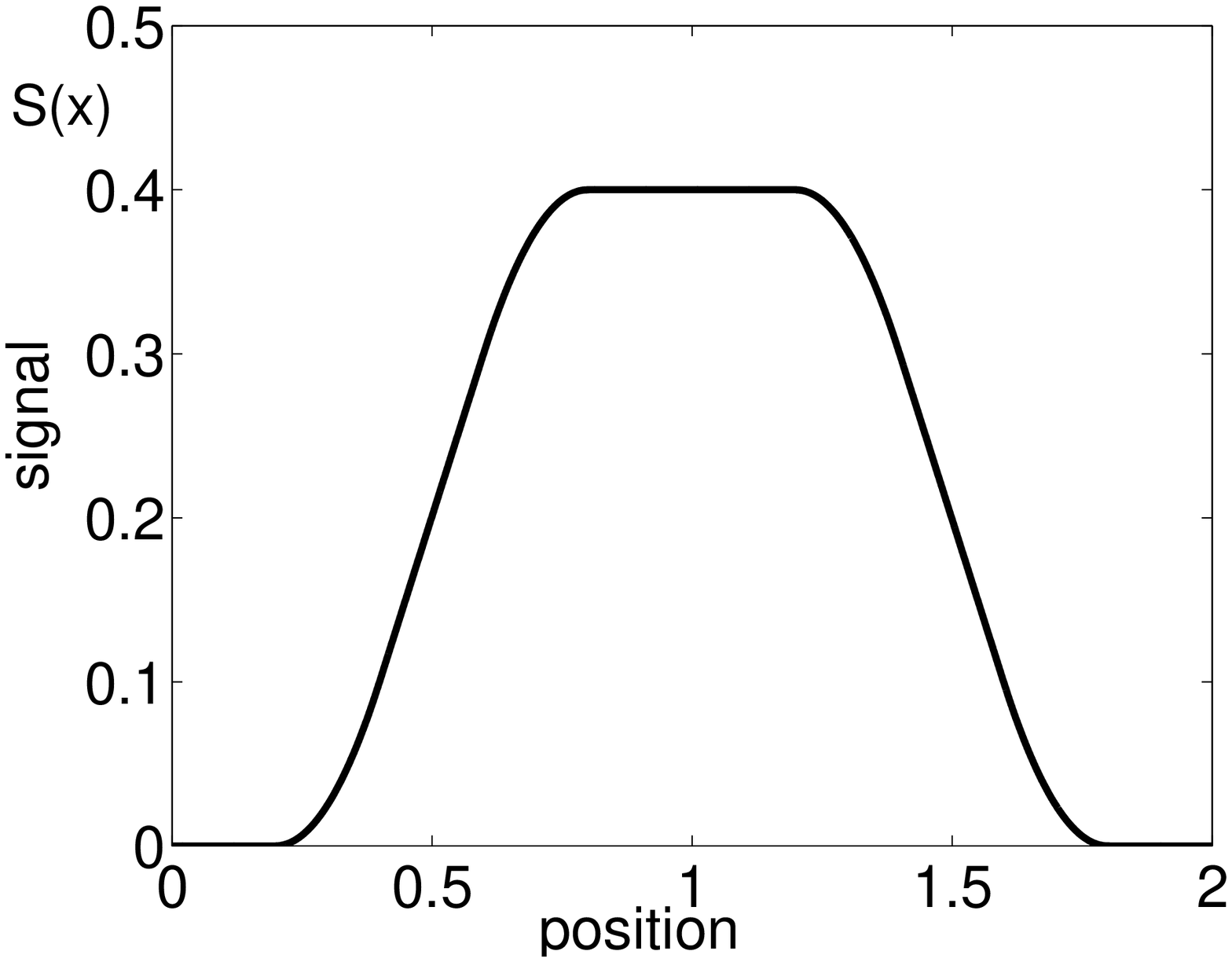}
{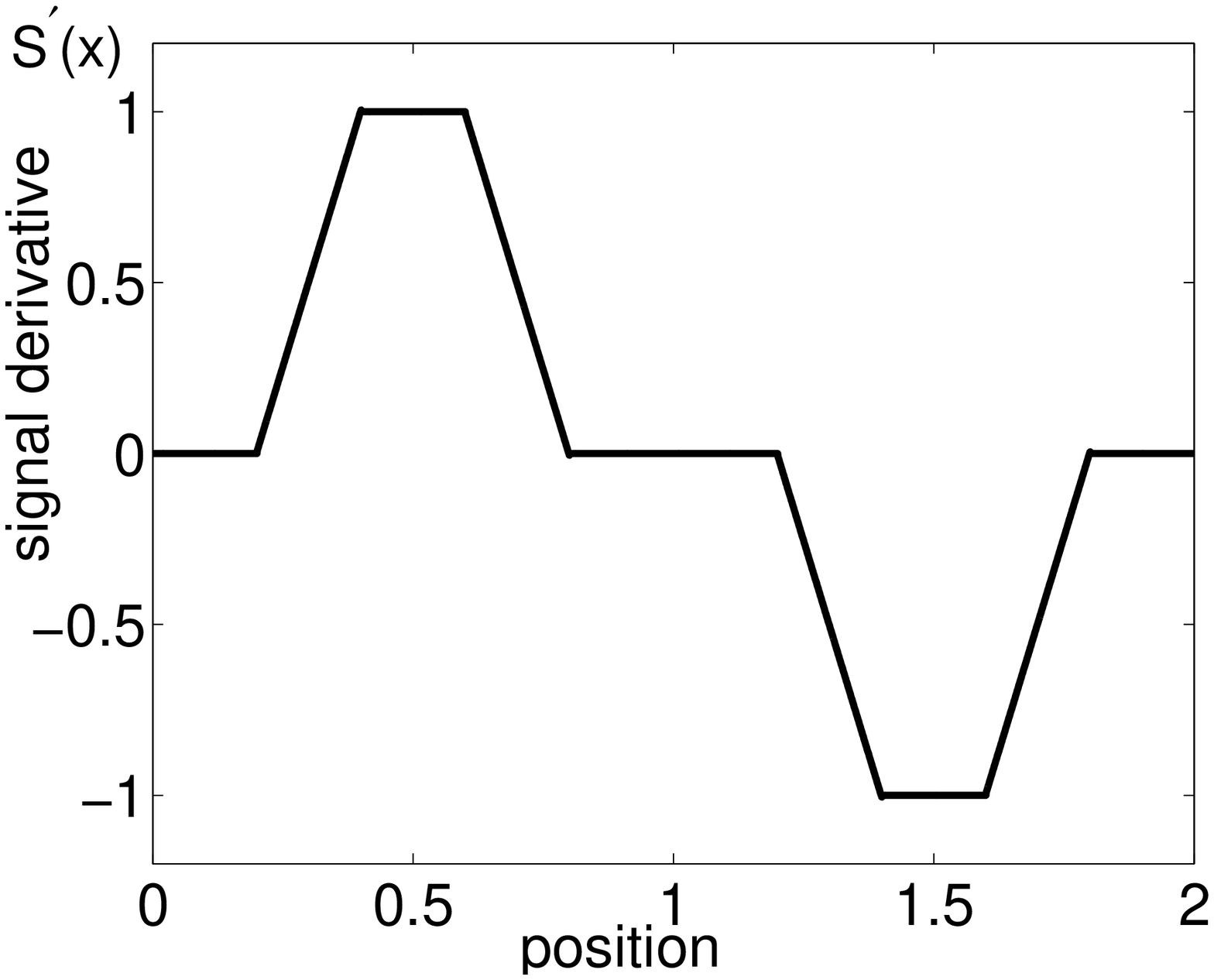}
{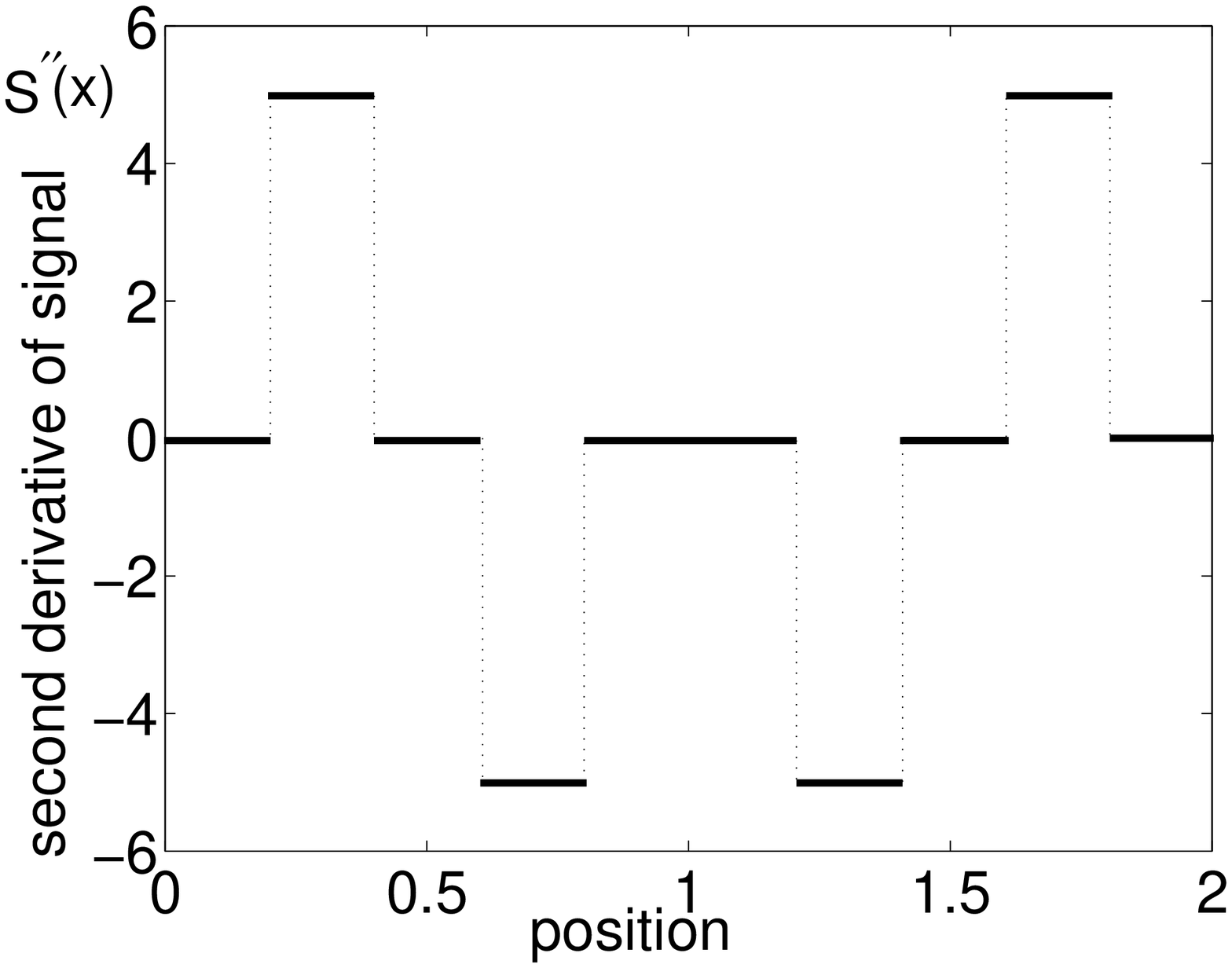}
{1.4 in}
\caption{ (a) {\it Graph of ``hat-profile" signal 
function $\overline{S}(x)$.} (b) {\it 
Graph of  $\overline{S}^\prime(x)$.}
(c) {\it Graph of $\overline{S}^{\prime\prime}(x).$} 
}
\label{figuresig2Z}
\end{figure}

In later sections, several numerical computations are presented.
Here we introduce the test family of signal functions which
we will use in these illustrative examples. 
In all examples, we consider the 
problem  (NJ) on the interval [0,2] with no-flux boundary 
conditions, where the signal belongs to the
one-parameter test family of signal functions given by
\begin{equation}
S_\alpha (x) = \alpha \overline{S}(x),
\qquad \mbox{for} \; \alpha \in [0,1],
\label{salphadefin}
\end{equation}
where  $\overline{S}(x)$ is a fixed signal function and
$\alpha$ scales the strength of the signal. 
The signal function $\overline{S}(x)$ is chosen in the following form 
(see also Figure \ref{figuresig2Z}):
$$
\begin{array}{|c|c|c|c|c|c|c|c|c|c|} \hline
\mbox{interval} & 
\xstrut \left[ 0, \frac{1}{5} \right]  &
\left[ \frac{1}{5},\frac{2}{5} \right] &
\left[ \frac{2}{5},\frac{3}{5} \right] &
\left[ \frac{3}{5},\frac{4}{5} \right] &
\left[ \frac{4}{5},\frac{6}{5} \right] &
\left[ \frac{6}{5},\frac{7}{5} \right] &
\left[ \frac{7}{5},\frac{8}{5} \right] &
\left[ \frac{8}{5},\frac{9}{5} \right] &
\left[ \frac{9}{5},2 \right] \\ 
\hline
\overline{S}(x) &
\xstrut 0 & \frac{(5x - 1)^2}{10} & x - \frac{3}{10} & 
\frac{4 - (5x-4)^2}{10} & \frac{4}{10} & \frac{4 - (5x - 6)^2}{10} & 
- x + \frac{17}{10} & \frac{(5x - 9)^2}{10}  & 0  \\ \hline
\overline{S}^\prime(x) &
0  & 5x - 1  & 1  & 4 - 5x  & 0 & 6 - 5x  & - 1  & 5x - 9  & 0 \\ \hline 
\overline{S}^{\prime\prime}(x) &
0  & 5 & 0  & - 5 & 0 & - 5 & 0  & 5  & 0 \\ \hline
\end{array}
$$
Since the maximal absolute value of the derivative $\overline{S}^\prime(x)$
is equal to 1, the assumption (\ref{sSxassume})
requires that  $\alpha \in [0,1]$, and $\alpha = 1$ means that the signal 
derivative $S^\prime_\alpha (x)$ is maximal possible in some subintervals of 
the domain $[0,2]$.  
For the $\bar{S}(x)$ chosen, the signal gradient $S_\alpha^\prime(x)$ is 
zero in the intervals $\left[ 0, \frac{1}{5} \right],$ 
$\left[\frac{4}{5},\frac{6}{5} \right]$
and $\left[\frac{9}{5}, 2 \right],$ so the behavior will be similar to the 
diffusion equation there (for any $\alpha$).
If $\alpha = 1$ in (\ref{salphadefin}), 
then the signal derivative $S_1^\prime(x)$ is maximal possible, 
equal to 1, in the  interval $\left[\frac{2}{5},\frac{3}{5}\right];$  
consequently,  the right moving individuals will never turn in
this interval and the corresponding coarse equation is a transport
equation (\ref{ntransport}) there.
Similarly, the signal gradient is minimal, equal to 
- 1, in the  interval $\left[\frac{7}{5},\frac{8}{5}\right];$ consequently, 
the left moving individuals will never turn in
this interval and the corresponding coarse equation is again the transport
equation there. 

\section{Projective integration}

\label{secprojint}

The next objective is to study the so-called projective integration of the 
system (NJ), or its equivalent forms (RL) and (NU).
To that end, we first summarize results from \cite{Gear:2003:PMS}
about the {\it projective forward Euler method}. 
Suppose that we want to solve the initial value problem for the linear
system of ordinary differential equations
\begin{equation}
\frac{\mbox{d}y}{\mbox{d}t} = {\cal L} y, \qquad y(0) = y_0,
\label{linODE}
\end{equation}
where $y$ is $n-$dimensional vector and ${\cal L}$ is a $n \times n$ matrix 
of real numbers. Given constants $k$ and $M$ and step size $\dt,$   
the projective forward Euler method (\Pkm) can be described as 
follows \cite{Gear:2003:PMS}:

{
\leftskip 3cm
\rightskip 2cm

\parindent -1.2cm

(\Pkm-1) Use the forward Euler method\footnote{In fact any other
integration scheme can be used here.} to integrate the system (\ref{linODE}) over 
$k$ time steps of the length $\dt$ to compute $y(t+k\dt)$ from $y(t);$

(\Pkm-2) perform one more integration step to compute $y(t+k\dt+\dt)$ from 
$y(t+k\dt);$

(\Pkm-3) perform an extrapolation over $M$ steps, using $y(t+k\dt+\dt)$ 
and $y(t+k\dt)$ to estimate $y(t+k\dt+\dt+M\dt)$ as 
$y(t+k\dt+\dt+M\dt)= (M+1) y(t+k\dt+\dt) - M y(t+k\dt).$ 

\leftskip 0cm
\rightskip 0cm
}

\noindent
Thus, the procedure (\Pkm-1) -- (\Pkm-3) integrates the system over the $(k+1+M)$ 
steps of the length $\dt.$ Next, we have the following result 
\cite{Gear:2003:PMS}.
\begin{lemma}
Method (\Pkm-1) -- (\Pkm-3) for solving $(\ref{linODE})$ is stable provided
that the error amplification given by 
\begin{equation}
\sigma (\lambda \dt) = \Big[ (M+1) (1 + \lambda \dt) - M \Big] 
(1 + \lambda \dt)^k
\label{amplificationfactor}
\end{equation} 
satisfies $|\sigma (\lambda \dt)| \le 1$ for all $\lambda$ in the spectrum of 
the matrix ${\cal L}$ of system $(\ref{linODE})$.
\label{stablemma}
\end{lemma}
\proof See \cite{Gear:2003:PMS} where a more general linear stability analysis
for systems of nonlinear ODEs is done. 

\begin{figure}[here]  
\picturesAB{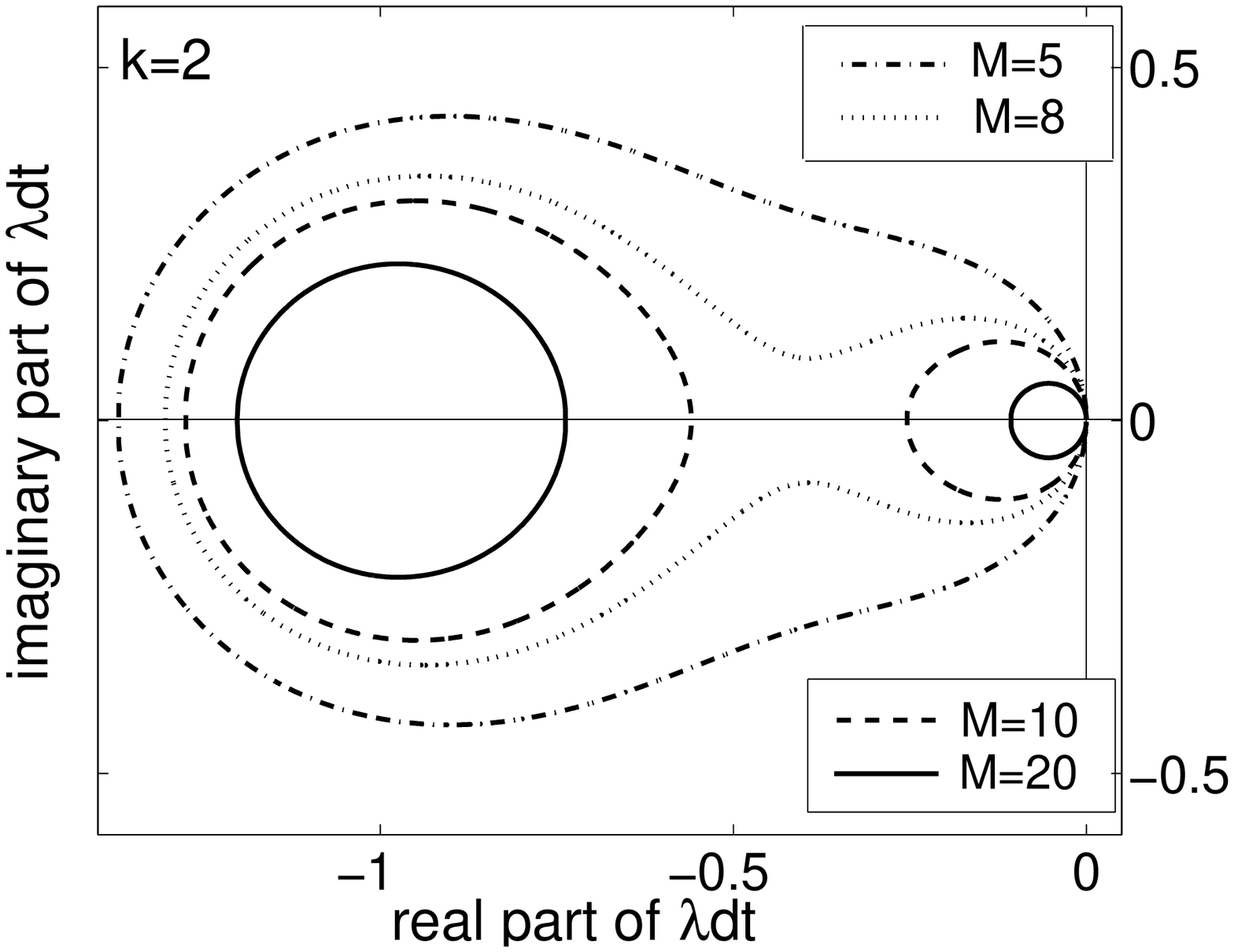}{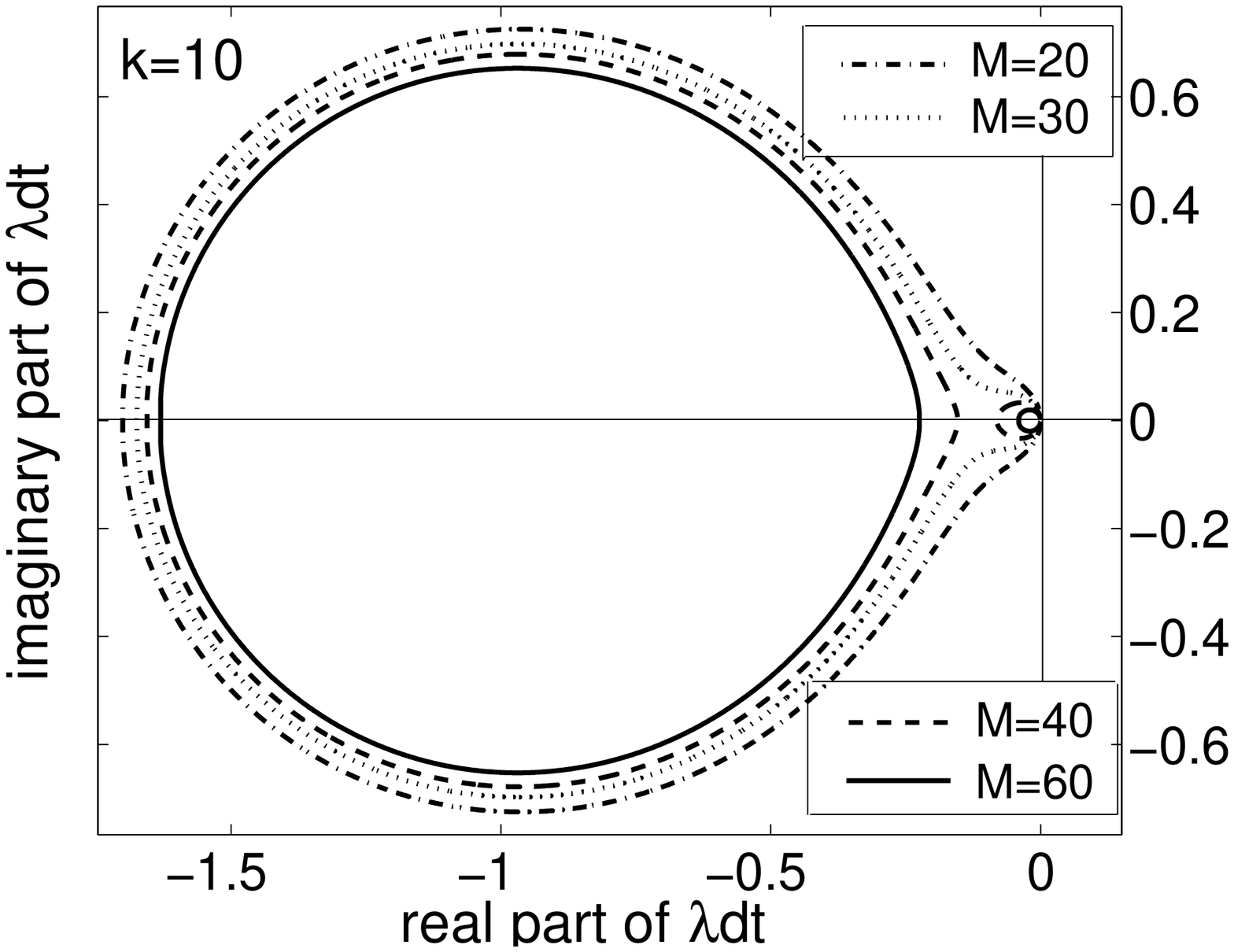}
{2.2 in}
\caption{(a) { \it The regions of absolute stability of \Pkm 
methods for $k=2$ and $M=5$ (dot-dashed line), $M=8$ (dotted line), $M=10$ 
(dashed line) and $M=20$ (solid line).}
(b) {\it The regions of absolute stability of \Pkm methods for $k=10$
and $M=20$ (dot-dashed line), $M=30$ (dotted line), $M=40$ (dashed line)
and $M=60$ (solid line). }}
\label{figstabregion}
\end{figure}

\begin{figure}[ht]  
\picturesAB{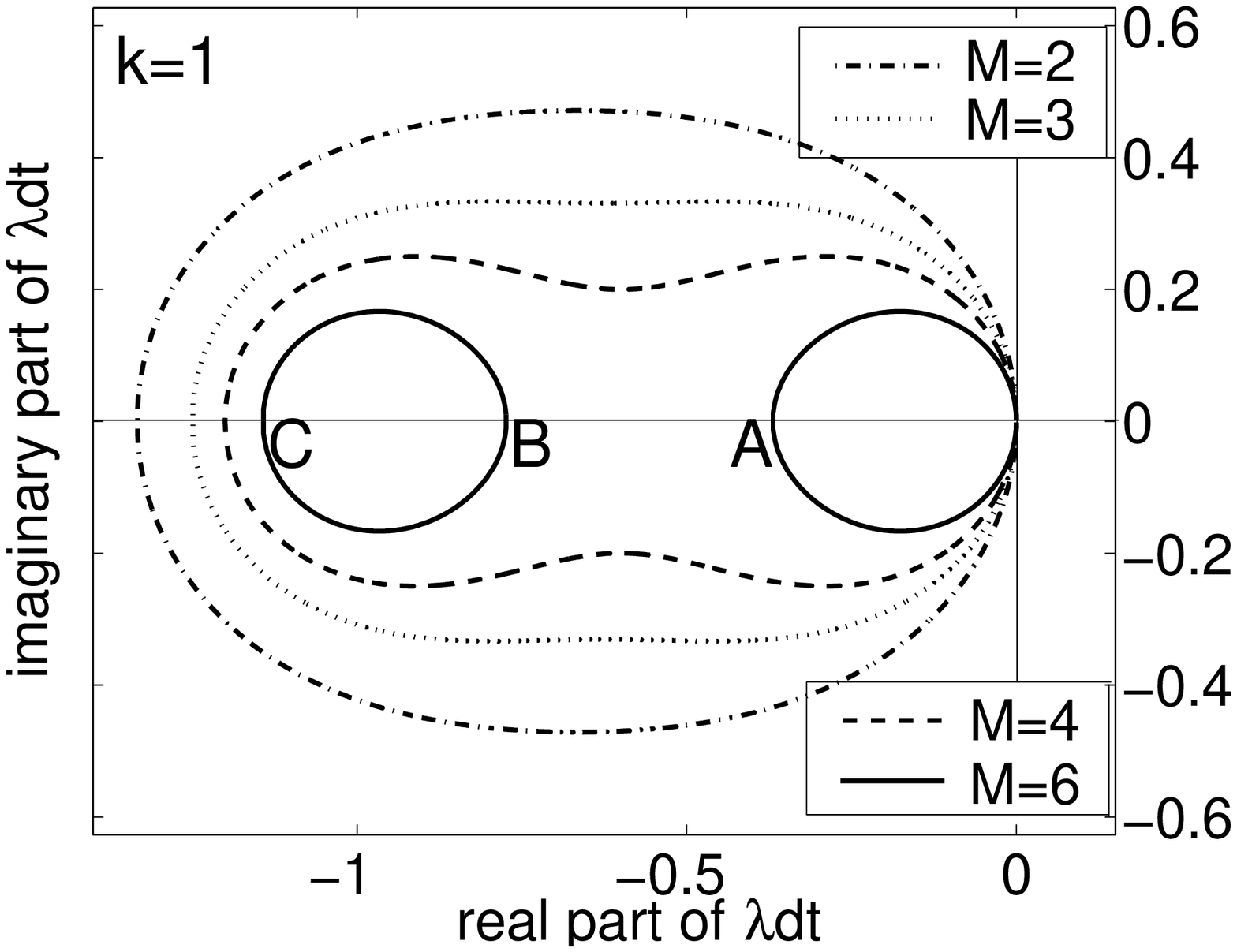}{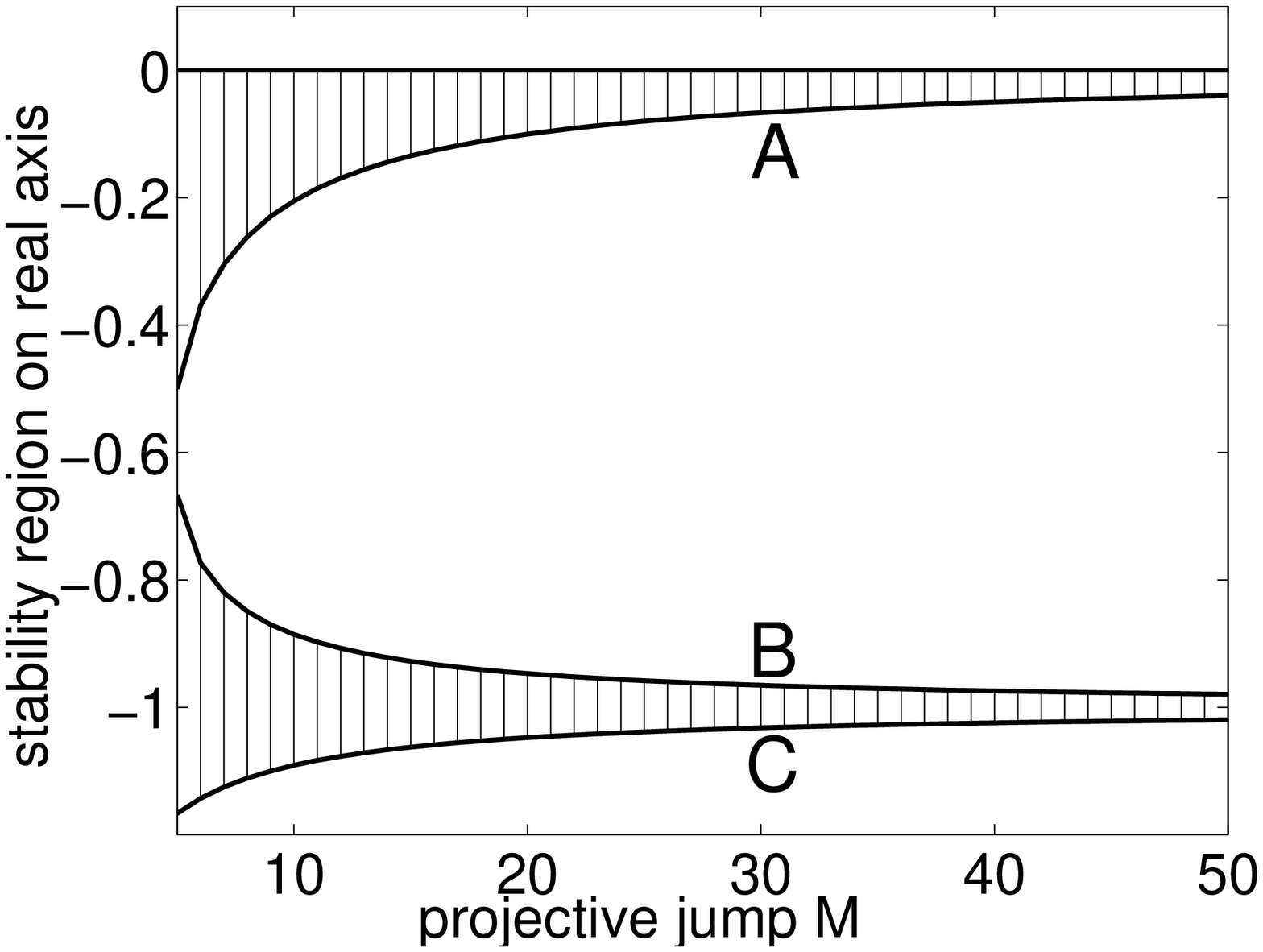}
{2.2 in}
\caption{(a) {\it  
The regions of absolute stability of \Pkm methods for $k=1$
and $M=2$ (dot-dashed line), $M=3$ (dotted line), $M=4$ (dashed line)
and $M=6$ (solid line).}
(b) {\it Intersection of stability region
from part (a) with real axis plotted as as a function of $M.$
The equations for boundary curves $A(M),$ $B(M)$ and $C(M)$ are
given in Lemma $\ref{stablemmareal}$.} 
}
\label{figstabregionreal}
\end{figure}

\noindent
The absolute stability region in the complex $\lambda\dt$-plane, which is plotted in 
Figure \ref{figstabregion} and Figure \ref{figstabregionreal}(a), is the area inside the curve 
$|\sigma (\lambda\dt)| = 1$. We see that the region splits into two parts 
for large $M.$ Consequently, the constant $M$ can be large if the
spectrum is concentrated into two widely-separated regions
corresponding to the  fast and slow components. We also
see that we increase the part of the stability
region corresponding to fast components if we increase the number of
inner integration steps $k$. The stability region for $k=1$ is given in 
Figure \ref{figstabregionreal}(a), for $k=2$ in 
Figure \ref{figstabregion}(a) and for $k=10$ in 
Figure \ref{figstabregion}(b).

In the following sections, we discretize the PDEs using the method of lines. 
Some of the systems we study will have
a real-valued spectrum for parameter values of interest. 
Consequently, the interesting part of the stability region 
from Figure \ref{figstabregionreal}(a) is its intersection with the
real axis. 
For large $M$, the real stability region comprises the  union of two intervals 
given by Lemma \ref{stablemmareal} for $k=1$, and plotted 
in Figure \ref{figstabregionreal}(b). 
\begin{lemma}
Suppose that the  eigenvalues of the matrix ${\cal L}$ are all real. Then
the procedure (\Pkm-1) -- (\Pkm-3) with $k=1$ and $M\ge5$ for solving 
$(\ref{linODE})$ is stable provided that 
\begin{equation}
\lambda \dt \in \; (C,B) \cap (A,0),
\qquad
\mbox{for all} \; \lambda \; \mbox{in the spectrum of} \; {\cal L},
\label{intervals}
\end{equation} 
where $C < -1 <  B < A < 0$ are given by
\begin{equation}
C = - 1 - \frac{1}{M+1},
\qquad
B = - \frac{M+2 + \sqrt{(M-2)^2 -8}}{2(M+1)},
\qquad \mbox{and} \qquad
A = - \frac{M+2 - \sqrt{(M-2)^2 -8}}{2(M+1)}.
\label{ABCconstants}
\end{equation} 
\vskip -4mm
\label{stablemmareal}
\end{lemma}
\proof This is an easy consequence of Lemma \ref{stablemma}. \hfill Q.E.D. 

\noindent
From Figure \ref{figstabregionreal}(b) we see that (in the case
of real spectrum) one can choose a large projective jump $M$ provided that 
the spectrum of ${\cal L}$ lies in two small intervals, separated by
a spectral gap. 
Later, we will see such linear systems arising in our simulations; the
natural question then is: if we know (or can estimate) the spectrum of
${\cal L},$ what is the maximal possible choice of $M$ such that the
\Pkm method is stable? The answer is given in the following
lemma.
\begin{lemma}
Suppose that eigenvalues of the matrix ${\cal L}$ are all real. 
Let $-2 < c < -1 < b < a < 0$ and {\rm $\, \dt > 0$}
be given constants such that 
{\rm
\begin{equation}
\lambda \dt \in \; (c,b) \cap (a,0),
\qquad
\mbox{for all} \; \lambda \; \mbox{in the spectrum of} \; {\cal L}.
\label{intervalsinverse}
\end{equation} 
}
Then (\Pkm-1) -- (\Pkm-3) is stable for all $M$ satisfying the 
inequality
\begin{equation}
M \le \min \left\{
\left|
\frac{1 + (1+a)^{k+1}}{a (1+a)^{k}}
\right|,
\left|
\frac{1 + (1+b)^{k+1}}{b (1+b)^{k}}
\right|,
\left|
\frac{1 - (1+c)^{k+1}}{c (1+c)^{k}}
\right|
\right\}.\label{maximalM}
\end{equation} 
\vskip -4mm
\label{lemmamaximalM}
\end{lemma}
\proof The amplification factor (\ref{amplificationfactor})
is given by the formula
$$
\sigma (\lambda \dt) =  M \lambda \dt (1 + \lambda \dt)^k +  
(1 + \lambda \dt)^{k+1}.
$$
In order to have a stable method, the following three inequalities
must be satisfied simultaneously:
$$
\sigma(a) \ge -1,
\qquad
\sigma(b) \ge -1,
\qquad
\mbox{and}
\qquad
\sigma(c) \le 1.
$$
Solving for $M$, we obtain (\ref{maximalM}). \hfill Q.E.D.

\noindent
Finally, let us note that the results of this section could be also
viewed as results of linear stability 
analysis of projective integration of general nonlinear systems of ODEs 
of the form $y^\prime = F(y),$  $y(0)=y_0,$ where $y$ is an $n-$dimensional 
vector and $F: \er^n \to \er^n$ \cite{Gear:2003:PMS}.

\subsection{Projective integration of chemotaxis systems (NJ), (RL) and (NU)}

Before we implement coarse projective integration, we illustrate the
use of projective integration and the factors affecting its implementation
and effectiveness through the use of discretizations of the chemotaxis
equations themselves. In this context, it is convenient to think that we
only have available as an ``inner simulator" a  black-box dynamic
integrator with a small, fixed time step (for example, a forward Euler
simulator of a discretization of the problem), and that we are attempting
to accelerate this black box code.  

Given a signal profile
$S(x),$  the speed $s$ and initial conditions, we will look for the solution 
of (NJ) in the finite interval $[0,2]$ with no flux boundary conditions
(\ref{boundarycondNJ}).
To do that, we will discretize (NJ) and rewrite it
as a system of ordinary differential equations of the form 
(\ref{linODE}) using the method of lines. 
The resulting system of ODEs is a starting point for our basic
projective integration algorithm, which is little different
than (\Pkm-1) -- (\Pkm-3). 
It is based on the sketch in Figure \ref{fig2}. 
Choosing a suitable time step $\dt,$ and constants $k$ and $M,$ 
the algorithm is given
in the following three steps (\Prkm-1) -- (\Prkm-3). 
Note that the 
steps (\Prkm-1) -- (\Prkm-2) correspond to the step (i) as outlined in Figure 
\ref{fig2}, and the last step (\Prkm-3) corresponds to step (ii) in 
Figure \ref{fig2}. 

\leftskip 1cm

\noindent 
(\Prkm-1) integrate system (NJ) over $k$ time steps of 
length $\dt$ to compute $N(t+k\dt)$ and $J(t+k\dt)$ from $N(t)$ and 
from  suitably initialized flux $J(t)$ -- 
see (\ref{fluxis0}) and (\ref{fluxisjprev});

\noindent 
(\Prkm-2) perform one more inner integration step to compute 
$N(t+k\dt+\dt)$ and $J(t+k\dt+\dt)$ from $N(t+k\dt)$ and $J(t+k\dt);$ 

\noindent 
(\Prkm-3) perform an extrapolation over $M$ steps, using $N(t+k\dt+\dt)$ and 
$N(t+k\dt)$ to compute $N(t+k\dt+\dt+M\dt)  = (M+1) N(t+k\dt+\dt) - M N(t+k\dt).$

\leftskip 0cm

\noindent
Note that we can approximate the time derivative of $N$ in step 
(\Prkm-2) by
$$
\frac{\partial N}{\partial t}
=
\frac{N(t+k\dt+\dt) - N(t+k\dt)}{\dt}
$$
and therefore the step (\Prkm-3) is equivalent to 
$$
N(t+k\dt+\dt+M\dt)  = N(t+k\dt+\dt) + M \dt \frac{\partial N}{\partial t}
$$
which is the forward Euler projective step. We see  that step (\Prkm-3)
is really equivalent to step (ii) in  Figure \ref{fig2}.
It is important to notice that integrating the full system (NJ) requires
initialization not only of the density $N$ (which is prescribed)
but also of the flux $J$,  which is not; this will be discussed further
below.
As we mentioned in Section \ref{seccoarseint}, the coarse/projective 
integration method is efficient provided that we can choose a large 
projective time $T$ in step (d) in Figure \ref{fig2} relative  to the time
$\Delta t$ of the steps (a) -- (c) from Figure \ref{fig2} and still
retain accuracy and stability.
Using the notation from Section \ref{seccoarseint},
we have
\begin{equation}
\Delta t = k \dt+\dt,
\qquad
T = M\dt,
\label{relationTM}
\end{equation}
and consequently, the gain $\ggain$ of the method (\ref{determgain})
can be expressed as 
\begin{equation}
\ggain = \frac{T + \Delta t}{\Delta t} = \frac{M + k + 1}{k+1}.
\label{gain}
\end{equation}
Our goal is to make this gain as large as possible. 
Moreover, in order to use the 
scheme (\Prkm-1) -- (\Prkm-3), we have to specify the spatial discretization 
of (NJ). 
We study two options in Section \ref{secdiscrRLNU}. 
Finally,  we also have to specify how we initialize the flux 
in step (\Prkm-1). There are  several possibilities for doing this, the easiest
of which is to use the initial flux $J(t)$ in step (\Prkm-1) given by
\begin{equation}
J(t) = 0.
\label{fluxis0}
\end{equation}
We can also use as an initial guess the value of the flux computed in 
the previous step (\Prkm-2) corresponding to a time  $M\dt$ ago, 
{\em i.e.,} before the projective jump.  Thus we could use 
\begin{equation}
J(t) = \mbox{``flux} \;  J(t - M \dt)
\; \mbox{which was computed in the previous step (\Prkm-2)"}
\label{fluxisjprev}
\end{equation}
A more sophisticated flux initialization is used in Section
\ref{secmontecarlo}, which deals with Monte Carlo simulations
(see also Figure \ref{figinitial}). 

\subsection{Discretization of (RL) and (NU)}

\label{secdiscrRLNU}

Various possibilities exist for discretizing the system (NJ) in the
spatial domain; we start with one which is based on the equivalent
form (RL) and on upwinding.  The advantage of upwinding is that it
provides a more stable scheme for problems with a significant
convection component, but  it introduces artificial
diffusion into the problem \cite{Strikwerda:1989:FDS}.
Another possibility to spatially discretize (NJ), (RL) 
or (NU) is to use central differences, which  leads to 
equation (\ref{odeNU}).

First, to solve the system (NJ) numerically, we transform it to the system
(RL) of two first order equations in diagonal form. 
We want to
solve (RL) over the interval $[0,2]$ with boundary conditions given
by (\ref{boundarycondNJ}). 
We choose a number $n$
and a mesh size $\dx = 2/n$, and we discretize the interval [0,2] with
$n+1$ mesh points
\begin{equation}
x_k = k \cdot \dx,
\qquad
\mbox{for}
\; k = 0, \dots, n.
\label{mesh}
\end{equation}
Next, we define
$$
R_{i}(t) = R (x_i,t),
\quad 
L_{i}(t) = L (x_i,t), 
\quad 
\mbox{and}
\quad 
S^\prime_i = S^\prime(x_i),
\qquad
i= 0, \dots, n.
$$ 
The zero flux boundary conditions (\ref{boundarycondNJ})
simply mean that $R_0 = L_0$ and $R_n = L_n;$
consequently, we have to compute the time evolution
of the $2n-$dimensional vector
\begin{equation}
w = \left( R_1, R_2, \dots R_{n-1}, R_n, L_0, L_1, L_2, \dots, 
L_{n-1} \right)^T. 
\label{notationRL}
\end{equation}
To discretize spatial derivatives in (RL), we use upwinding,
that is,
$$
\frac{\partial R}{\partial x} (x_i,t) \approx
\frac{R_{i}(t)- R_{i-1}(t)}{\dx},
\qquad
\frac{\partial L}{\partial x} (x_i,t) \approx
\frac{L_{i+1}(t)- L_{i}(t)}{\dx}.
$$
Then, the solution of (RL) with boundary conditions (\ref{boundarycondNJ})
is approximated by the solution of a system of ordinary differential equations
\begin{equation}
\frac{\mbox{d}w}{\mbox{d}t} = {\cal A} w, \qquad w(0) = w_0,
\label{odeRL}
\end{equation}
where $w_0$ is a given initial condition and matrix ${\cal A}$ is
defined by
{\small
$$ 
{\cal A} 
=
\left( \begin{array}{cccccccccc}
-1 - \varepsilon + S^\prime_1 & 0 & .. & 0 & 0 &
                        \varepsilon & 1 + S^\prime_1 & 0 & .. & 0 \\
\varepsilon & -1 - \varepsilon + S^\prime_2 & .. & 0 & 0 & 
                        0 & 0 & 1 + S^\prime_2 & .. & 0 \\ 
. & . &  .. & . & . &
                        . & . &  . &  .. & . \\
 0 & 0 & .. & -1 - \varepsilon + S^\prime_{n-1} & 0 &
                        0 & 0 & 0 & .. & 1 + S^\prime_{n-1} \\
0 & 0 & .. & \varepsilon & - \varepsilon &
                        0 & 0 & 0 & .. & 0 \\ 
0 & 0 & .. & 0 & 0 &
                     -\varepsilon & \varepsilon & 0 & .. & 0 \\ 
1 - S^\prime_1 & 0 & .. & 0 & 0 &
                     0 & -1 - \varepsilon - S^\prime_1 & \varepsilon & .. & 0 \\ 
0 & 1 - S^\prime_2 & .. & 0 & 0 &
                     0 & 0 & -1 - \varepsilon - S^\prime_2 & .. & 0 \\ 
. & . & .. & . & . & 
                        . & . &  . &  .. & . \\
0 & 0 & .. & 1 - S^\prime_{n-1} & \varepsilon &
                    0 & 0 & 0 & .. & -1 - \varepsilon - S^\prime_{n-1} 
\end{array}
\right)
$$}
where 
\begin{equation}
\varepsilon \equiv \frac{s}{\dx}.
\label{defvarepsilon}
\end{equation}
Since we have  approximated the original PDE system as a system of ordinary
differential equations of the form (\ref{linODE}), the results
in  Lemma \ref{stablemma}, Lemma \ref{stablemmareal}, 
and Lemma \ref{lemmamaximalM} can be applied. 
Alternatively, we can discretize the chemotaxis system in its equivalent 
form (NU) using standard central differences to approximate spatial 
derivatives in (NU). 
We use $n+1$ mesh points (\ref{mesh}) and we define
$$
N_{i}(t) = N (x_i,t),
\quad 
U_{i}(t) = U (x_i,t), 
\quad 
\mbox{and}
\quad 
S^\prime_i = S^\prime(x_i),
\qquad
i= 0, \dots, n,
$$ 
\begin{equation}
z = \left( N_0, N_1, N_2, \dots, N_{n-2}, N_{n-1}, N_n, U_0, U_1, U_2, \dots, 
U_{n-2}, U_{n-1}, U_{n} \right)^T, 
\label{notationNU}
\end{equation}
{\small
$$ 
{\cal D} 
=
\left( \begin{array}{cccccccc}
-2 \varepsilon^2 & 2 \varepsilon^2 -\varepsilon S^\prime_1 & 0 & 0 
& .. & 0 & 0 & 0 \\
\varepsilon^2  & -2 \varepsilon^2 & 
\varepsilon^2 - \varepsilon S^\prime_2 & 0 &.. & 0 & 0 & 0  \\
0 & \varepsilon^2 + \varepsilon S^\prime_1 & -2 \varepsilon^2 & 
\varepsilon^2 - \varepsilon S^\prime_3 &.. & 0 & 0 & 0  \\
0 & 0 & \varepsilon^2 + \varepsilon S^\prime_2 & -2 \varepsilon^2 & .. 
& 0 & 0 & 0 \\
. & . & . & . & .. & . & . & . \\
0 & 0 & 0 & 0 & .. &  -2 \varepsilon^2 & \varepsilon^2 - 
\varepsilon S^\prime_{n-1} & 0 \\
0 & 0 & 0 & 0 & .. &  \varepsilon^2 + \varepsilon S^\prime_{n-2} &
-2 \varepsilon^2 & \varepsilon^2   \\
0 & 0 & 0 & 0 & .. & 0 
& 2 \varepsilon^2 + \varepsilon S^\prime_{n-1} & -2 \varepsilon^2  
\end{array}
\right),
\quad
{\cal B} 
=
\left( \begin{array}{cc}
0 & I \\
{\cal D} & - 2 I
\end{array}
\right),
$$}%
where $I$ is $(n+1) \times (n+1)$ identity matrix and $\varepsilon$
is given by (\ref{defvarepsilon}). 
Then, the solution of 
(NU) with boundary conditions (\ref{boundarycondNJ})
is approximated by the solution of a system of ordinary differential equations
\begin{equation}
\frac{\mbox{d}z}{\mbox{d}t} = {\cal B} z, \qquad z(0) = z_0,
\label{odeNU}
\end{equation}
where $z_0$ is a prescribed initial condition.

\subsection{Efficiency of projective integration}

\label{secefficiency}

\begin{figure}[here]  
\picturesAB{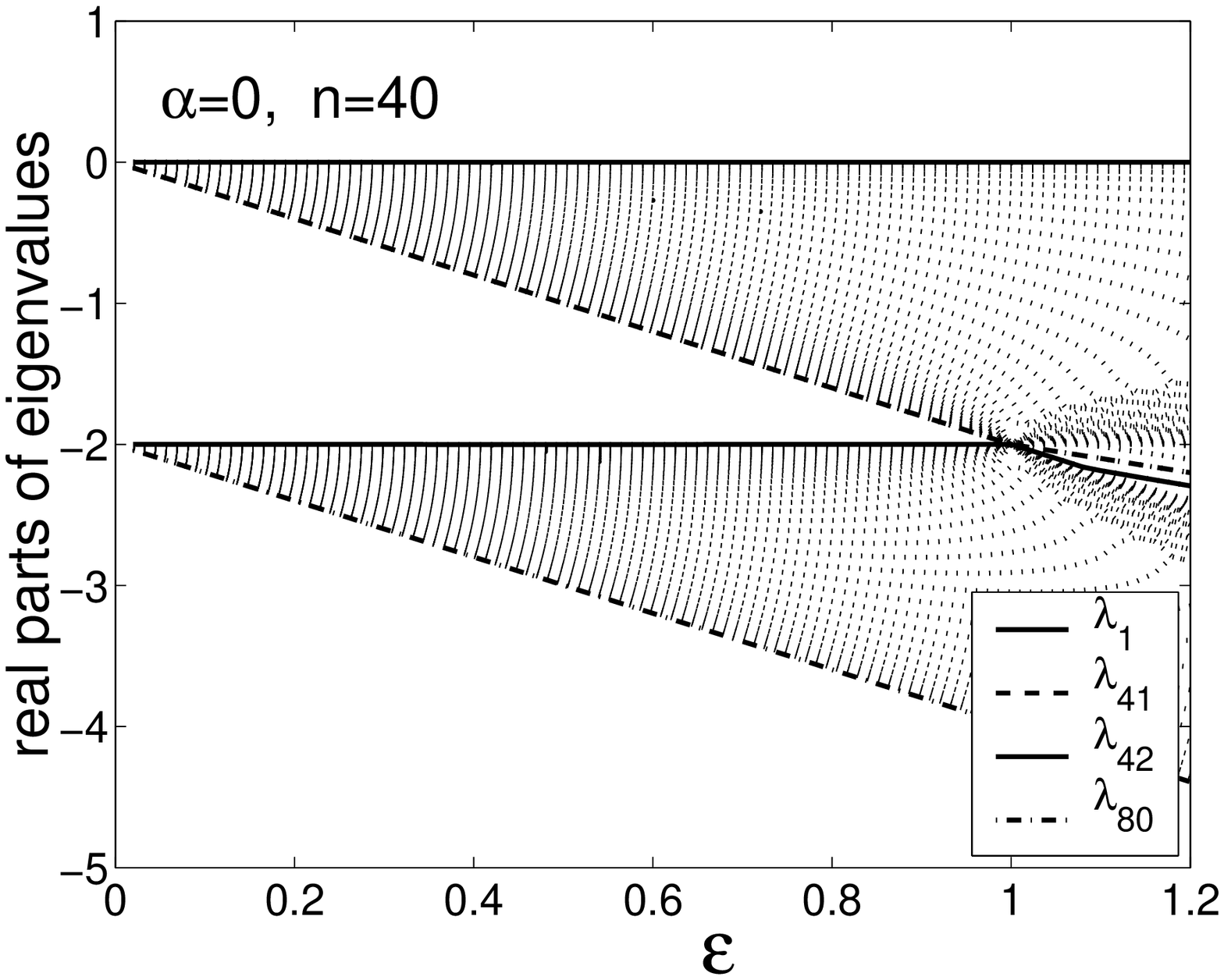}{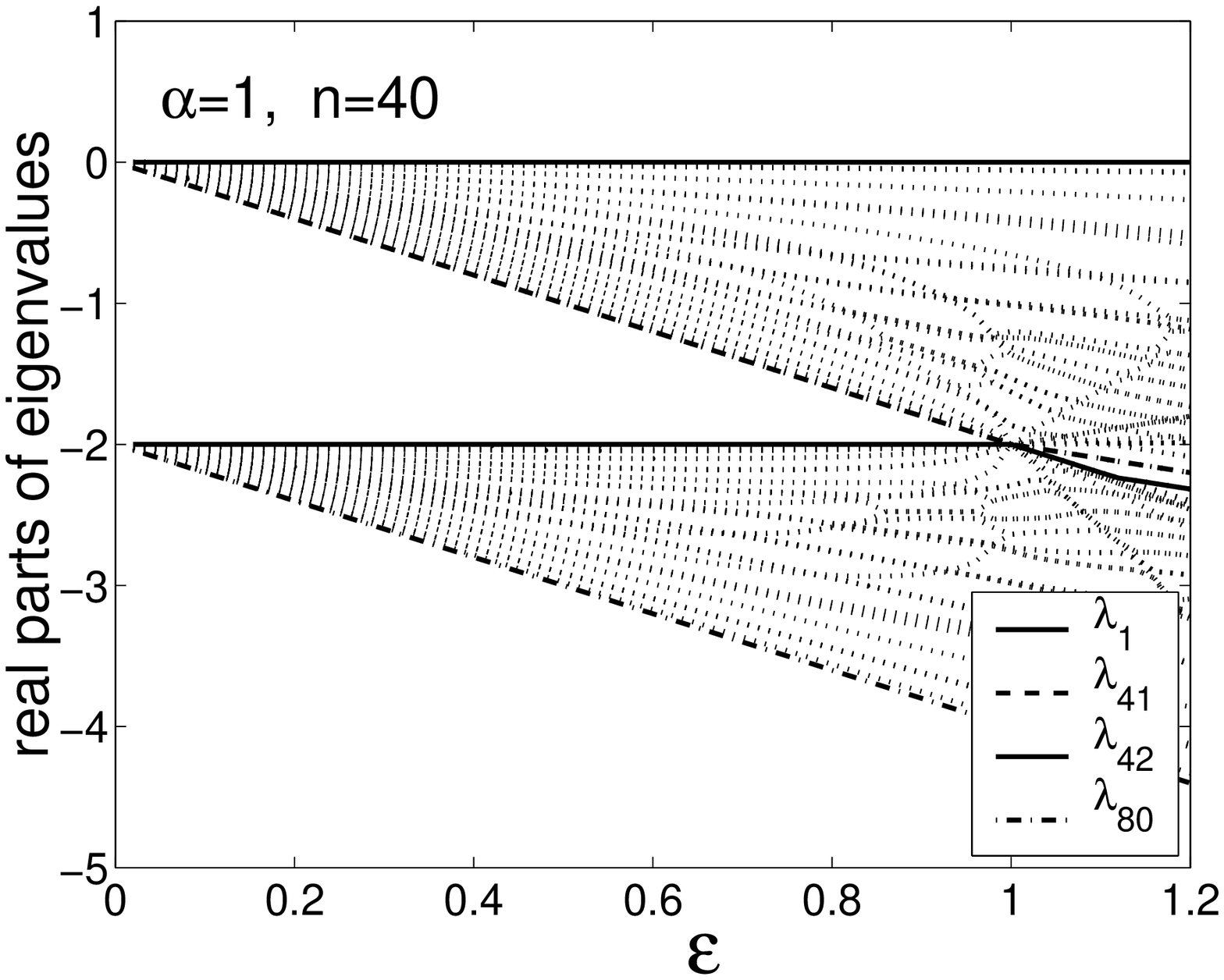}{2.2 in}
\caption{(a) {\it  
Graph of real parts of eigenvalues of matrix ${\cal A}$ for $n=40$
and no signal gradient in the environment, {\em i.e.,} 
$S^\prime_0 = S^\prime_1 = \dots S^\prime_n = 0.$
The eigenvalues are real for $\varepsilon \in [0,1]$
and there is a spectral gap between $\lambda_{41}$ and $\lambda_{42}.$
}
(b) {\it Graph of real parts of eigenvalues of matrix ${\cal A}$ 
for signal given by $(\ref{salphadefin})$ with $\alpha=1$ and for $n=40$.} 
}
\label{figeigenA}
\end{figure}

\begin{figure}[ht]  
\picturesAB{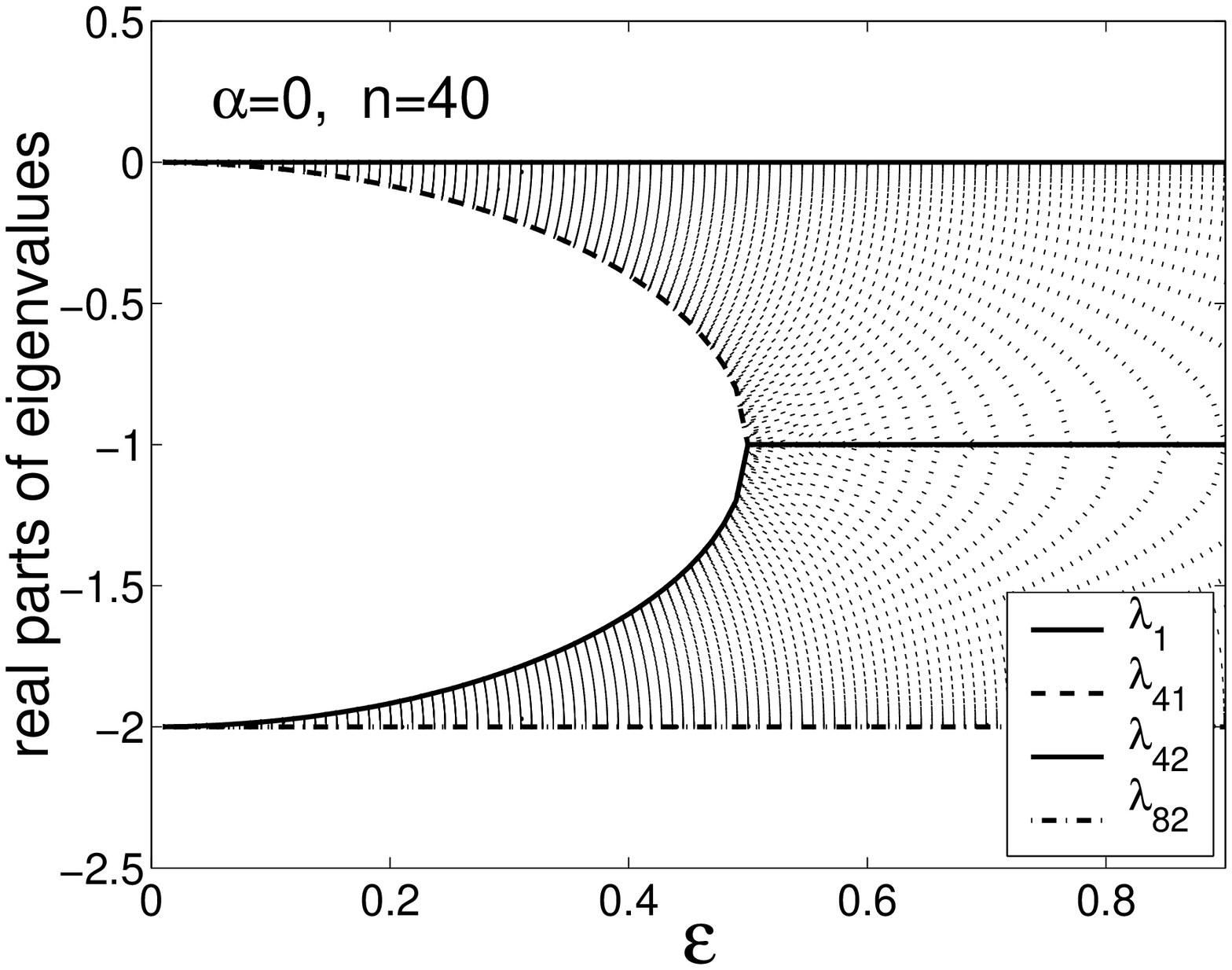}{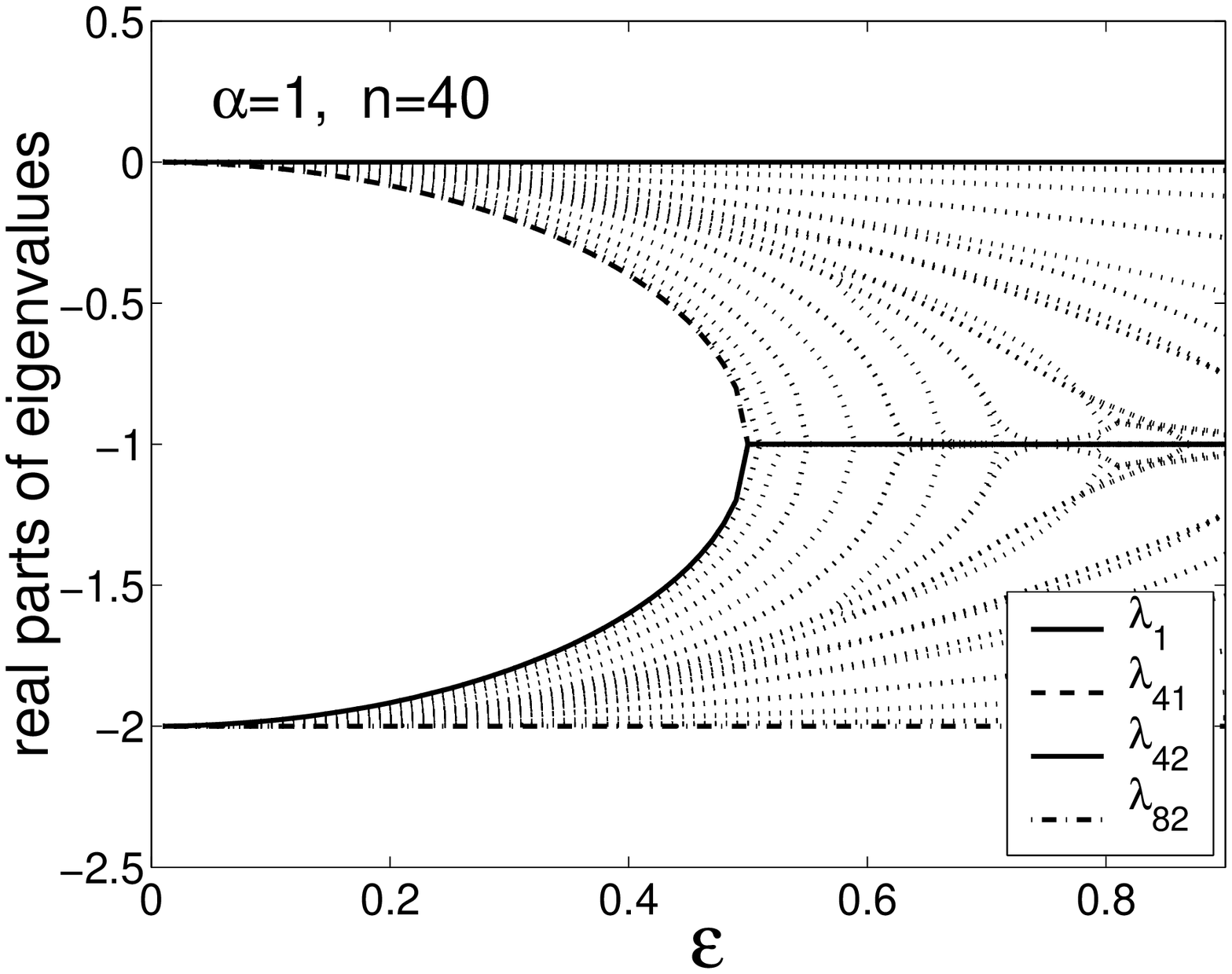}{2.2 in}
\caption{(a) {\it  
Graph of real parts of eigenvalues of matrix ${\cal B}$ for $n=40$
and no signal gradient in the environment, {\em i.e.,} 
$S^\prime_0 = S^\prime_1 = \dots S^\prime_n = 0.$
The eigenvalues are real for $\varepsilon \in [0,0.5]$
and there is a spectral gap between $\lambda_{41}$ and $\lambda_{42}.$
}
(b) {\it Graph of real parts of eigenvalues of matrix ${\cal B}$ 
for signal given by $(\ref{salphadefin})$ with $\alpha=1$ and for $n=40$.} 
}
\label{figeigenB}
\end{figure}

First, suppose that there is no signal gradient 
in the domain of interest, {\em i.e.,} we put
$S^\prime_0 = S^\prime_1 = \dots S^\prime_n = 0$
in matrices ${\cal A}$ and ${\cal B}$. 
Choosing $n=40,$
the real parts of the eigenvalues of ${\cal A}$ and ${\cal B}$ as a function
of $\varepsilon$ are plotted in Figures \ref{figeigenA}(a)
and \ref{figeigenB}(a), respectively. 
We see that there is a clear
spectral gap for small $\varepsilon.$
The eigenvalues of ${\cal A}$ are all real for $\varepsilon
\in [0,1]$ and they satisfy
\begin{equation}
\lambda \in \Big[ - 2 \varepsilon, 0 \Big] 
\; \bigcup \; \Big( -2 - 2 \varepsilon, -2 \Big).
\label{eigenA}
\end{equation}
The eigenvalues of ${\cal B}$ are all real for $\varepsilon
\in [0,0.5]$ and for no signal in the environment and they satisfy
\begin{equation}
\lambda \in \left[- 1 + \sqrt{1 - 4 \varepsilon^2}, 0 \right] 
\; \bigcup \; \left[-2, - 1 - \sqrt{1 - 4 \varepsilon^2} \right].
\label{eigenB}
\end{equation}
The spectral gap between $-2 \varepsilon$ and $-2$ in the case of 
matrix ${\cal A}$ is independent of the signal as can be seen 
from Figure \ref{figeigenA}(b), where we use the signal profile 
(\ref{salphadefin}) with $\alpha=1.$ 
We see that some eigenvalues 
changed, but that the spectral gap between $\lambda_{41}$ and 
$\lambda_{42}$ survived. 
The imaginary parts of the eigenvalues do
not grow significantly with $\alpha,$ and consequently
the values of the real parts determine the stability of the scheme;
we can use results from Lemma \ref{lemmamaximalM} for matrix 
${\cal A}$ and small $\varepsilon \in [0,1).$
To do that, we specify the time step $\dt.$ Since we want $\lambda \dt$ 
close to -1 for eigenvalues corresponding to fast modes, we put
\begin{equation}
\dt = 0.5
\label{choiceofdt}
\end{equation}
Considering our scaling (\ref{scalinghat}), 
we see that (\ref{choiceofdt}) means that $\dt$ is equal to time 
$1/(2\gamma_0).$ Next, if $k$ is at least 2, then the ``component"
of the stability region around $-1$ is more extended
then its second component around $0$
(see Figure \ref{figstabregion}). 
Consequently, using Lemma 
\ref{lemmamaximalM}, the size of interval containing the 
slow eigenvalues determines the
gain $\ggain$ of the method. 
Using (\ref{eigenA}) for the matrix ${\cal A}$, we 
have
\begin{equation}
\ggain =  
\frac{1 + (1 - \varepsilon)^{k+1}}{\varepsilon (1 - \varepsilon)^{k}(k+1)},
\qquad
\mbox{which is approximately}
\;
\frac{2}{\varepsilon(k+1)}
\;
\mbox{for small} 
\;
\varepsilon.
\label{gainA}
\end{equation}
Note that the gain $\ggain$, given by (\ref{gainA}), is {\it independent} of 
the signal strength $\alpha$ and it can be very large for small $\varepsilon.$ 
On the 
other hand, as we will see in Section \ref{secaccuracy}, the choice of small 
$\varepsilon$ will decrease the {\it accuracy} of the upwind  discretization ${\cal A}$ 
due to the strong artificial diffusion of the scheme.

\begin{figure}[here]  
\picturesABC
{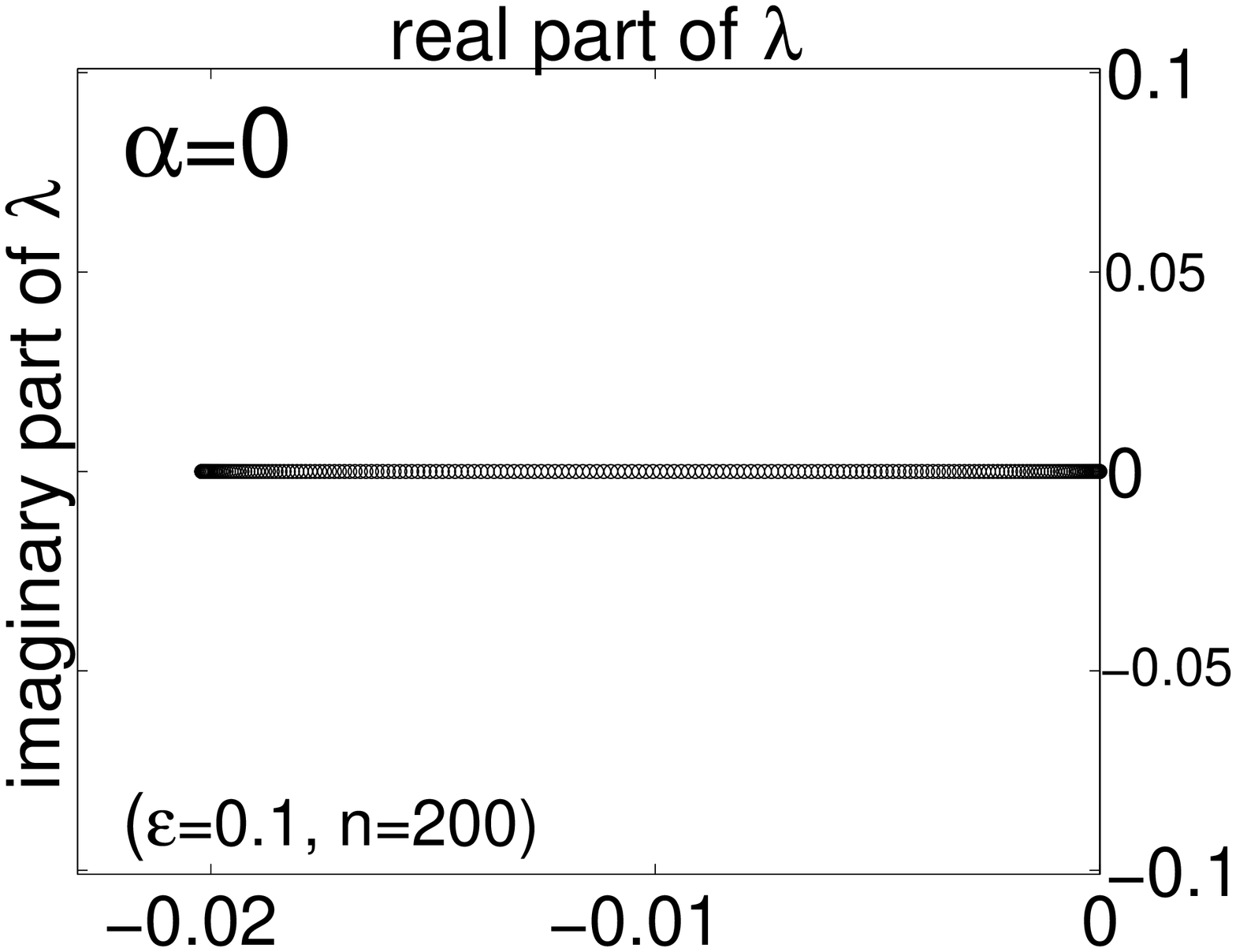}
{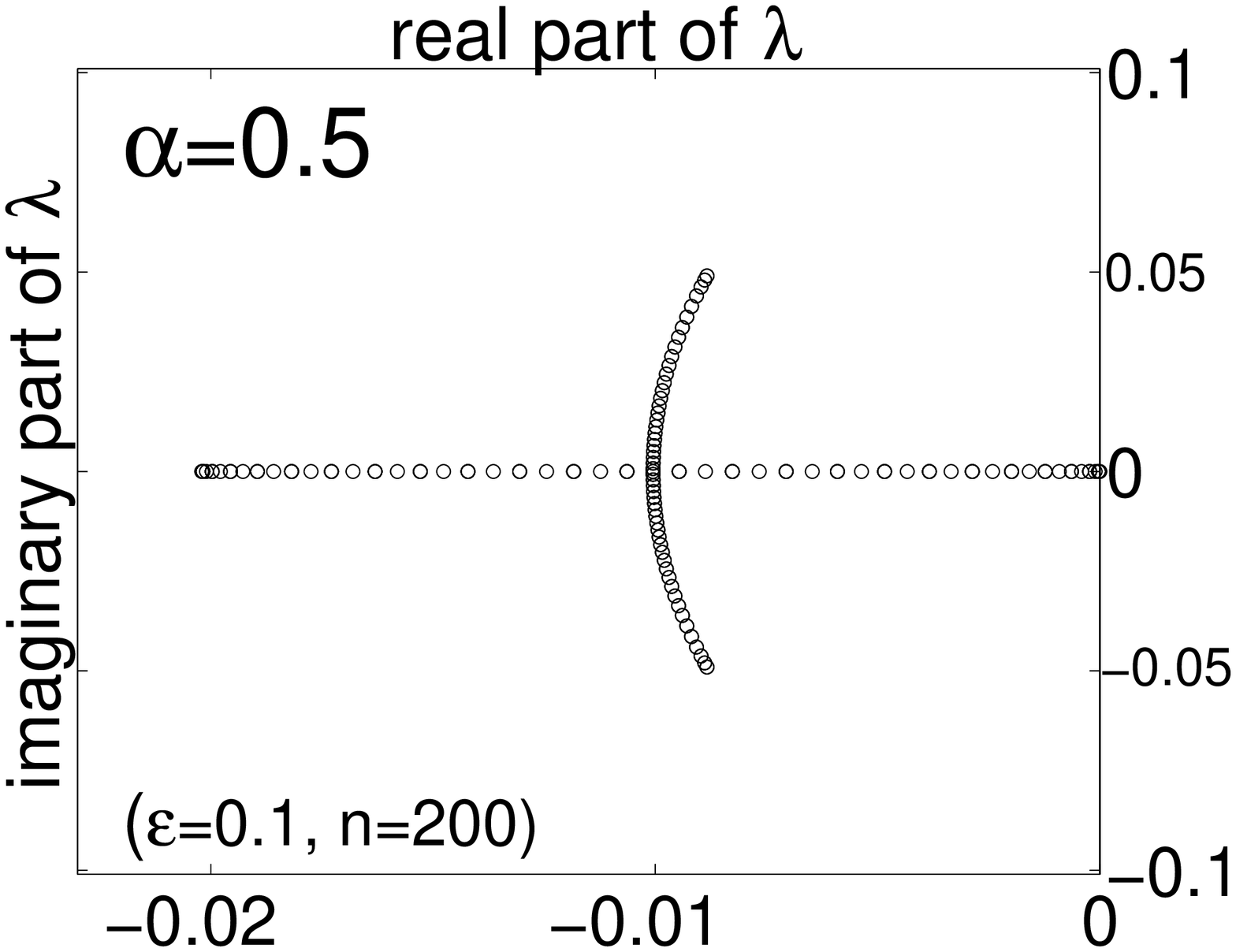}
{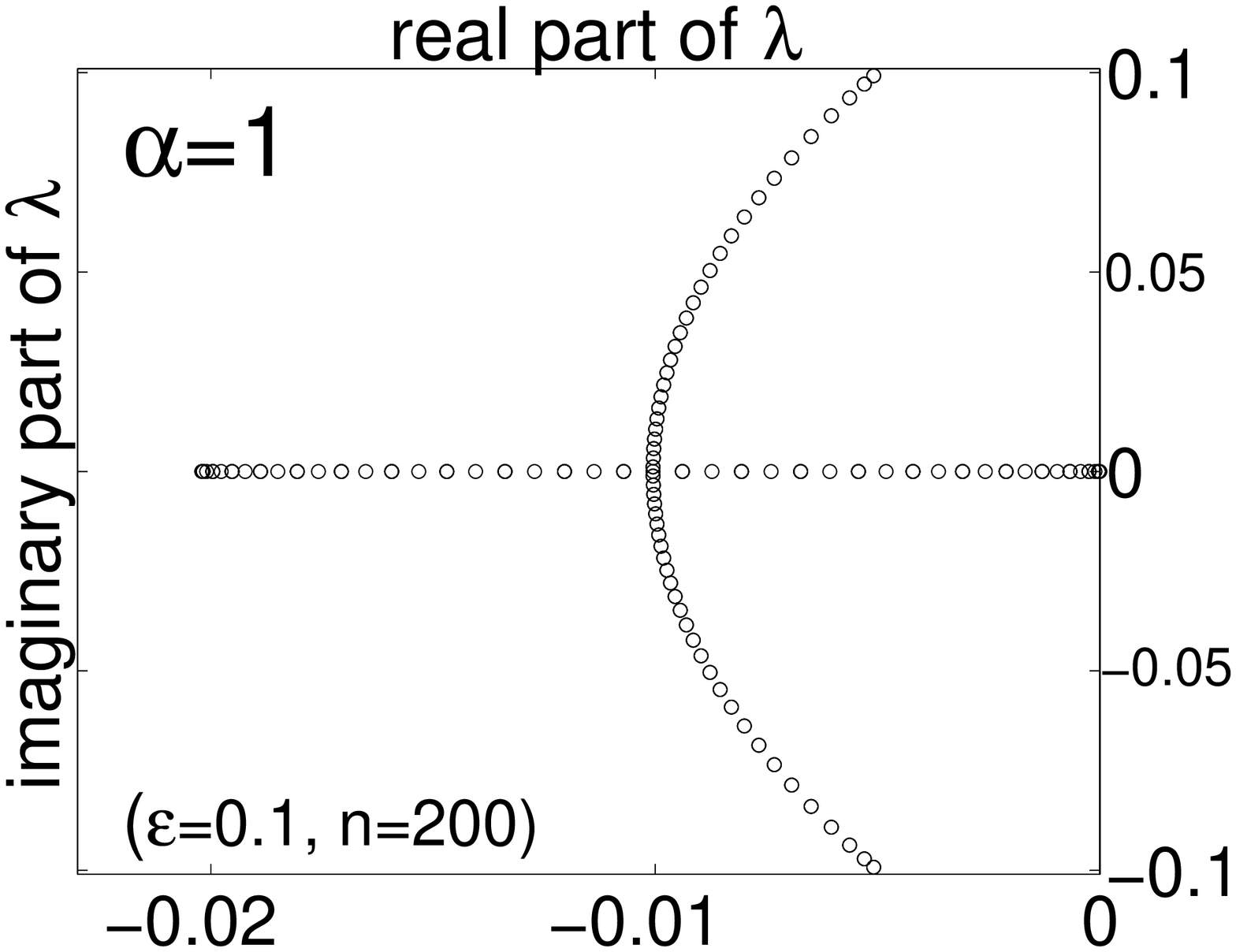}
{1.4 in}
\caption{{\it  
A plot of ``slow" eigenvalues of matrix ${\cal B}$ for $n=200$
and $\varepsilon = 0.1.$ We plotted only eigenvalues
close to zero for different strength of the signal $\alpha$
from (\ref{salphadefin}), namely:} (a) $\alpha=0,$
(b) $\alpha=0.5$ and (c) $\alpha=1.$}
\label{figeigenBcomplex}
\end{figure}

Next consider the matrix ${\cal B}$.  The real parts of its
eigenvalues as functions of $\varepsilon$ are plotted in Figure
\ref{figeigenB}.
We see that the ``boundary" eigenvalues 
$-2,$ $- 1 + \sqrt{1 - 4 \varepsilon^2},$
$- 1 + \sqrt{1 - 4 \varepsilon^2}$, $0$ of ${\cal B}$ are 
signal independent. 
The eigenvalues are all real for $\varepsilon \in [0,0.5]$
and for no signal in the environment. 
However, if we increase the signal
strength $\alpha$ some eigenvalues become complex, as can
be seen from Figure \ref{figeigenBcomplex}, where we plot
the ``slow" eigenvalues close to zero in the complex plane for
$n=200$, $\varepsilon = 0.1$, and for different signal strengths.
Choosing $\dt$ by (\ref{choiceofdt}) and $k$, we can (for small
signals) apply the results of Lemma \ref{lemmamaximalM} to compute 
maximal possible  projective jump $M$ and, hence, to compute the gain $\ggain$ 
by (\ref{gain}) for small $\varepsilon$ and for small signals.
Using (\ref{eigenB}) and Lemma \ref{lemmamaximalM} for the matrix ${\cal B}$
for small signal gradients, 
we have
\begin{equation}
\ggain = 
\frac{1 + (1 - 4 \varepsilon^2)^{(k+1)/2}}
{(- 1 + \sqrt{1 -  4 \varepsilon^2}) (1 - 4 \varepsilon^2)^{k/2}(k+1)},
\qquad
\mbox{which is approximately}
\;
\frac{1}{\varepsilon^2(k+1)}
\;
\mbox{for small} 
\;
\varepsilon.
\label{gainB}
\end{equation}
We see that discretization (\ref{odeNU}) results in a very large gain $\ggain$ 
for small $\varepsilon$ and for small signal gradients. 
On the other hand, if
we increase the signal gradients, then the result (\ref{gainB}) is no longer
true, because complex eigenvalues can appear outside the stability region
(compare Figure \ref{figeigenBcomplex} and Figure \ref{figstabregion}).
For example, we see from Figure \ref{figeigenBcomplex}(c) that the slow
eigenvalues lie in the complex interval $[-0.02,0] \times [0.1 i,0.1 i]$ 
for $\alpha = 1.$ 
Consequently, the absolute values of the imaginary parts of the 
eigenvalues are much larger than the absolute values of the
real parts and the result (\ref{gainB}) is not applicable for large  signals.

\subsection{Accuracy of projective integration}

\label{secaccuracy}

As we see in (\ref{gainA}), (\ref{gainB}) and (\ref{defvarepsilon}), 
choosing a larger $\dx$ will make $\varepsilon$ smaller and we will have 
a larger gain $\ggain$ for the projective integration. 
On the other hand, a smaller 
$\dx$ will increase the accuracy of the numerical method obtained by 
(\ref{odeRL}) or (\ref{odeNU}). 
The right choice of $\dx$ depends
on the underlying signal. 
If we have signals with sharp second derivatives
and if we want to capture the detailed transient behavior accurately, we have to use
a sufficiently small $\dx.$ 
However, if we want to make use
of the spectral gaps (\ref{eigenA}) or (\ref{eigenB}), we must 
assure that $s \ll \dx$ to have $\varepsilon \ll 1.$ 

Two types of errors arise in these computations: 
(1) the error between the projective integration 
of (\ref{odeRL}) or (\ref{odeNU}) and  the corresponding solutions 
of (\ref{odeRL}) or (\ref{odeNU}), respectively; 
and (2) the error between solutions of (\ref{odeRL}) or (\ref{odeNU})
and the exact solution of (NJ).  The error in part (1) is sufficiently
small as will be seen in Section \ref{secnumexamples}; it is easy to
estimate this error here, since the exact solution of (\ref{odeRL}),
(\ref{odeNU}) or even of (NJ) can be found through careful,
error-controlled computations.  For microscopic simulations though,
when the corresponding macroscopic equation is not known, estimating
these errors becomes an important task; fortunately, numerical
analysis techniques for {\it on line} {\it a posteriori} error
estimates have been extensively developed for continuum problems, and
can be naturally incorporated in equation-free computation
\cite{Estep:2004:SCD}.  For example, comparing results of the same
computation with half the projective time step can be used to estimate
the error of the scheme and control projective time step selection;
comparable techniques for adaptive spatial meshing can also be used.
It is, however, important to note one ``twist" to traditional {\it a
posteriori} numerical error estimates: errors due to the estimation
scheme, {\em e.g.,} due to fluctuations in stochastic simulations;
this can be controlled through variance reduction schemes, either by
brute force computation of several replica simulations or possibly
through biasing for variance reduction
\cite{Melchior:1995:VRS}.  Beyond adaptive time steps, adaptive mesh
sizes and possibly variance reduction, we will discuss at the end of
the paper the adaptive check of the {\it level} at which a macroscopic
description closes, {\em i.e.,} the number of macroscopic variables
required, or the dimension of the ``slow manifold".

\begin{figure}[ht]  
\picturesAB{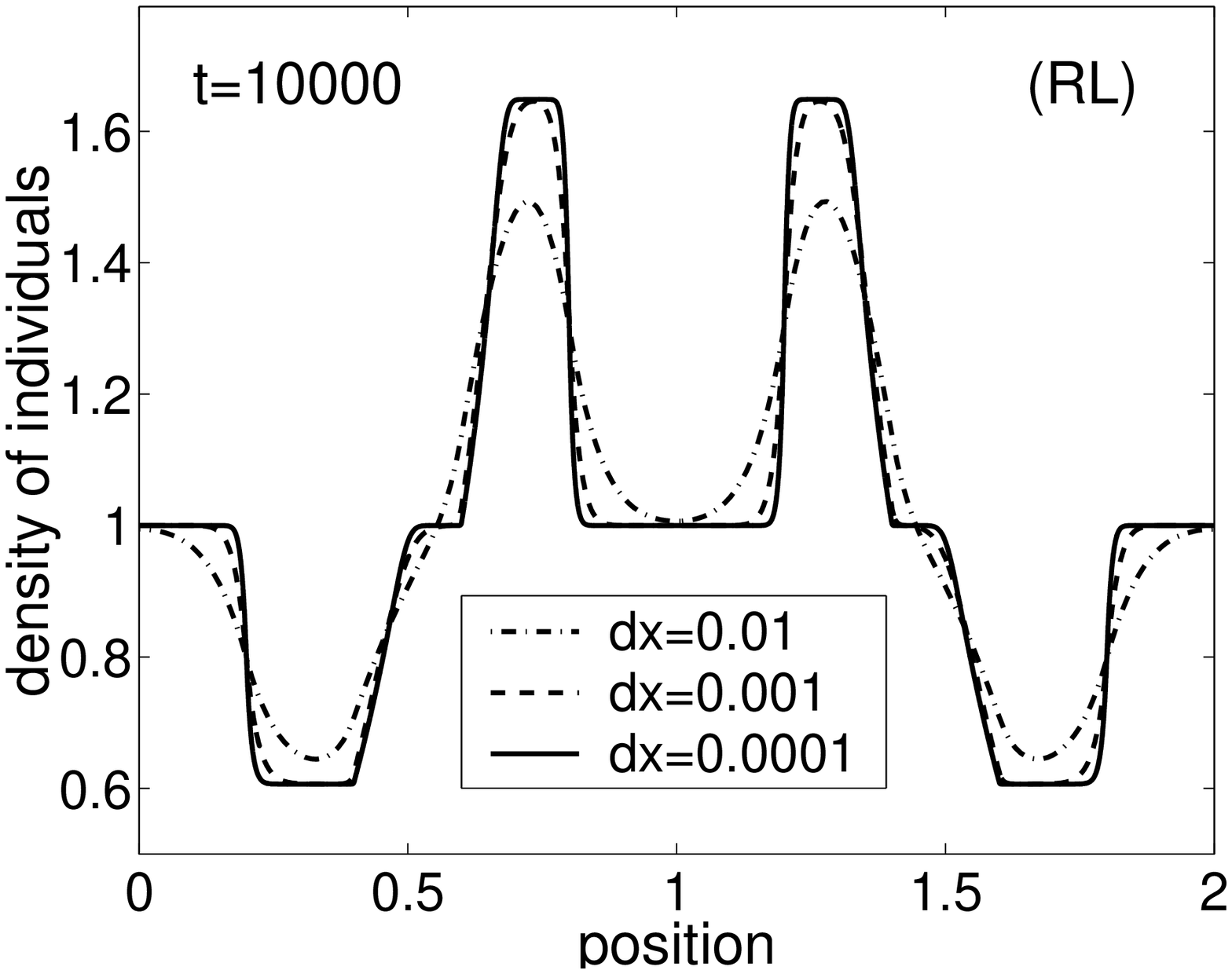}{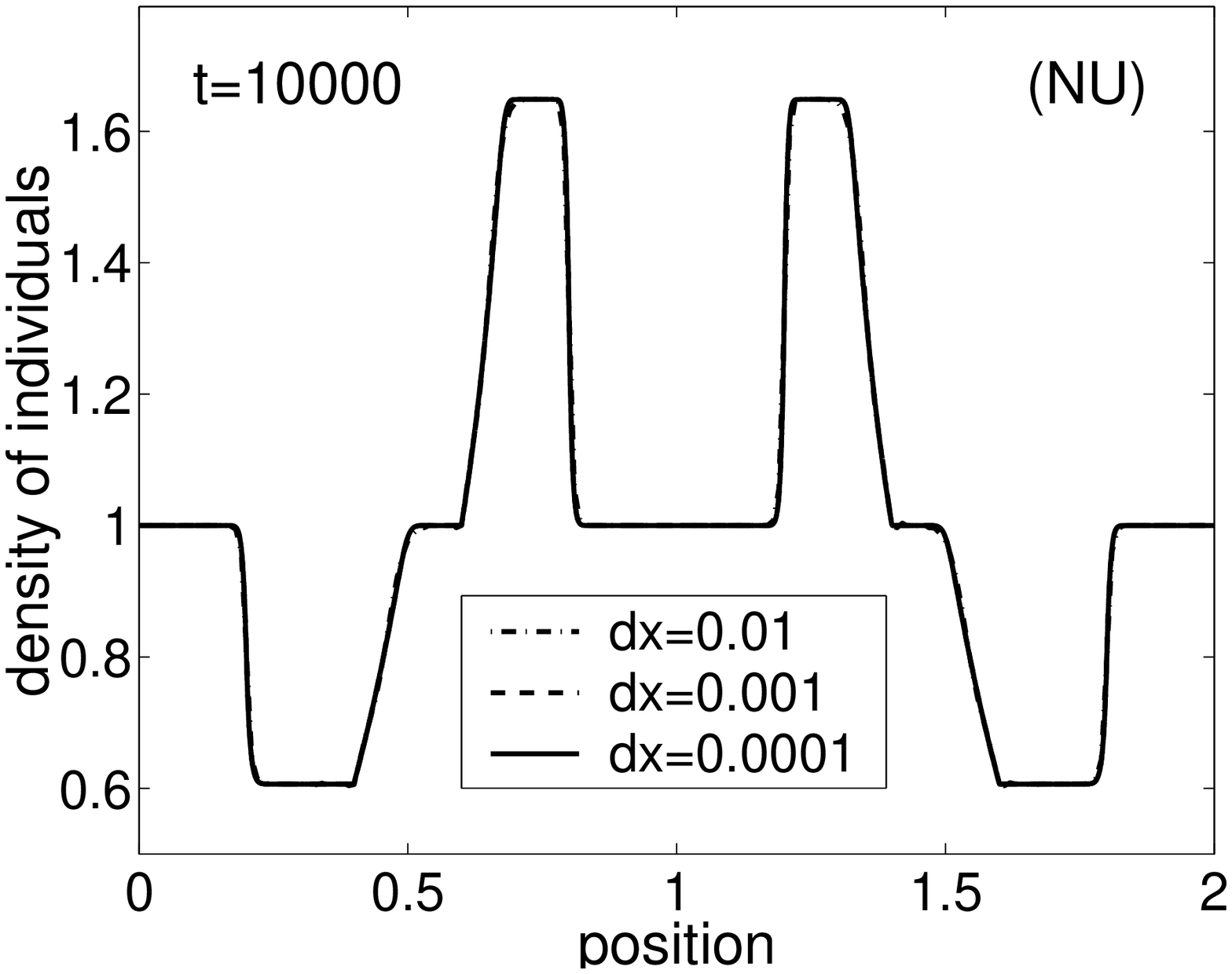}{2.2 in}
\caption{(a) {\it  
Graph of a solution of (RL) given by $\dot{w} = {\cal A} w$ at time $t=10^4$ 
for different choices of $\dx.$
We used $s = 0.0001$, signal given by
$(\ref{salphadefin})$ with $\alpha = 0.1,$ $\dt$ given by 
$(\ref{choiceofdt})$ and initial conditions $(\ref{initcondsigcoms})$. 
Consequently, $\dx=0.01,$  $\dx = 0.001$ and $\dx = 0.0001$ correspond 
to $\varepsilon = 0.01,$
$\varepsilon = 0.1,$ and $\varepsilon = 1,$ respectively.
}
(b) {\it Graph of a solution of (NU) given by $\dot{z} = {\cal B} z$ 
at time $t=10^4$ 
for different choices of $\dx.$ The parameters are the same as in (a).
} 
}
\label{accuracy}
\end{figure}

We will now discuss errors of type (2),
{\em i.e.,} errors between solutions of (\ref{odeRL}) or (\ref{odeNU}) 
and the exact solution of (NJ). 
We can numerically estimate those errors by comparing the solution of 
(\ref{odeRL}) or (\ref{odeNU}) for different $\dx.$ 
Representative results can be found
in Figure \ref{accuracy} where we used $s = 0.0001$, a signal given by
(\ref{salphadefin}) with $\alpha = 0.1$ and $\dt$ given by (\ref{choiceofdt}).
As is well known,  the upwinding (\ref{odeRL}) discretization introduces artificial
diffusion to the problem which makes the solution more dependent on $\dx.$
The central differences discretization (\ref{odeNU}) is more accurate here
\cite{Strikwerda:1989:FDS}.

\subsection{Numerical examples}

\label{secnumexamples}

Here we present illustrative numerical results. In view of (\ref{sSxassume}), 
we choose
\begin{equation}
s = \frac{1}{10000}, \qquad \dx = 0.01, \qquad
\varepsilon = \frac{s}{\dx} = 0.01;
\label{numparam1}
\end{equation}
and we consider 201 mesh points (\ref{mesh}) in the interval $[0,2]$.
The time step $\dt$ is given by (\ref{choiceofdt}) and the initial 
condition is 
\begin{equation}
N(x,0) = 1, \qquad J(x,0) = 0.
\label{initcondsigcoms}
\end{equation}
We know from Section \ref{secaccuracy} that the discretization (\ref{odeNU}) 
gives rise to a sufficiently accurate solution of (NJ) for the
choice of parameters (\ref{numparam1}), so we start with the discretization 
(\ref{odeNU}) first. 
We learned in Figure \ref{figeigenBcomplex}
that we can have a large gain $\ggain$ of using projective inegration
of (\ref{odeNU}) if the signal gradient is small; consequently, 
we consider the signal (\ref{salphadefin}) with $\alpha = 0.1.$
\begin{figure}
\centerline{
\epsfxsize=3in\epsfysize=2.3in\epsfbox{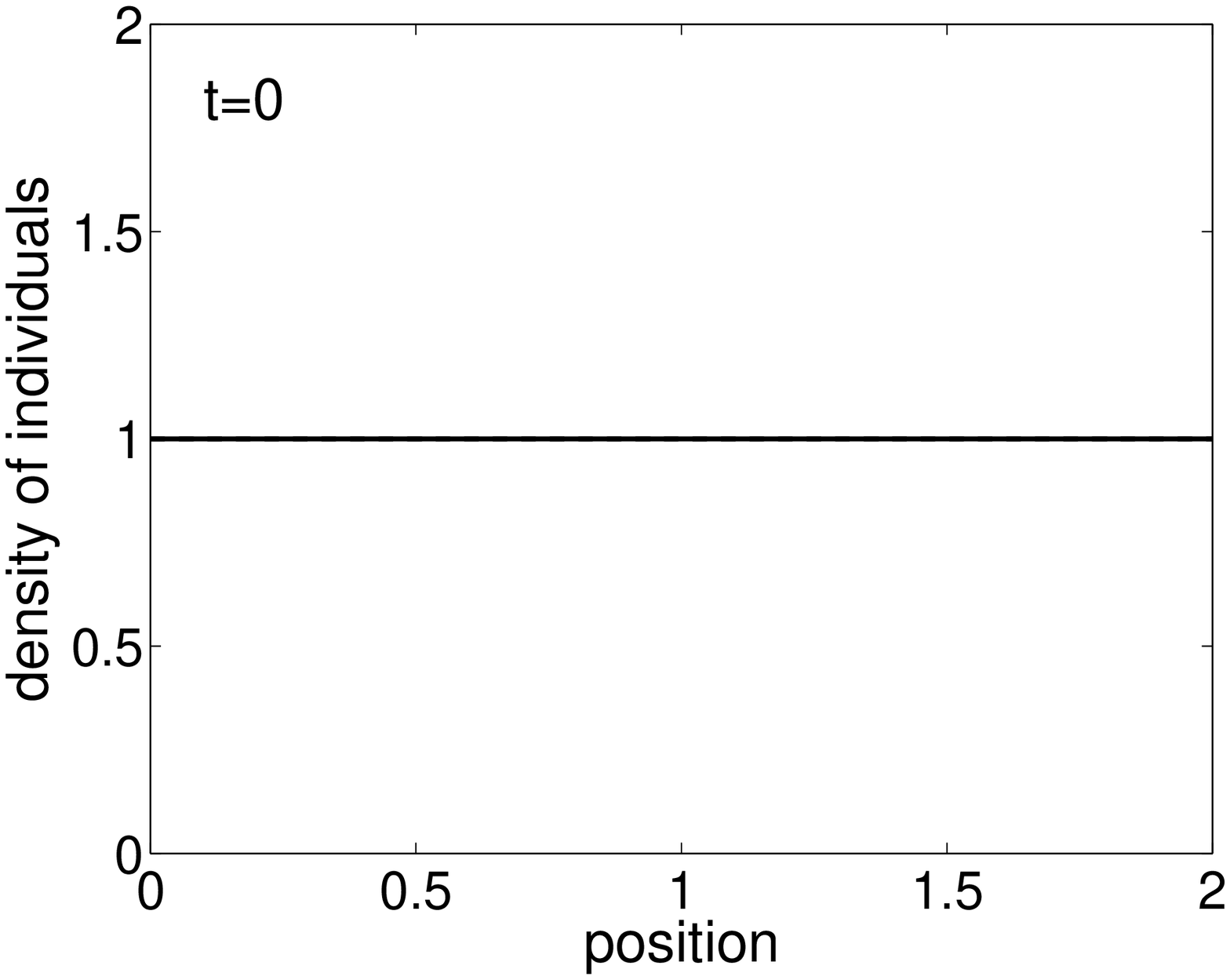}
\epsfxsize=3in\epsfysize=2.3in\epsfbox{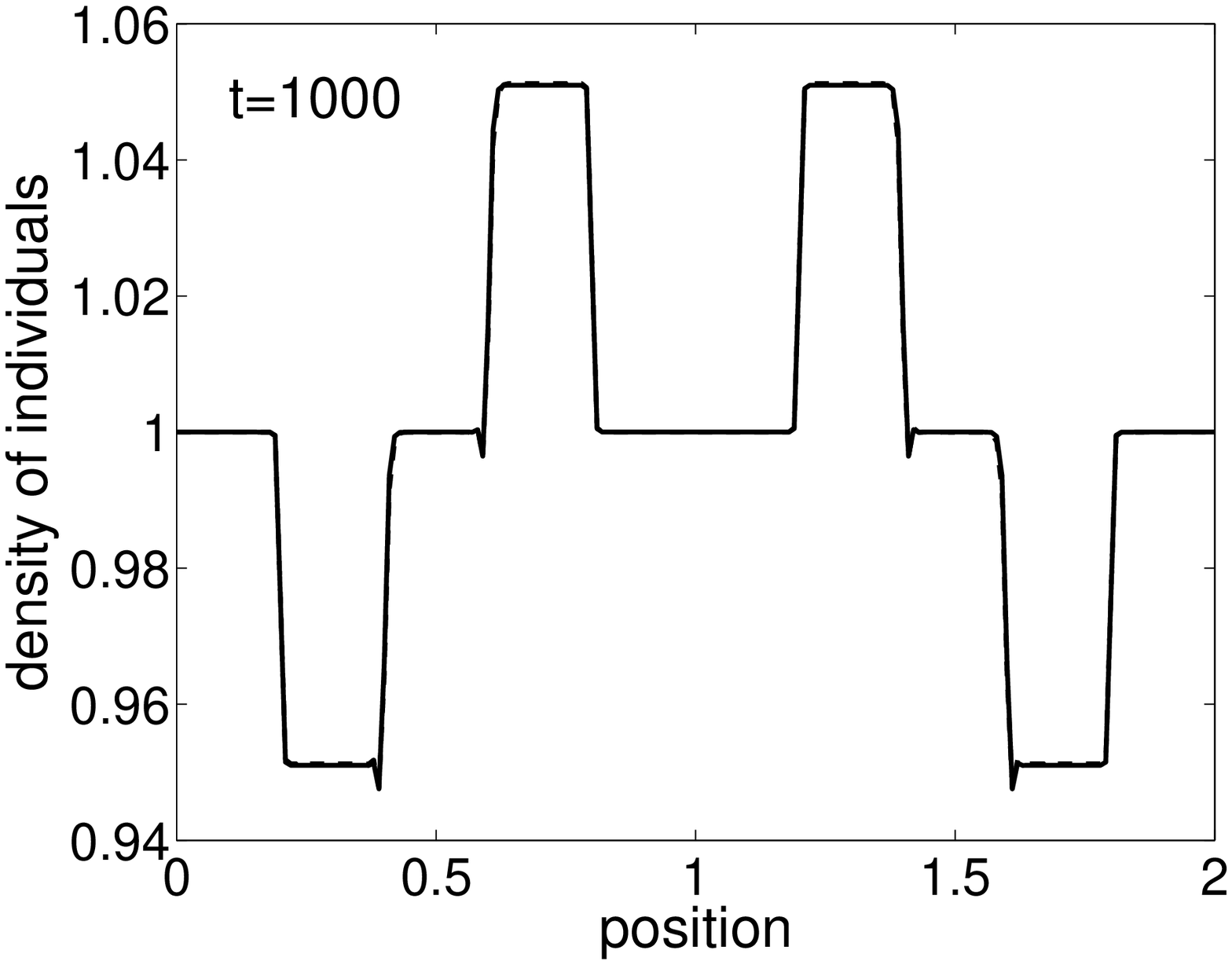}
}
\centerline{
\epsfxsize=3in\epsfysize=2.3in\epsfbox{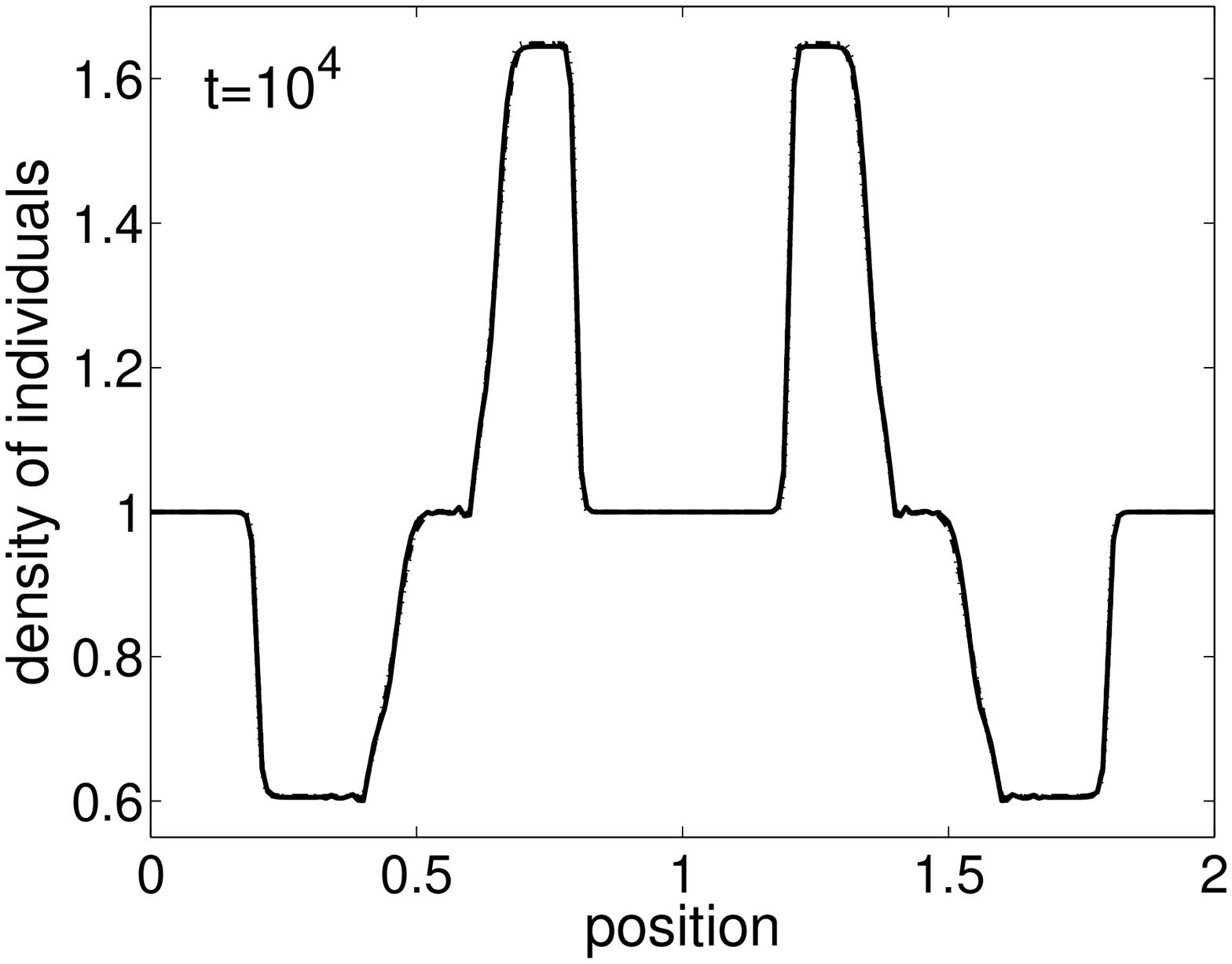}
\epsfxsize=3in\epsfysize=2.3in\epsfbox{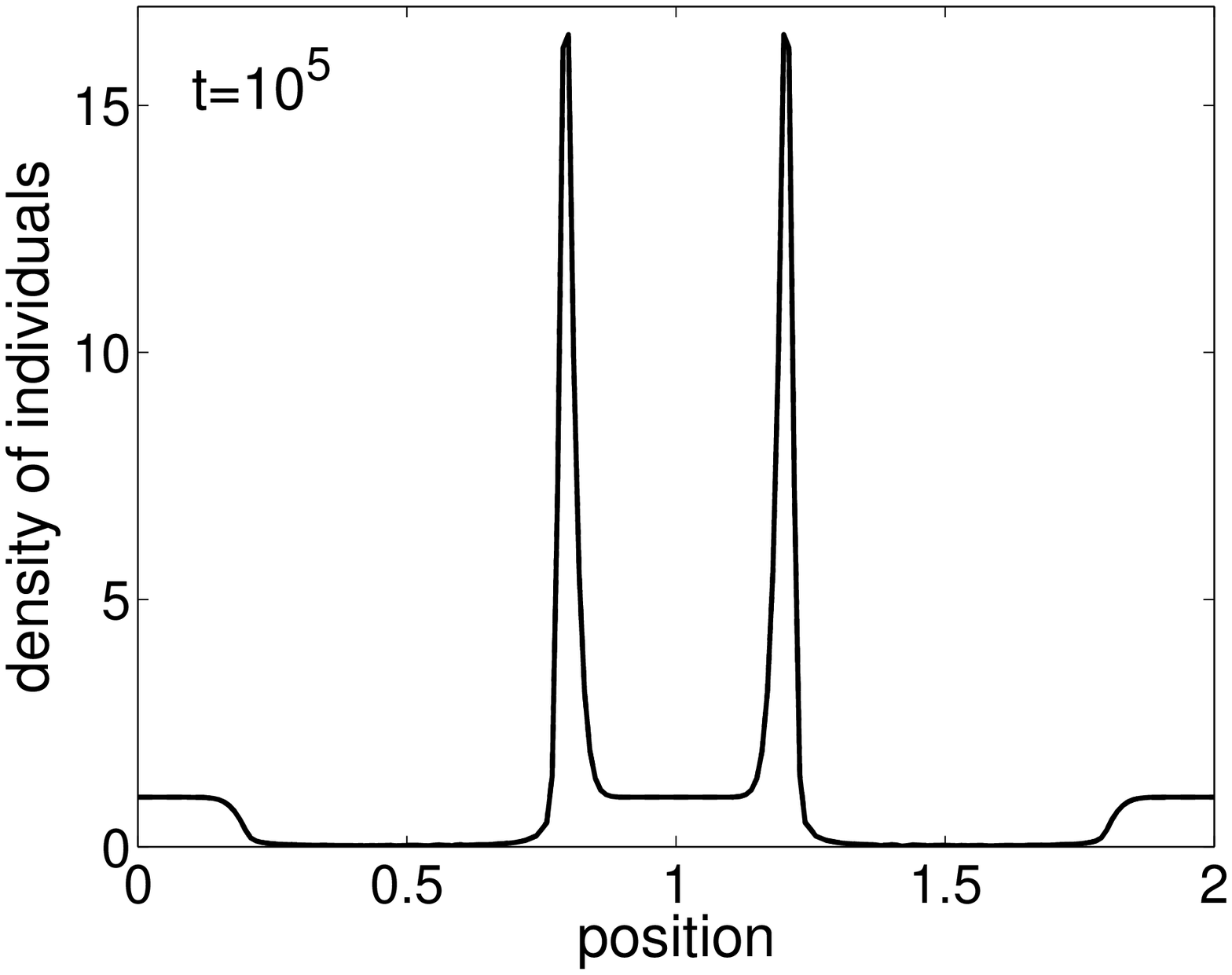}
}
\centerline{
\epsfxsize=3in\epsfysize=2.3in\epsfbox{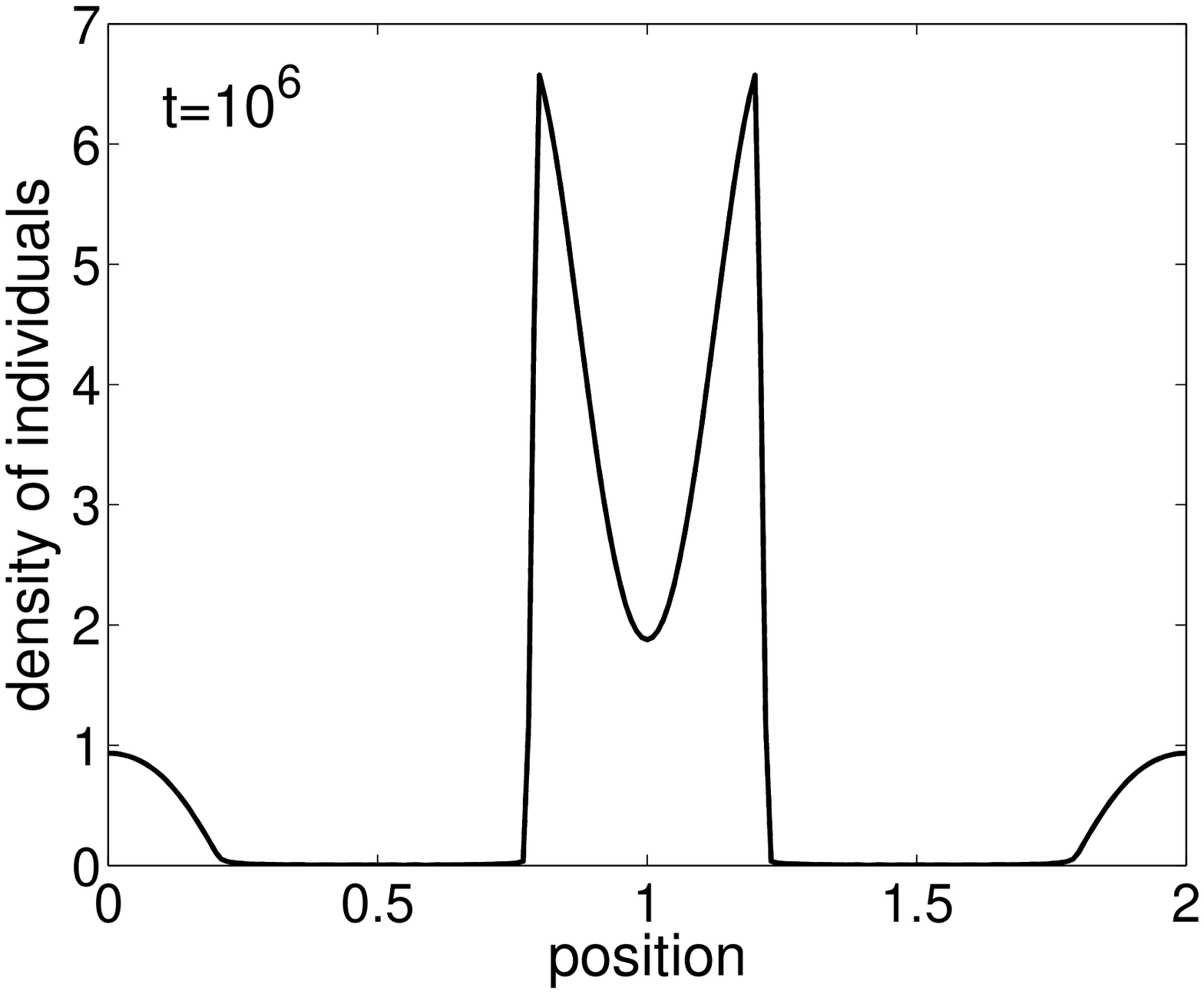}
\epsfxsize=3in\epsfysize=2.3in\epsfbox{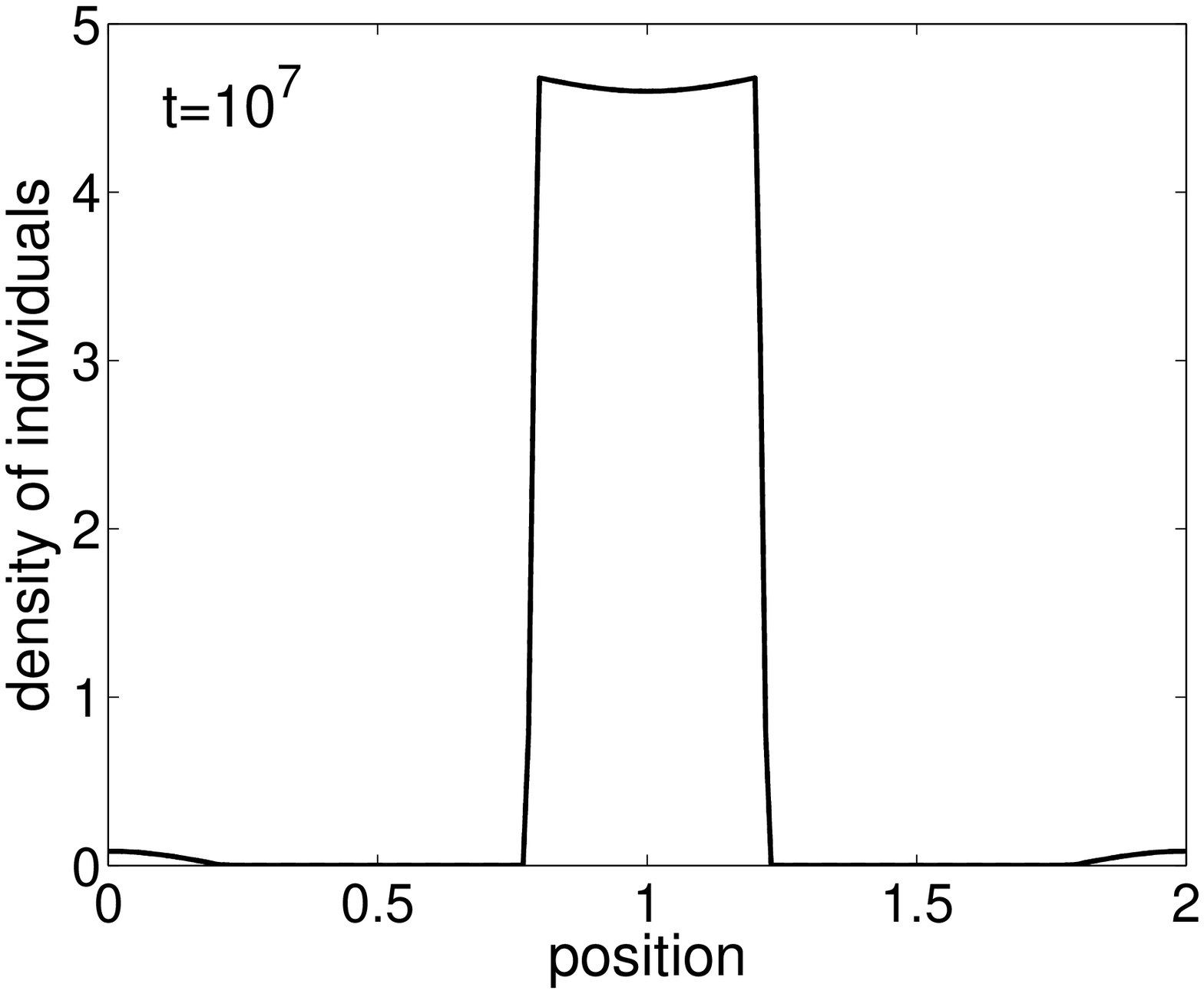}
}
\caption{{\it The time evolution of the density
of individuals for $s = 0.0001$ and $\alpha = 0.1$. 
We plot  the solution of the system $(\ref{odeNU})$ 
(dashed line) and the solution obtained by the projective algorithm 
{\rm (\Pkm-1) -- (\Pkm-3)} for $(\ref{odeNU})$ with $k=1$ and $M=398$ 
(solid line).
The gain $\ggain$ is $200$.
We use $(\ref{numparam1})$, $(\ref{choiceofdt})$
and $(\ref{initcondsigcoms})$.}}
\label{figprojectiveNU}
\end{figure}
The numerical results for $k=1$ and $M=398$ are given in Figure 
\ref{figprojectiveNU}. 
Here the gain is $\ggain = 200$ using definition 
(\ref{determgain}). 
In Figure \ref{figprojectiveNU}, we compare
the solution of system (\ref{odeNU}) with the projective integration
of (\ref{odeNU}). 
We see that the errors between the projective integration
of the system of ordinary differential equations (\ref{odeNU})
and the solution of (\ref{odeNU}) are small.
Since the discretization (\ref{odeNU}) gives a reasonably accurate
solution of (NJ), we can view also Figure \ref{figprojectiveNU}
as a plot of the exact solution of (NJ). 
Consequently, what we presented appears to be capable of significantly
speeding up an explicit forward Euler method for 
small signal gradients (see also 
\cite{Gear:2003:PMS,Lebedev:1997:EDS,Medovikov:1998:HOE}).

\begin{figure}[ht]  
\picturesAB{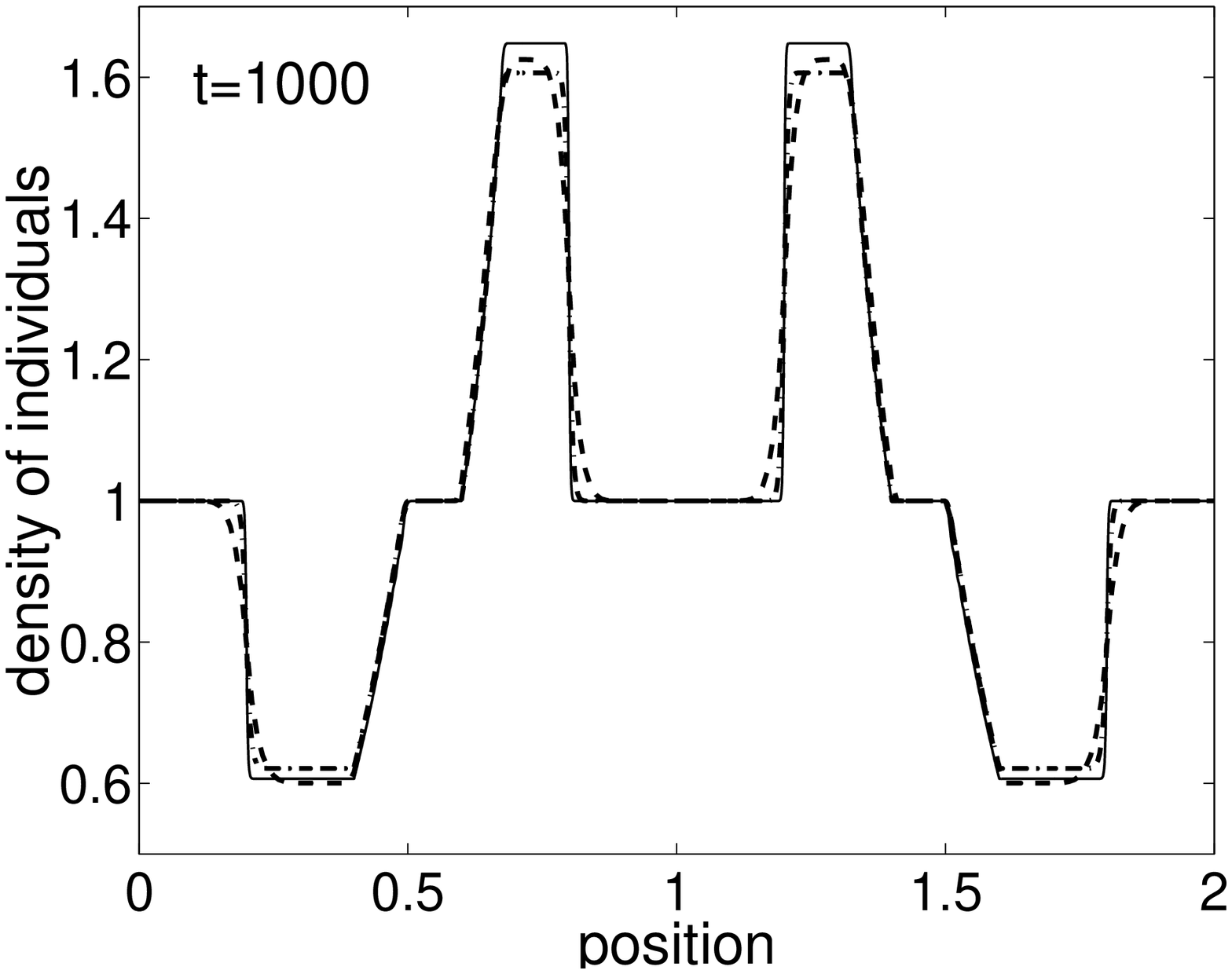}
{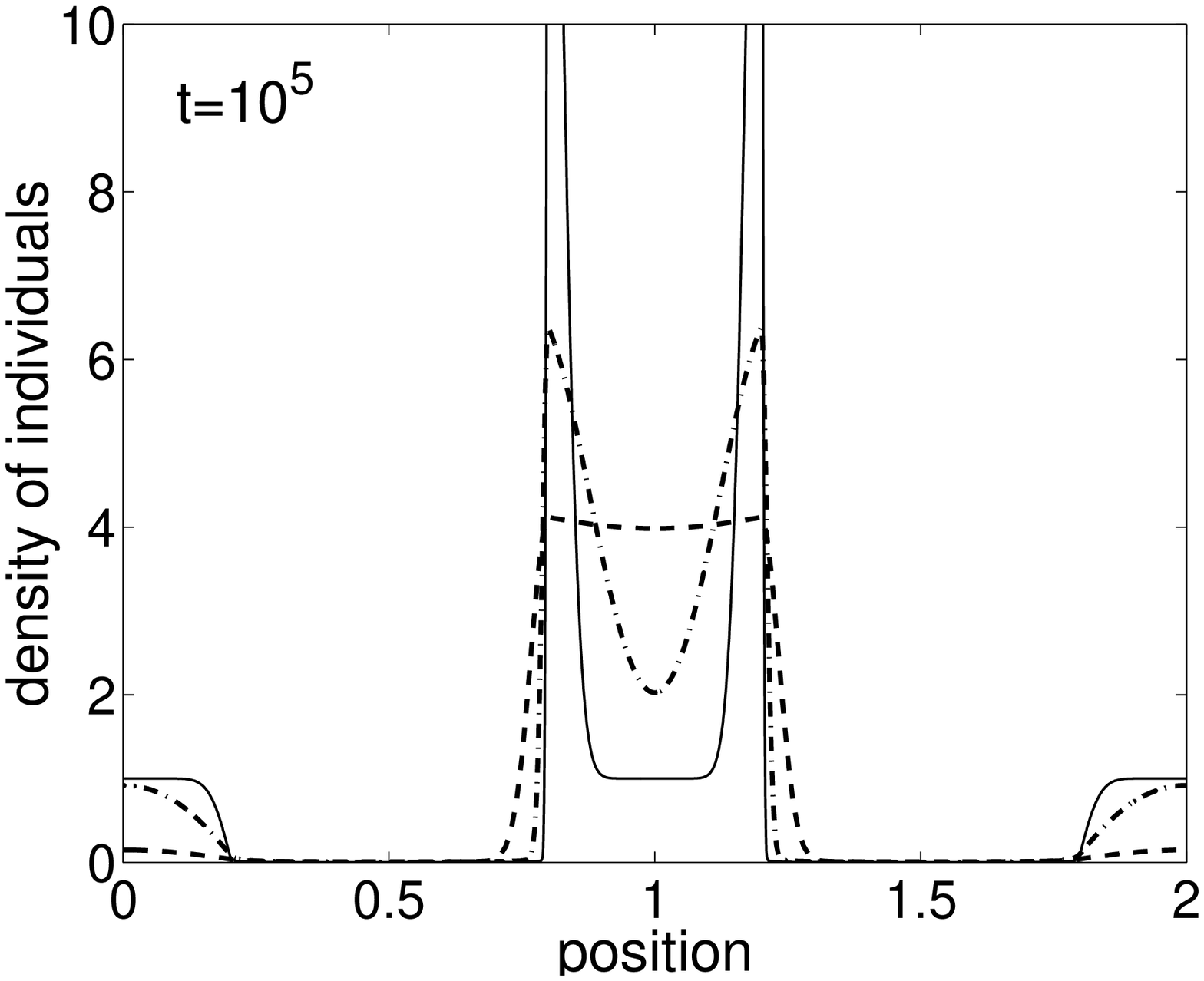}{2.2 in}
\caption{(a) {\it  
Density of individuals for $s = 0.0001$ and $\alpha = 1$ at time
$t=1000.$ We plot the solution of (NJ) given by accurate numerical
method (solid line), the solution obtained by projective algorithm 
{\rm (\Pkm-1) -- (\Pkm-3)} for $(\ref{odeRL})$
with $(\ref{numparam1})$, $k=1$ and $M=198$,
{\em i.e.,} with $\ggain = 100$ (dashed line).
We also plot the solution obtained by projective algorithm 
{\rm (\Pkm-1) -- (\Pkm-3)} for $(\ref{odeRL})$ 
with $(\ref{numparam2})$, $k=1$ and $M=18$,
{\em i.e.,} with $\ggain = 10$ (dot-dashed line). In all computations,
we use $(\ref{choiceofdt})$ and $(\ref{initcondsigcoms})$.
}
(b) {\it The same plots as in (a) for time $t=10^5.$} 
}
\label{figprojectiveRL}
\end{figure}

The second numerical example in this section is based on the upwind
discretization (\ref{odeRL}).  We know from Section \ref{secaccuracy}
that the discretization (\ref{odeRL}) provides a less accurate
solution of (NJ) than (\ref{odeNU}) for parameters (\ref{numparam1})
due to the artificial diffusion of the upwinding scheme.  On the other
hand, the gain $\ggain$ of the projective integration method
(\ref{odeRL}) is independent of the signal stregth $\alpha.$
Consequently, we will present here results for $\alpha=1,$ {\em i.e.,}
when the signal is maximal possible.  If we compare the results
obtained by projective integration of (\ref{odeRL}) and the
corresponding plots of solutions of (\ref{odeRL}), we again obtain
small errors (results not shown) similar to those in Figure
\ref{figprojectiveNU}.
This would again support
the numerical results from \cite{Gear:2001:PIM} concerning the
accuracy of projective integration of ordinary differential equations.
Instead, we compare the results of projective integration
for two different choices of $\varepsilon$ with an accurate solution
of (NJ). 
We use either (\ref{numparam1}) or
\begin{equation}
s = \frac{1}{10000}, \qquad \dx = 0.001, \qquad
\varepsilon = \frac{s}{\dx} = 0.1.
\label{numparam2}
\end{equation}
Moreover, we use the initial condition (\ref{initcondsigcoms}) and 
$\dt$ given by (\ref{choiceofdt}); the results are shown in 
Figure \ref{figprojectiveRL}.
We see that the long time behavior is highly influenced by the artificial
diffusion of the scheme. 
Projective integration with small $\varepsilon$
has large gain $\ggain$, but it will reach the steady state much faster
than the exact solution of (NJ). 
Note, that it is not an inaccuracy
in projective integration {\it per se}; the inaccuracy is created by 
the inaccurate spatial discretization, based on upwinding with small $\varepsilon.$

Next, we will discuss how the ideas described so far in this paper
can be used in Monte Carlo simulations of chemotaxis.

\section{Coarse projective integration: a kinetic Monte Carlo example}

\label{secmontecarlo}

In the previous sections we studied the gain $\ggain$ of projective
integration for the system (NJ) of deterministic partial differential
equations.  Here we present results of Monte Carlo simulations of the
underlying random walks  using {\it coarse projective integration} 
\cite{Gear:2001:PIM,Gear:2002:CIB,Kevrekidis:2003:EFM}
that will make use of the previous analysis.  While in principle we
simulate the evolution of the particle density profile over the entire
spatial domain, we will demonstrate how to perform the computations
required for coarse projective integration on a relatively small
portion of the full domain. This is based on the presumed smoothness
{\it in physical space} of the evolving density profile, which
constitutes the underpinning of equation-free methods such as the {\it
gap-tooth scheme}
\cite{Gear:2003:GTM,Samaey:2003:GTS,Kevrekidis:2003:EFM} as described
below.  Here we are able to speed up the kinetic Monte Carlo
simulation about a thousand times.

Suppose that we have $2 n_0$ random walkers in the interval $[0,2]$,
and suppose that we have only a kinetic Monte Carlo simulator to model
the evolution of the system.  As before, the interesting macroscopic
quantity is the density of random walkers $N$ which can be obtained as
follows.  We choose a macroscopic mesh size $\dx$ and we discretize
the interval $[0,2]$ using mesh (\ref{mesh}).  Then we obtain the
(probability) density $N_{i+1/2}(t)$ at point $\frac{x_i+x_{i+1}}{2}$
as the number of particles in the interval $[x_{i},x_{i+1}]$ divided
by $n_0 \dx.$
We thus create a histogram of particles, which can of course be noisy.

If we have randomly walking {\it noninteracting} particles, the
histograms obtained by using $10^6$ or $10^9$ random walkers appear
roughly the same; the former is just less ``noisy" than the latter.
Consequently, we can obtain relatively accurate results quickly by
simply decreasing the number of particles.  However, in many
interesting biological problems, cells change their environment, they
consume nutrients, secrete waste, etc.  Consequently, cells {\it
interact} through environmental chemicals and then the number of cells
is prescribed by the biological setup, and we cannot change it without
changing the computed solution.

Therefore, in the examples of this section we will suppose that we do
not know that the particles are noninteracting; we will suppose that
there is a fixed number $2 n_0$ of individuals in the domain of
interest - the interval [0,2] - which are moving according to the
rules of the random described in Section \ref{secdirchem}.  We will
show that in the case of a fixed number of particles the coarse
integration method leads to an even larger gain $\ggain$ than the
projective integration method used earlier, where as before, the gain
$\ggain$ is defined by (\ref{generalgain}).
In the numerical  example, we choose
\begin{equation}
2 n_0 = 10^8,
\qquad
s = \frac{1}{10000}, 
\qquad \dx = 0.01, 
\label{numparam3}
\end{equation} 
{\em i.e.}, we consider 201 equi-spaced mesh points (\ref{mesh}) in the interval $[0,2]$
on which we track the evolution of the macroscopic density, 
and the parameters are the same as in (\ref{numparam1}).

Monte Carlo simulations are performed as follows.  Each particle is
described by two variables -- position $x \in [0,2]$ and velocity $\pm
s.$ We use a small microscopic time step $\mbox{d}t=0.01$, {\em i.e.},
the unbiased turning frequency divided by 100, and during each time step
the particle moves with speed $s$ in the chosen direction.  At the end
of each time step, a random number chosen from a uniform distribution
on $[0,1]$ is generated and compared with the probability of the turn
$\gamma \mbox{d}t = (1 \pm S^\prime(x)) \mbox{d}t$.  If a turn occurs,
the cell will move in the opposite direction during the next time
step.  To apply the previous results, we choose a macroscopic time
step $\dt$ given by (\ref{choiceofdt}) and do  kinetic Monte
Carlo simulations in the interval  $[t,t + \dt]$, which  means that we use the
Monte Carlo simulator for $\dt/\mbox{d}t$ microscopic time steps
$\mbox{d}t.$

\begin{figure}
\picturesAB{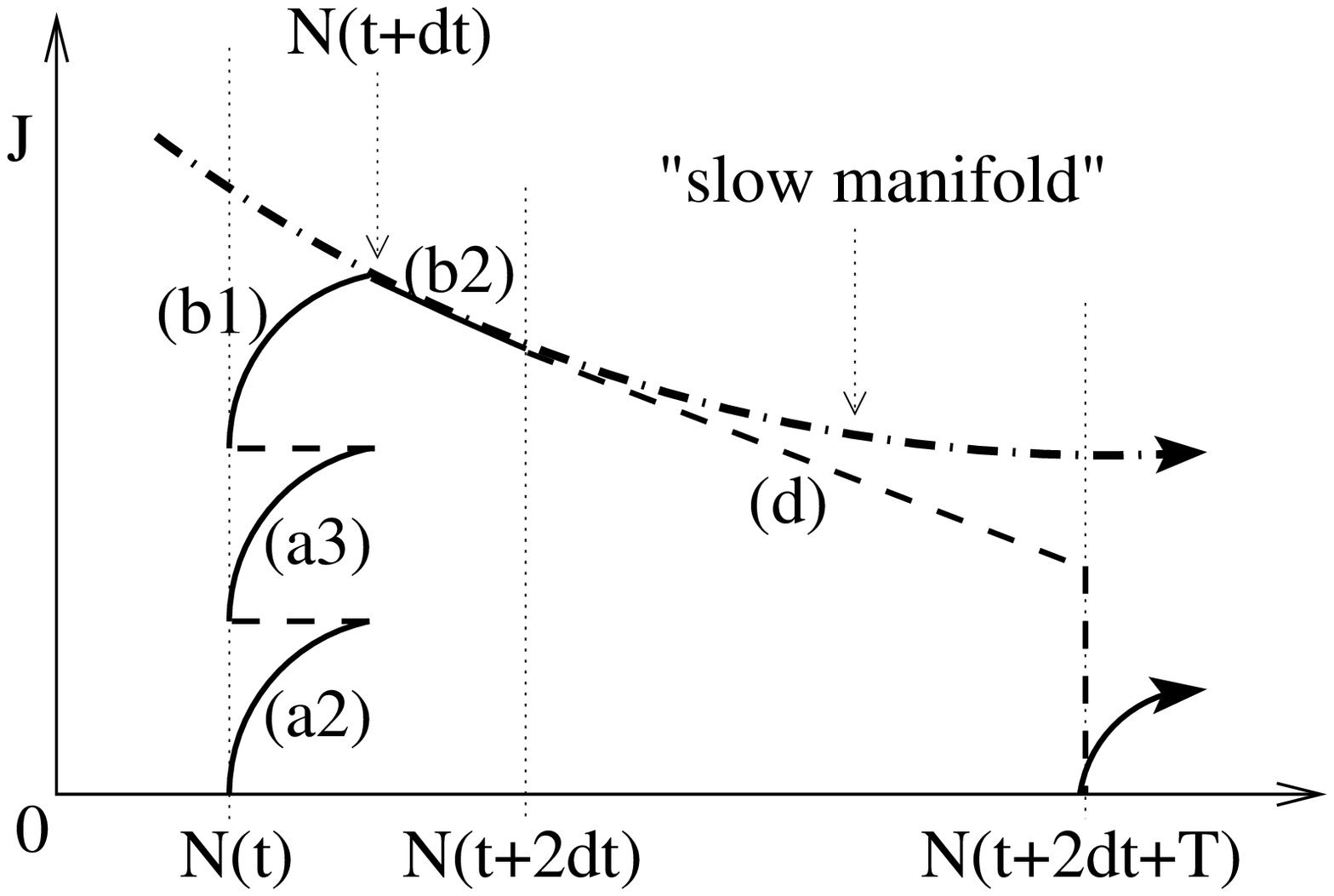}{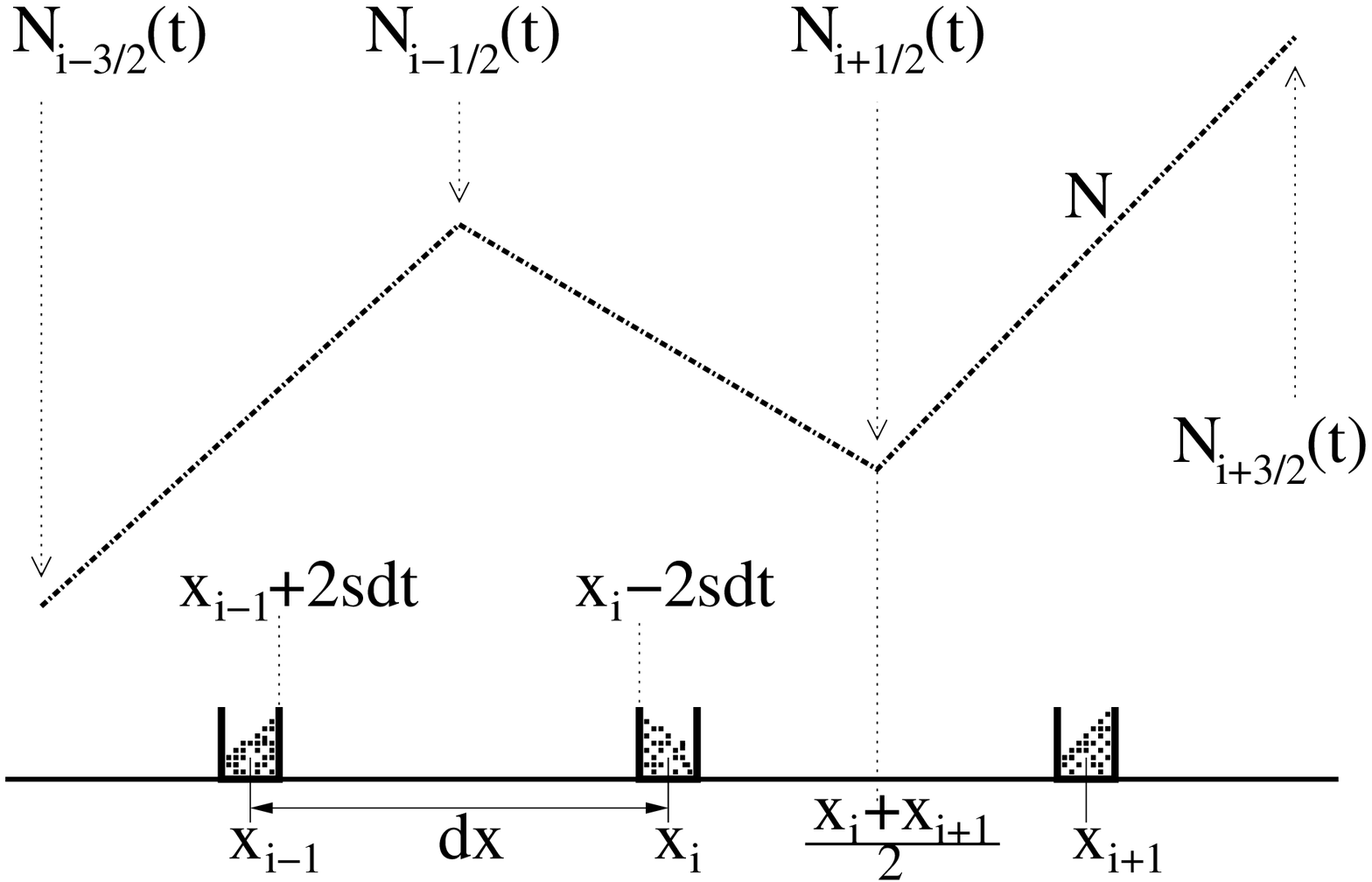}{2 in}
\caption{(a) {\it
Schematic of steps (a1) -- (d) of the coarse integration
algorithm. Monte Carlo simulation is denoted by solid lines. Dashed
lines denote relatively fast steps, {\em i.e.,} resetting the values
of density to $N(t)$ in steps (a2) and (a3) and the projective step
(d). The dot-dashed line represents the evolution on the slow manifold 
of the system.
We assume that the slow manifold can be accurately parametrized by
density.  } (b) {\it Schematic of three macroscopic mesh points
$x_{i-1}$, $x_i$ and $x_{i+1}$ and kMC computational domains around
them.  Only particles close to the mesh points need be considered; the
remaining particles will not leave/enter the interval $[0,x_i]$ during
steps (a1) -- (b2) and consequently, they do not have to be simulated.
At the top, is the (piecewise linear) estimated density profile which
is used in step (a1).  We place particles in the small computational
domains that their number is consistent with this density profile.  } }
\label{figukazkacoga}
\end{figure}

Since the histograms are noisy, we will work
with the integral of the density -- {\em i.e.,} with the 
cumulative density function defined by
\begin{equation}
\cdf(x,t) = \int_0^x N(x,t) \mbox{d}x. 
\label{CDFdef}
\end{equation}
Discretizing the interval $[0,2]$ using mesh (\ref{mesh}), we obtain
\begin{equation}
\cdf_i(t) \equiv \cdf(x_i,t) = \dx \sum_{k=1}^i N_{k-1/2} (t),
\qquad \quad
\mbox{and}
\qquad \quad
\cdf_i(t) - \cdf_{i-1}(t) = N_{i-1/2} (t) \dx.
\label{CDFeq}
\end{equation}
In particular, the number of particles in the interval 
$[0,x_i],$ $i=1, \dots, n$  is given by
\begin{equation}
n_0 \cdf_i(t) = n_0 \dx \sum_{k=1}^i N_{k-1/2} (t),
\qquad i=1, \dots, n.
\label{CDFeq2}
\end{equation}
In order to use coarse integration, it is important to compute the
change of $\cdf_i(t)$ during the time interval $[t, t + \dt]$;
equivalently, we want to know the change of the number of particles in
$[0, x_{i}]$ during the time step $[t, t + \dt]$.  Given that the
speed of the particles is $s$, only particles which are in the small
interval $[x_i-s\dt,x_i+s\dt]$ at time $t$ can enter or leave the
interval $[0,x_i]$. Consequently,
only a small number of particles around each mesh point
have to be simulated (compare Figure \ref{figukazkacoga}(b));
of course we are implicitly assuming that the discretization
mesh is fine enough so that interpolation between mesh points
provides an accurate estimate of the evolving density profile.
Using (\ref{choiceofdt}) and previous results, we choose 
\begin{equation}
\dt = 0.5, 
\qquad \quad
T = 99,
\label{choiceofdtandT}
\end{equation}
and  compute the cumulative density at time $t + 2 \dt + T = t + 100$
from the cumulative density function at time $t$ by the following 
algorithm (compare with Figure \ref{figukazkacoga}(a)
and Figure \ref{fig2})

\leftskip 1cm

\noindent
{\bf (a1)} Given a macroscopic initial cumulative density 
$\cdf(t)$ at mesh points (\ref{mesh}), we
compute the density $N_{i-1/2}(t)$ by the formula
$$
N_{i-1/2}(t) = \frac{\cdf_{i}(t) - \cdf_{i-1}(t)}{\dx},
\qquad
i = 1,\dots,n.
$$
We put $n_0 N_{i-1/2}(t) \dx$ particles in each interval 
$[x_{i-1},x_{i}]$ and distribute them so that the 
resulting probability density function is a continuous 
piecewise linear function with value $N_{i-1/2}(t)$ at point
$\frac{x_{i-1}+x_{i}}{2}$, $i=1,\dots,n$. Thus 
(see Figure \ref{figukazkacoga}(b))
$$
N(x,t) = N_{i-1/2}(t) + \frac{N_{i+1/2}(t) - N_{i-1/2}(t)}{\dx}
\left(
x-\frac{x_{i-1}+x_{i}}{2}
\right)
\quad
\mbox{for}
\;
x \in \left[ \frac{x_{i-1}+x_{i}}{2}, \frac{x_{i}+x_{i+1}}{2} \right].
$$
Moreover, we assign alternating velocities to the particles, so that the
initial flux is effectively zero.
As we mentioned earlier, we do not have to simulate all particles in 
$[\frac{x_{i-1}+x_{i}}{2}, \frac{x_{i}+x_{i+1}}{2}];$
instead, we consider only particles which are inside a small interval
$[x_{i} - 2 s \dt,x_{i} + 2 s \dt]$ around the macroscopic 
mesh point $x_i$ (this could be thought as analogous to
the gap-tooth scheme \cite{Gear:2003:GTM}, except that one does not
have to formulate and impose  effective smoothness boundary conditions,
see Figure \ref{figukazkacoga}(b)).

\noindent
{\bf (a2)} Evolve the system using the microscopic Monte Carlo
simulator for time $\dt.$ Then return the particles to their initial
position as given in (a1) but with a velocity equal to the values
computed in (a2) (this and the following are {\it preparatory} steps
to bring the microscopic initialization close to the slow
manifold).\footnote{This step annihilates the correlations between the
present and the initial velocities of a particle, and at the
macroscopic level the time required is essentially that in which a
hyperbolic equation rather a parabolic equation is needed at the
macroscopic level. This was already known to Einstein ({\em cf.}
\cite{Othmer:1976:SFP}).}
 
\noindent
{\bf (a3)} Repeat (a2) again, {\em i.e.,} evolve the system using the microscopic Monte 
Carlo simulator for time $\dt.$ Then return the position of particles to their 
initial values as given in (a1) keeping the velocities equal to computed 
velocities in (a3) (this can be repeated a few times).

\noindent
{\bf (b1)} Using the positions and velocities  produced at the 
end of step (a3), evolve the system using the 
microscopic Monte Carlo simulator for time $\dt.$ Compute the number
of particles in the interval $[0,x_i]$ at time $t + \dt$
for $i=1,2,\dots,n.$

\noindent
{\bf (b2)} Evolve the system using the microscopic Monte Carlo
simulator for another time  step 
$\dt.$ Compute the number of particles in the interval $[0,x_i]$
at time $t+ 2 \dt$ for $i=1,2,\dots,n.$

\noindent
{\bf (c)} Using data from (b1) and (b2), compute cumulative densities 
$\cdf(t+\dt)$ and $\cdf(t+ 2 \dt)$ at mesh points $x_0,$ $x_1,$ \dots, $x_n$
(this is the {\it restriction} step in equation-free computation).

\noindent
{\bf (d)} Estimate the time derivative 
\begin{equation}
\frac{\partial \cdf_i}{\partial t} = 
\frac{\cdf_i(t+ 2 \dt) - \cdf_i(t+\dt)}{\dt}
\label{timedercdf}
\end{equation}
and take an extrapolation (projective) step  
\begin{equation}
\cdf_i (t + 2 \dt + T) = \cdf_i (t + 2 \dt) + 
T \frac{\partial \cdf_i}{\partial t}.
\label{cdfproj}
\end{equation}
Then use $\cdf_i(t + 2 \dt + T)$ as the new initial condition in step 
(a1).

\leftskip 0cm

\noindent
The steps (a1)--(d) of the algorithm  are illustrated in Figure 
\ref{figukazkacoga}(a) where the slow manifold in density-flux 
space is shown as a dot-dashed line.  Note that the steps (a1) -- (a3)
correspond to the step (a) from Figure \ref{fig2}.  They are {\it
preparatory} steps used to initialize the flux close to the slow
manifold (since we assume that the flux equilibrates quickly); they
qualitatively correspond to evolving the macroscopic PDE for a short
time {\it constraining} the density profile to be the one we want to
prescribe as our macroscopic initial condition.
Such constrained 
evolution preparatory procedures (like ``umbrella sampling") are 
standard in computational chemistry 
\cite{Torrie:1974:MCF,Ryckaert:1977:NIC}.
A more detailed description of such initialization algorithms
in the case of legacy simulators can be found in 
\cite{Gear:2004:CDM,Gear:2004:PSM}.

The steps (b1) -- (b2) correspond to step (b) from Figure \ref{fig2}.
Moreover, (b1) corresponds to the step (\Prkm-1) and (b2) to the step
(\Prkm-2) from projective integration algorithm of (NJ).  Similarly,
steps (c) and (d) can be also found in Figure \ref{fig2}; moreover,
steps (c) and (d) together form step (\Prkm-3) of the projective
integration algorithm of (NJ).

If there is a small number of cells in one of the computational
domains, then the straightforward application of the algorithm (a1) -- (d) 
could give unrealistic results. 
For example, suppose that there 
are only two cells in the interval $[x_{i-1},x_i]$ at time
$t$, that the first cell moves to the interval $[0,x_{i-1}]$
during the time interval  $[t,t+2\dt]$, the second
cell moves outside the interval $[0,x_{i}]$ and that no other cell
crosses mesh points $x_{i-1}$ and $x_i$ during time interval  $[t,t+2\dt]$.
Then the time  derivative of the cumulative density function (\ref{timedercdf}) 
would be negative at point $x_i$ and positive at $x_{i-1}$. 
Moreover, the projected solution (\ref{cdfproj}) satisfies
$\cdf_{i-1}(t+2\dt+T) > \cdf_{i}(t+2\dt+T)$; consequently there is a
negative number of particles in the interval $[x_{i-1},x_i]$ at time
$t+2\dt+T.$ To avoid this problem we have to consider more
realizations for each computational domain containing a small number
of particles, and compute an average over this ensemble of
realizations.  Practically, if the number of particles $n_i$ in the
small computational domain around $x_i$ is less than a given number
$m$, we choose to repeat (a1) -- (d) for $m/n_i$ microscopic
realizations in this computational subdomain.

\begin{figure}  
\centerline{
\epsfxsize=3in\epsfysize=2.3in\epsfbox{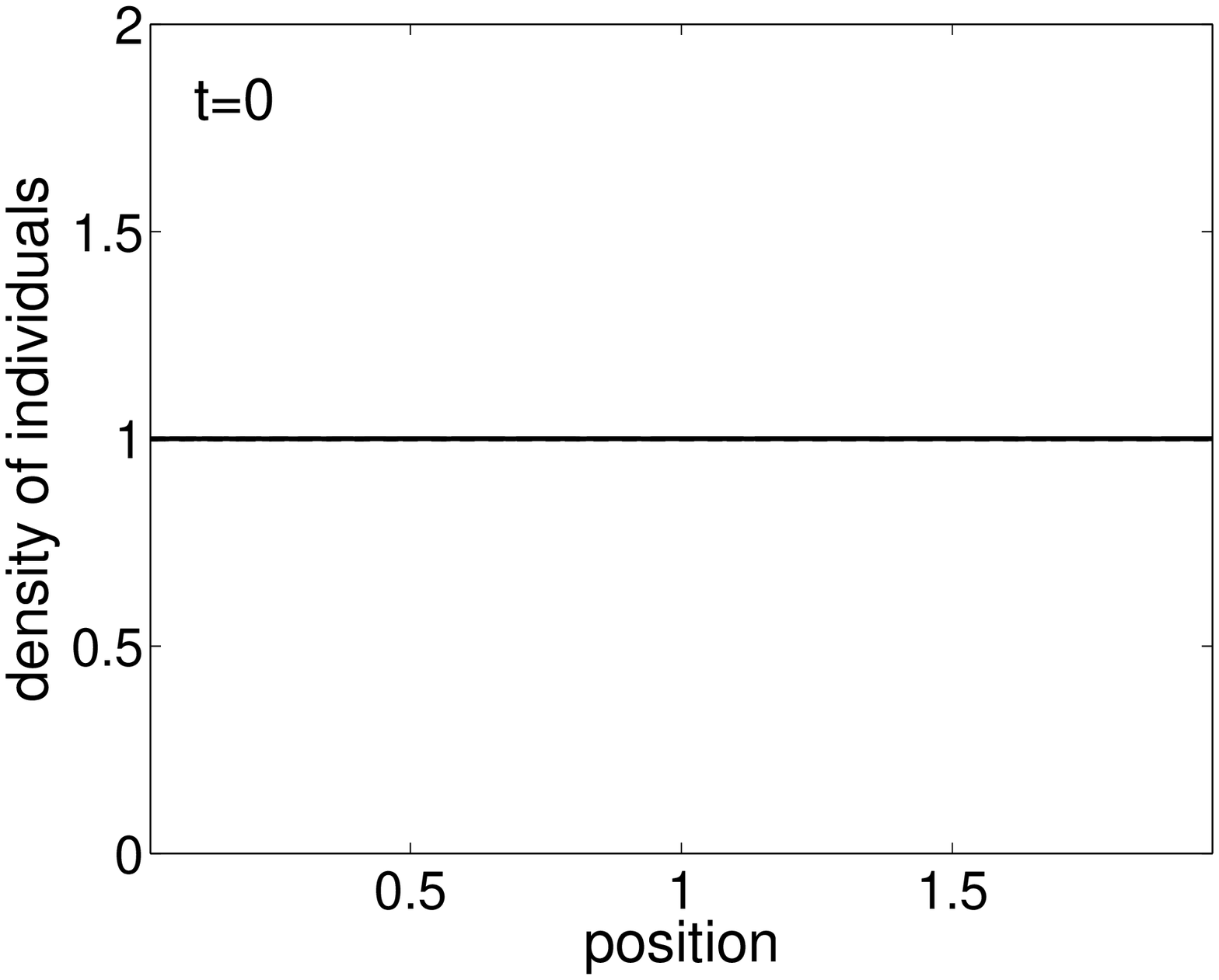}
\epsfxsize=3in\epsfysize=2.3in\epsfbox{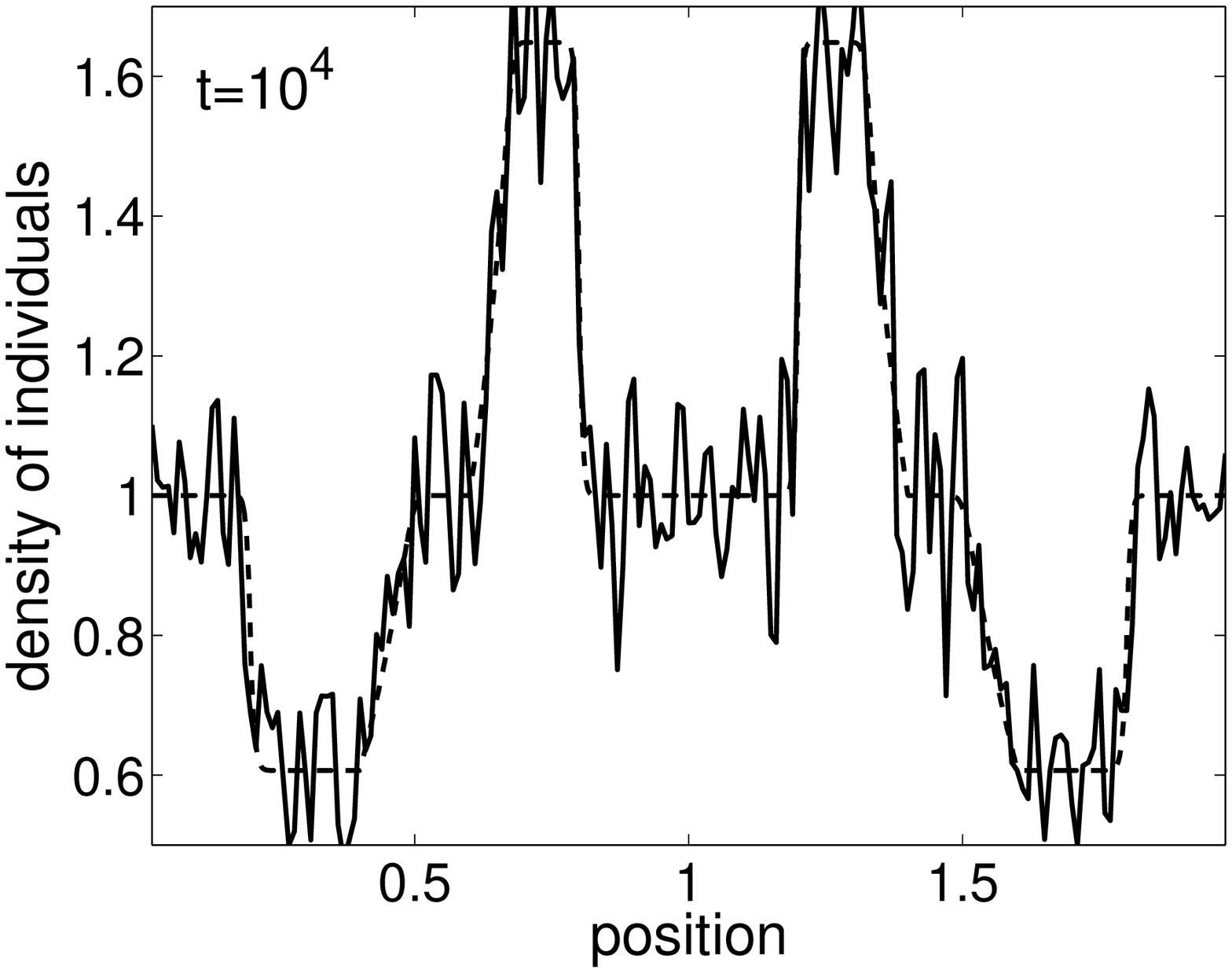}
}
\centerline{
\epsfxsize=3in\epsfysize=2.3in\epsfbox{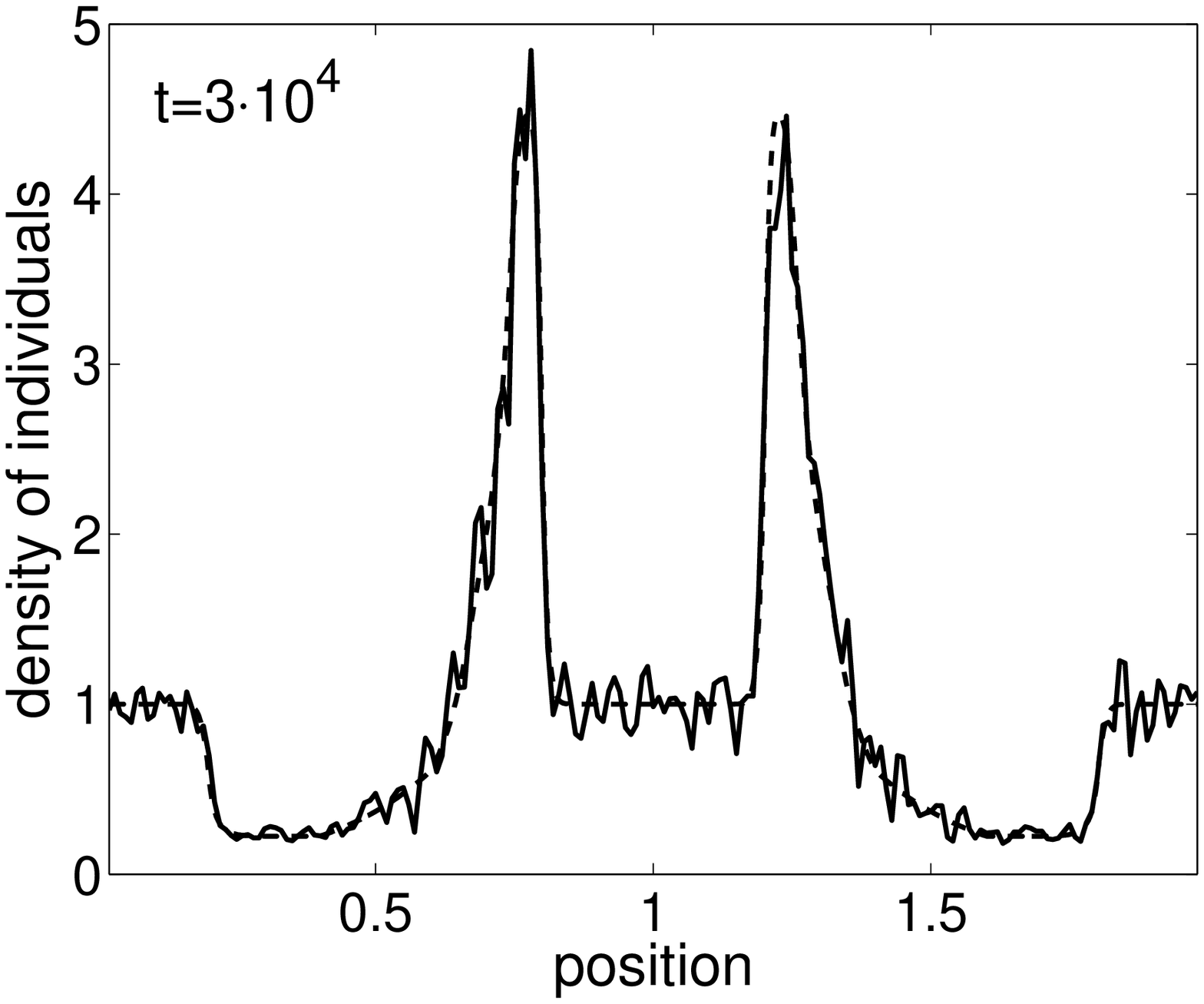}
\epsfxsize=3in\epsfysize=2.3in\epsfbox{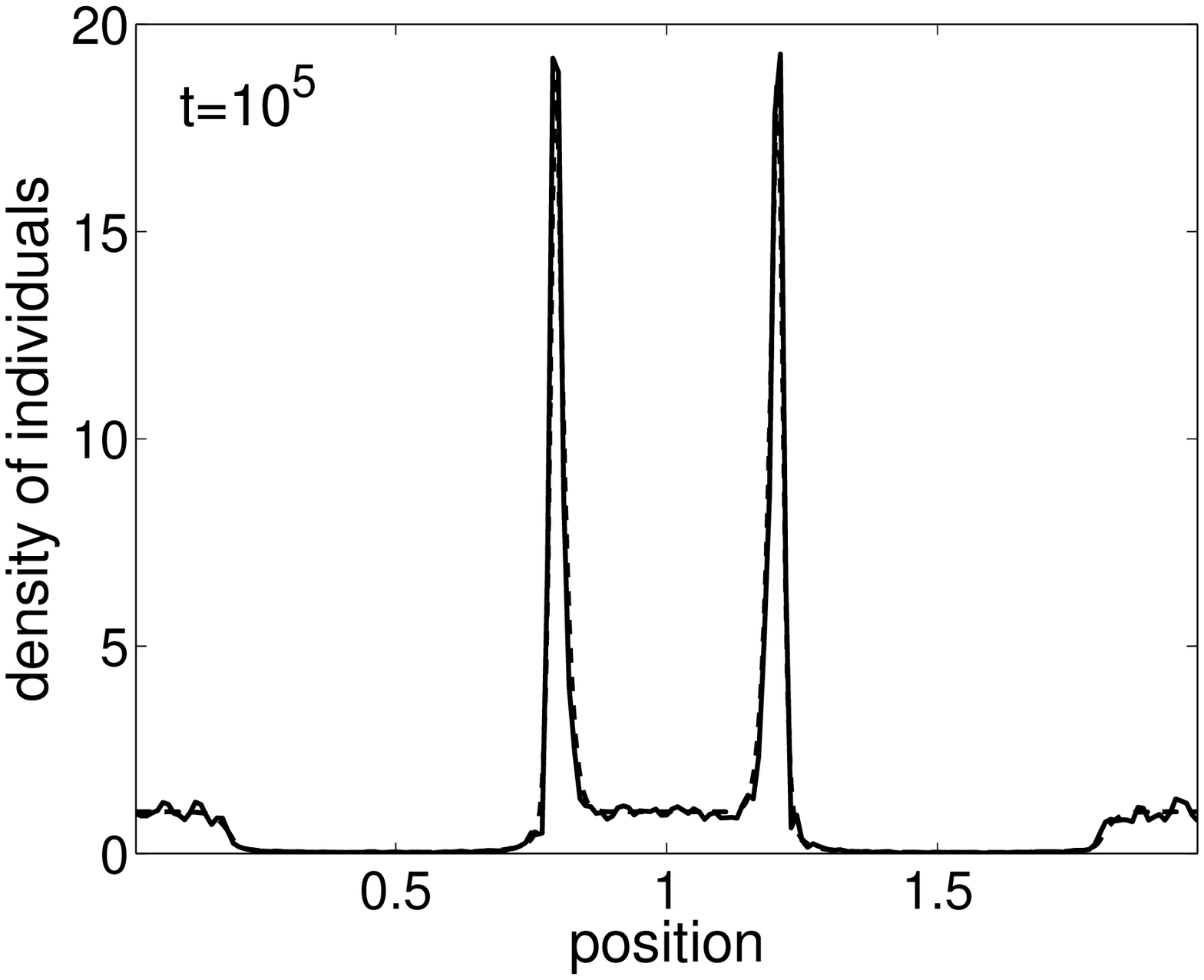}
}
\centerline{
\epsfxsize=3in\epsfysize=2.3in\epsfbox{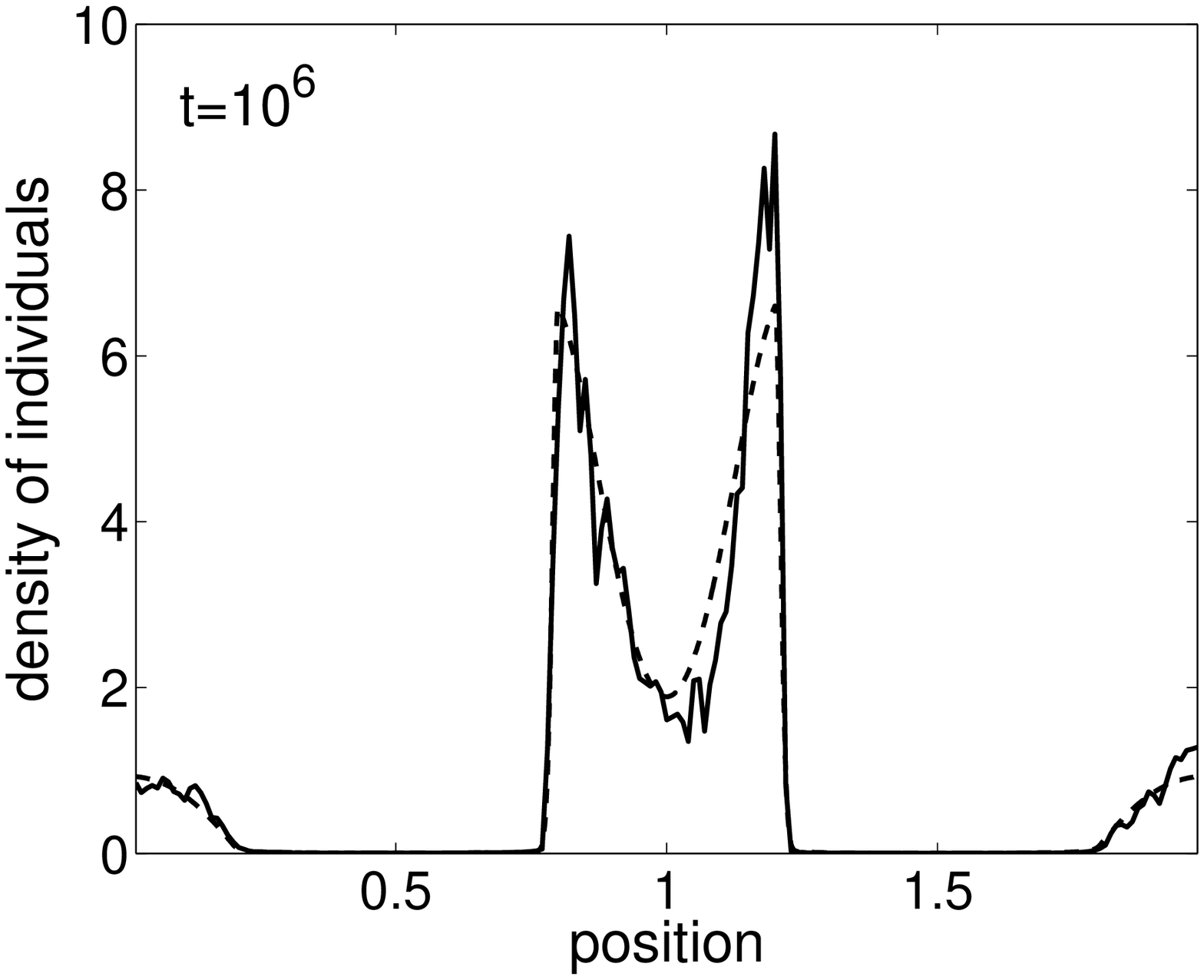}
\epsfxsize=3in\epsfysize=2.3in\epsfbox{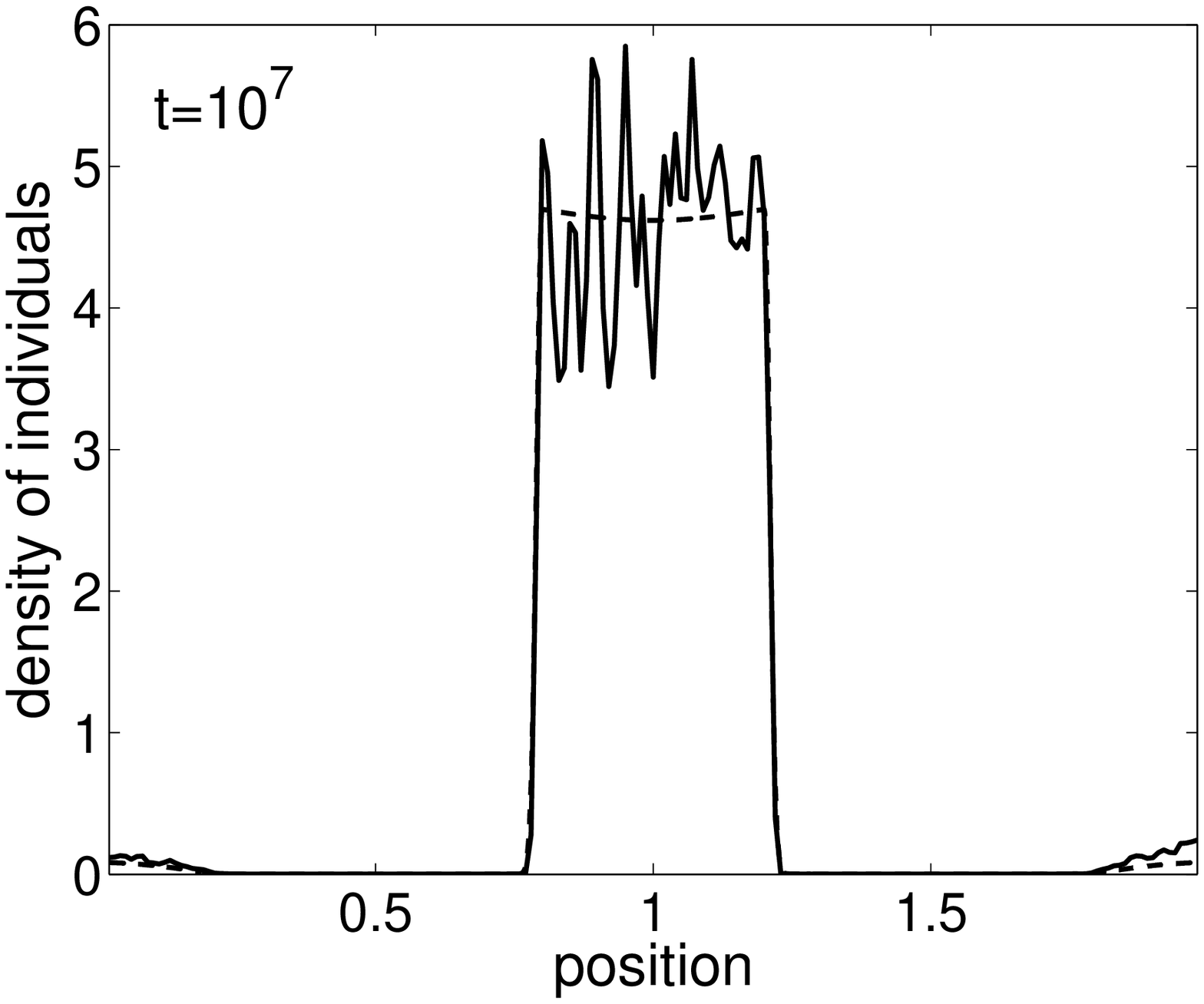}
}
\caption{ {\it The time evolution of the density of individuals 
for $(\ref{numparam3})$, $(\ref{choiceofdtandT})$,
$\alpha=0.1$ and initial conditions $(\ref{initcondsigcoms})$. 
We plot the density given by 
coarse integration $(a1)$ -- $(d)$ and obtained
by formula $(\ref{formNC})$ from the computed cumulative density
function $C(t)$ (solid line). Here we have gain $\ggain = 1250$. 
We also plot the solution of the corresponding
macroscopic moment equations (dashed line).
}}
\label{figukazkares}
\end{figure}

Numerical results for $\dt=0.5,$ $T=99$, 
signal strength $\alpha=0.1$ and $m=10000$
are given in Figure \ref{figukazkares}. 
There are two sources 
of gain for this method. 
First, we have the gain of the projective step. 
In one step (a1) -- (d), we compute the evolution of the system over 
time $T + 2 \dt = 100$ and we run the Monte Carlo simulator for time
$4 \dt = 2$ in steps (a2) -- (b2). 
Consequently, the gain factor of the projective step
is $(T+2\dt)/4\dt = 50$. 
The second part of the gain comes from the fact that important
particles (for the estimation of the evolution of a smooth
macroscopic density) are only those particles which are leaving/entering the interval
$[0,x_{i}]$ at the endpoint. 
From Figure 
\ref{figukazkacoga}(b), we see that only particles which are at time $t$ with distance
less than $2 s \dt = 0.0001$ from the endpoint can leave/enter the
interval $[0,x_{i}]$ during steps (b1) -- (b2).  Consequently, only
the fraction $2 n_0 \frac{4 s \, \delta t}{\delta x} = \frac{2
n_0}{50}$ of particles have to be simulated, and another factor of 50
appears in the gain.

Therefore, the combined gain of the coarse integration and
reduced spatial simulation (based on macroscopic density
smoothness) is $50 \times 50 = 2500$. 
However, 2500 is not
the actual gain $\ggain$ because some computational time was lost by
considering multiple microscopic realizations of domains which
contained a small number of particles. 
In any case,
we add less than $199 m$ particles to the simulation where
199 is the number of ``inner" computational domains 
and $m=10000$ is the minimal number of particles
in each of them. 
Consequently, we actually simulated more
cells than $\frac{2n_0}{50} = 2 \cdot 10^6$ but, at any
time, the number of simulated cells did not exceed 
$\frac{2 n_0}{50} + 199 M \sim 4 \cdot 10^6.$ 
So, in the
worst possible case, we slow down the computation by a factor of
2, which means that the total gain $\ggain$
of the method is at least $\ggain = 50 \times 50/2 = 1250.$

\begin{figure}[ht]  
\picturesABC{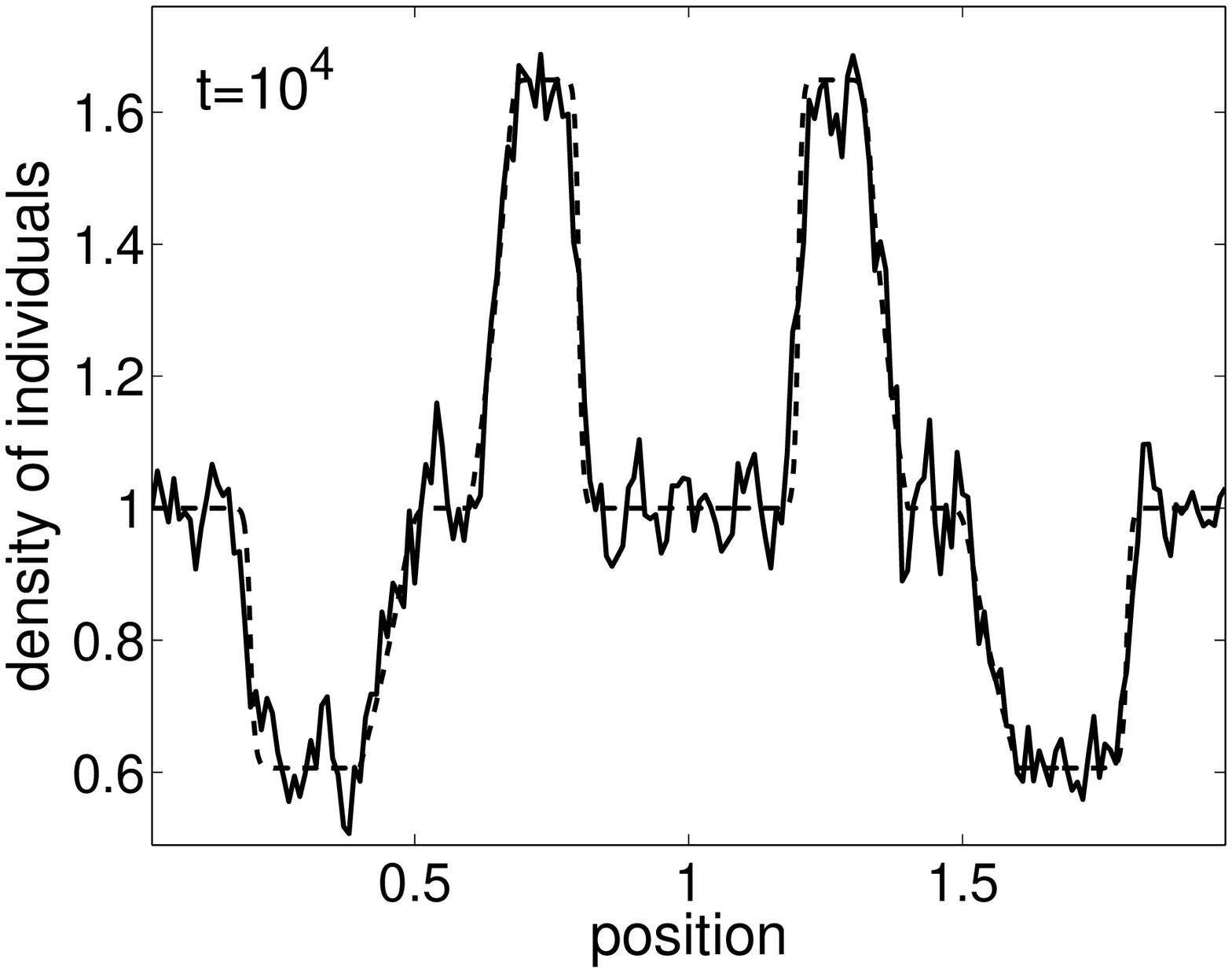}
{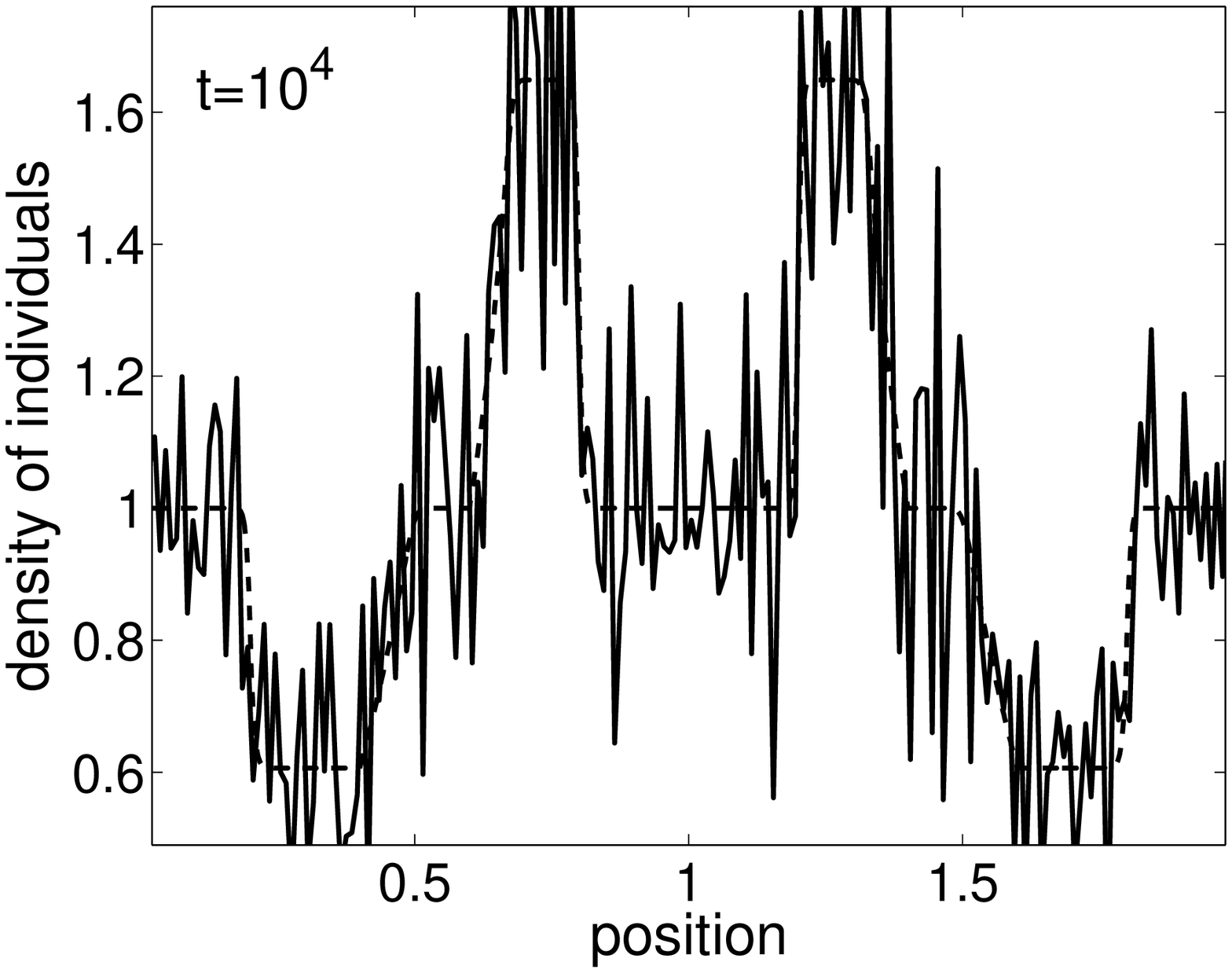}
{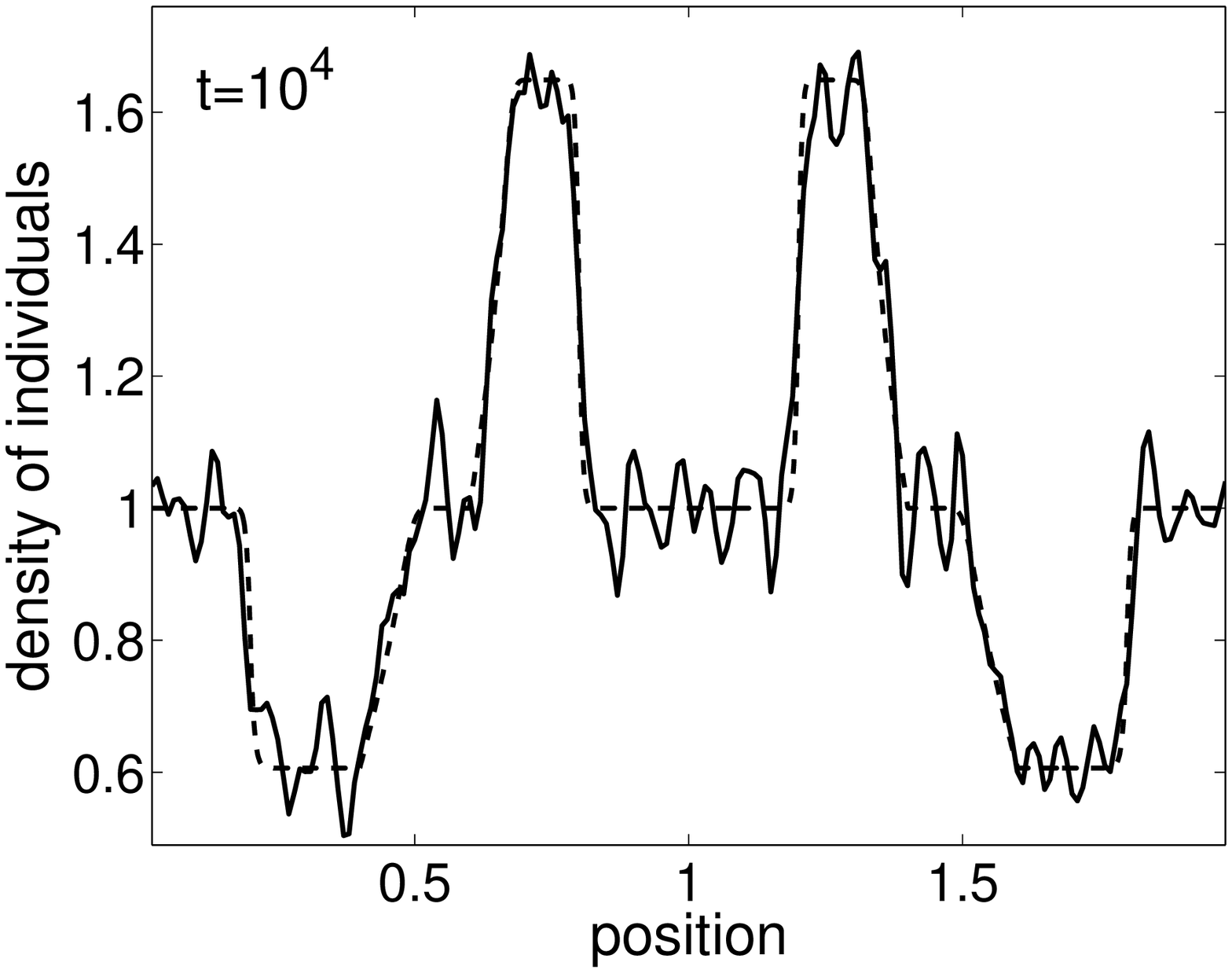}
{1.4in}
\caption{
{\it Plots of the density of individuals at time $t=10^4$ 
for the same cumulative distribution function as
in Figure \ref{figukazkares}, using different expressions for  for
computing the discretized density.}
(a) {\it formula} (\ref{formNC1});
(b) {\it formula} (\ref{formNC2});
(c) {\it formula} (\ref{formNC3}).
{\it We also plot the solution of the corresponding
macroscopic moment equations (dashed line).}
}
\label{figukazkaviz}
\end{figure}

In Figure \ref{figukazkares}, we present the time evolution
of the solution given by method (a1) -- (d) (solid line) compared to the
solution of the macroscopic PDE equations (dashed line). 
Since the algorithm (a1)--(d) computes cumulative density functions
$C(t)$ and we visualize the density $N(t)$ in Figure \ref{figukazkares},
the results are noisy and the plots depend on the formula which
is used to generate the density curves from the computed cumulative density
data. 
To be precise, in Figure \ref{figukazkares}, we show 
a plot of the function
\begin{equation}
N(x_i,t) = \frac{C_{i+1}(t) - C_{i-1}(t)}{2 \dx}.
\label{formNC} 
\end{equation}
Another representation  of the results is given in Figure
\ref{figukazkaviz},  where we show results
for time $t=10^4$ using different formulas for the density function $N$,
namely
\begin{eqnarray}
& \mbox{(a)} &  N(x_i,t) = \frac{C_{i+2}(t) - C_{i-2}(t)}{4 \dx}, 
\label{formNC1} \\
& \mbox{(b)} & N \left( \frac{x_i+x_{i+1}}{2},t \right) 
= \frac{C_{i+1}(t) - C_{i}(t)}{\dx},
\label{formNC2} \\ 
& \mbox{(c)} & 
N(x_i,t) = \frac{C_{i+2}(t) + C_{i+1}(t) - C_{i-1}(t)- C_{i-2}(t)}{6 \dx}.
\label{formNC3} 
\end{eqnarray}
Comparing plots in Figure \ref{figukazkaviz} and the corresponding
plot from Figure \ref{figukazkares}, we see how the visualization
of the results depends on the formula for estimating the
discretized density function $N(t)$ from the discretized 
cumulative density function $C(t).$ In particular, if we
use (\ref{formNC3}) instead of (\ref{formNC}) in Figure
\ref{figukazkares}, then the results will look less noisy.
Alternative discretizations of the particle density (namely, the use
of orthogonal polynomials to represent the inverse cumulative
distribution function, or ICDF) can be found in the literature
\cite{Gear:2001:PIM,Setayeshgar:2003:ACI}.  Techniques for estimating
smooth field profiles from noisy particle data have been proposed,
among other places, in the computational materials science literature
({\em e.g.,} the thermodynamic field estimator \cite{Li:1998:CCM}).

One can also decrease the noise in the computation, and the resulting
macroscopic field estimates, by considering multiple microscopic
realizations, {\em i.e.,} by increasing the value of $m$.  However, if
we increase $m$, then the gain $\ggain$ will decrease (obviously the
``wall clock" time of the overall computation remains the same if one does
these computations in parallel).

Certainly, there is a relationship between the 
initial number of particles $2n_0$, the minimal number of 
particles in each small computational domain $m$ and
the gain $\ggain$ of the method. 
If we
have a large number of particles $2 n_0$, then we can
choose a large $m$ without significantly decreasing the gain
of the method. 
On the other hand, if we have a stochastic
problem with a small number of particles $2 n_0$, then 
it may not be appropriate to consider a closed
PDE as a good model of a single system realization.
In our example, $m$ was chosen in such 
a way that the gain $\ggain$ of the method was decreased only by a factor
of two,  and thus the Monte Carlo simulation was accelerated
by a factor of at least 1250. 
Increasing $m$ would further decrease
the gain $\ggain$ and reduce the magnitude of the fluctuations.

\section{Discussion}

\label{secdiscussion}

In Section \ref{secmontecarlo} we analyzed an example in which a
simple coarse integration scheme was ``wrapped around" a kinetic Monte
Carlo simulation.  The short (in time) bursts of kMC simulation were
performed over only part of the full computational domain; this
provides another important factor in decreasing the overall
computational cost for such complex problems.  The idea of reduced
spatial as well as temporal simulation (the so called ``gap-tooth"
scheme and its combination with projective integration in ``patch
dynamics") is based on smoothness in the evolution of macroscopic
observables and constitutes a hallmark of equation-free computation.
Let us note that the computation of long term dynamics of our system
took several days on a IBM SP 375MHz Power3 processor using algorithm
(a1) -- (d).  Consequently, a computation using the kinetic Monte
Carlo simulator would take several years and was not attempted.  We
estimated the accuracy of coarse projective computations by comparing
to solutions of accurate macroscopic partial differential equations,
which in this example happened to be known.  When we do not have
population level equations, we must use standard {\it a posteriori}
error estimates to check accuracy and adaptively control the error of
our results as discussed below.
%
%

As we saw in Figure \ref{figeigenBcomplex} for matrix ${\cal B}$, the
length of the possible projective step $T$, as determined by stability
considerations, decreases with increasing strength of the signal
$\alpha$.  The same is true for algorithm (a1) -- (d).  If we increase
$\alpha$, then we have to decrease $T$ in order to have a stable
scheme.  In order to achieve stability for larger $T$ we could use a
similar strategy to that used for the matrix ${\cal A}$: we could
introduce artificial diffusion into the scheme which would make the
scheme stable, independently of the strength of the signal $\alpha.$
It is not difficult to design a coarse integration scheme with
artificial diffusion present; however, such a scheme would predict 
incorrect  dynamics for  the system. 

A better solution to the problem of large signal gradients is to note
that large signal gradients are typically localised only in small
parts of the domain of interest.  In fact, the problem with the coarse
integration scheme begins when a large signal gradient is present and
particles become highly localized in space.  Then the mesh is not fine
enough in certain small domains of interest (around peaks) but it is
sufficiently fine in the remainder of the interval $[0,2].$ Similarly,
the projective step is good for most of computational subdomains, but
it would lead to instabilities because of strong signal gradients for
a few of the computational domains.  One could conceivably adapt the
mesh, leading to a nonuniform mesh, finer in regimes with large signal
gradients and coarser otherwise.  Then we may need to make different
projective jumps in different parts of the domain of interest; issues
of this nature have been studied for nonuniform meshes in traditional
continuum numerical analysis using adaptive mesh refinement (AMR)
methods \cite{Berger:1984:AMR}, and in hybrid situations AMAR methods
\cite{Garcia:1999:AMA}.  Efficient implementations of such adaptive
techniques may be the key to significant acceleration of our
illustrative Monte Carlo scheme, since they would allow us to obtain
relatively accurate results for even larger sets of signal functions
and for problems where the signal is also altered by the cells.

 Detailed methods have been developed for adapting the computation to
the time and space scales of the problem in continuum numerical
analysis.  Adaptive stepsize selection in numerical integration, as
well as adaptive mesh refinement in spatial discretizations is an
indispensible part of modern software, and is typically based on {\it
a posteriori} error estimates of the solution accuracy computed on
line.  These methods can be naturally incorporated in equation-free
algorithms to control, for example, projective integration time steps
to control accuracy.  It should be noted that in addition to adaptive
time-step selection (for coarse projective integration) and adaptive
mesh selection (for gap-tooth algorithms), there is an additional type
of adaptivity that arises in equation-free computation. This is the
adaptive detection of the {\it level} of modeling, which may involve
augmenting or decreasing the number of variables needed for closure.
At a very qualitative level, adaptation of this ``level of
description" comes from the estimation of the gap between ``fast" and
``slow" system variables, which can be attempted using matrix-free
eigensolvers. By initializing the microscopic distribution using {\it
more} variables than the current level of modeling, one can try to
estimate the characteristic relaxation times of the additional
variables to functionals of the ones we need.  This allows one to
detect (while the level of description is still successful) whether
variables that are  treated as ``fast" are becoming ``slow", and should
be included as independent variables in the modeling.  A good
illustration of this is the evolution of stresses in a microscopic
simulator of fluid flow: for a Newtonian fluid stresses rapidly become
proportional to velocity gradients, while in non-Newtonian fluids this
is not true, and one must use more independent variables to model such
flows.  This could be considered analogous to closing bacterial
chemotaxis equations with only a single field (density) which can be
done for long time dynamics in some parameter regimes versus needing
two independent variables (right- and left- fluxes) to successfully
close system in some other cases.  In our case, the flux quickly
becomes functional of density, as can be seen directly from simulations. 

\noindent
A summary of the steps of our computational approach is as follows. 

\leftskip 2cm

\noindent
$\bullet \;$ identify the appropriate level of closure \hfill\break
$\bullet \;$ apply the equation-free computational algorithm \hfill\break
$\bullet \;$ do {\em a posteriori} error estimation 

\leftskip 0cm

\noindent
As we discussed above, we have to first
identify the level of closure, i.e. identify the slow dynamics of the
system which we want to model. Then we can do coarse projective
integration by making use of the spectral gap between fast and slow
modes of the system. As we saw, it can be natural or desirable to
combine coarse projective integration with gap-tooth methods, {\em
i.e.}  exploit the smoothness in physical space to only perform the
computations on relatively small subdomains. As a result one can do
transient calculations much faster than by direct simulations. If a
modeller is interested in steady states and the transient dynamics are
unimportant, then he or she can use other computational equation-free
techniques (such as application of Newton-GMRES method) to obtain
steady state behavior faster or do even bifurcation analysis
\cite{Gear:2002:CIB,Siettos:2003:CBD}. The final step
is {\it a posteriori} error analysis as suggested above. This is
an important issue if one wants to use our computational approach
for the problems where macroscopic equations are unavailable.

As we discussed, the large gain of the coarse projective integration
is governed by the large spectral gap between fast and slow eigenvalues
of the system. Our biological model system had such a large spectral gap 
because the mean running distance of individuals was much smaller than
the size of the domain of interest. The method has a potential to speed 
up other models of biological dispersal with similar properties.

\newpage
\bibliographystyle{amsplain}
\bibliography{bibrad}

\end{document}